\documentclass[tradiabstract]{aa} 

\usepackage{graphicx,txfonts}
\usepackage{natbib}

\newcommand{\Lsun} {L$_{\odot}$}
\newcommand{\Msun} {M$_{\odot}$}

\newcommand{\simgreat} {\mathbin{\lower 3pt\hbox{$\rlap{\raise
        5pt\hbox{$\char'076$}}\mathchar"7218$}}}

\newcommand{\simless}{\mathbin{\lower 3pt\hbox {$\rlap{\raise
        5pt\hbox{$\char'074$}}\mathchar"7218$}}}

\begin{document}

\title{Dark dust and single-cloud sightlines in the ISM}

\authorrunning{Siebenmorgen et al.}
\titlerunning{Dark dust and single-cloud sightlines in the ISM}

\author {R.~Siebenmorgen\inst{1}, J. Kre{\l}owski\inst{2}, J. Smoker\inst{3}, G. Galazutdinov\inst{4,5}, and S. Bagnulo\inst{6}}

\institute{ European Southern Observatory, Karl-Schwarzschild-Str. 2,
  85748 Garching, Germany. {\tt email: Ralf.Siebenmorgen@eso.org} \and
  Rzesz{\'o}w University, al. T. Rejtana 16c, 35-959, Rzesz{\'o}w,
  Poland \and European Southern Observatory, Alonso de Cordova 3107,
  Vitacura, Santiago, Chile \and Instituto de Astronomia, Universidad
  Catolica del Norte, Av. Angamos 0610, Antofagasta, Chile \and
  Pulkovo Observatory, Pulkovskoe Shosse 65, Saint-Petersburg 196140,
  Russia \and Armagh Observatory and Planetarium, College Hill, Armagh
  BT61 9DG, UK }

\date{Received: January 15, 2019 / Accepted:
  June 15, 2020}

\abstract{The precise characteristics of clouds and the nature of dust
  in the diffuse interstellar medium can only be extracted by
  inspecting the rare cases of single-cloud sightlines. In our
  nomenclature such objects are identified by interstellar lines, such
  as K\,{\sc i}, that show at a resolving power of $\lambda /\Delta
  \lambda \sim 75,000$ one dominating Doppler component that accounts
  for more than half of the observed column density.  We searched for
  such sightlines using high-resolution spectroscopy towards reddened
  OB stars for which far-UV extinction curves are known.  We compiled
  a sample of 186 spectra, 100 of which were obtained specifically for
  this project with UVES. In our sample we identified 65 single-cloud
  sightlines, about half of which were previously unknown. We used the
  CH/CH$^+$ line ratio of our targets to establish whether the
  sightlines are dominated by warm or cold clouds. We found that CN is
  detected in all cold (CH/CH$^+ >1$) clouds, but {is frequently
    absent} in warm clouds. We inspected the WISE ($3 - 22\, \mu$m)
  observed emission morphology around our sightlines and excluded a
  circumstellar nature for the observed dust extinction.  We found
  that most sightlines are dominated by cold clouds that are located
  far away from the heating source. For 132 stars, we derived the
  spectral type and the associated spectral type-luminosity distance.
  We also applied the interstellar Ca\,{\sc ii} distance scale, and
  compared these two distance estimates with GAIA parallaxes. These
  distance estimates scatter by $\sim 40$\,\%. By comparing spectral
  type-luminosity distances with those of GAIA, we detected a hidden
  dust component that amounts to a few mag of extinction for eight
  sightlines. This dark dust{\it } is populated by $\simgreat 1 \mu$m
  large grains and predominately appears in the field of the cold
  interstellar medium.}

\keywords{ISM: clouds -- ISM: lines and bands -- Line: profiles -- Stars: early-type -- (ISM) dust, extinction}

\maketitle


 \section{Introduction}
 The main physical and chemical characteristics of the interstellar
 medium (ISM) are obtained by studying six aspects. a) Interstellar
 extinction (reddening), which is caused by solid particles of
 submicron size \citep{Trumpler}. b) Interstellar polarisation
 \citep{HG80}, which is caused by non-spherical dust particles when
 they are oriented by the magnetic field, for instance. c) Spectral
 lines of interstellar atomic gas; see the UV survey by \cite{Field},
 for example, who proved that heavy elements in the ISM are strongly
 depleted in comparison to their abundances in stellar
 atmospheres. That is, they are hidden in dust grains. d) Rotational
 emission features that have been detected for many ($\sim 200$)
 complex molecules with a dipole momentum. e) Bands of simple
 molecules or radicals, such as {\scriptsize $^{\bullet}$}CH, CH$^+$,
 {\scriptsize $^{\bullet}$}OH, OH$^+$, {\scriptsize $^{\bullet}$}SH,
 {\scriptsize $^{\bullet}$}NH, {\scriptsize $^{\bullet}$}CN, C$_2$,
 C$_3$.  f) Diffuse interstellar bands (DIBs) that were discovered by
 \cite{Heger22}. A census of interstellar and circumstellar diatomics
 and molecules is given by \cite{McGuire18}.  Currently, the list of
 known DIBs exceeds 550 entries \citep{Fan19}, and most of them are
 very shallow. Finally, g) cosmic rays, which ionise parts of the ISM.
 Interstellar absorption spectra differ from cloud to cloud, and large
 variations in the dust properties are observed,
 e.g. \citet{Krelowski92}, \citet{FM07}, \citet{CK}, \citet{K19}.
 These differences show that all components mentioned above are
 interdependent, that is, variations in extinction and/or polarisation
 curves are accompanied with ratios of various strengths of atomic
 and/or molecular features.

Individual Doppler components can be resolved into interstellar atomic
and/or molecular lines by means of high-resolution spectroscopy, while
extinction and continuum polarisation are necessarily averaged along
any sightline. Profiles of interstellar features observed at extremely
high resolving power of $\lambda /\Delta \lambda \sim 10^6$ with the
Ultra-High Resolution Facility \citep[UHRF,][]{Diego95} or the
McDonald \citep{Tull95} instruments reveal multiple Doppler
components, e.g. \cite{Crane95}, \cite{Barlow95}, \cite{Price00},
\cite{Crawford02}, \cite{Welty01}, \cite{Welty03}.  Strictly speaking,
only very few and nearby sightlines can be considered genuine
single-cloud sightlines that do not display fine structures in the
Doppler profiles of interstellar lines. Clouds are not static
entities, however. Dynamically, they may be colliding, merging, or be
disrupted. Velocity shears, shells, bubbles, and filamentary
structures are observed.  Cloud structures are impacted by cosmic rays
that change the ionisation structures of the atoms and
molecules. Magneto-hydrodynamical waves generate small-scale cloud
structures \citep{FalleHartquist02}. Shocks propagating through the
ISM can ablate or destroy the clouds \citep{Klein94}. In some clouds,
self-gravitation is at work and forms cloudlets \citep{Wada08}, which
are observed as time-variable interstellar absorption lines, e.g.
\citet{Falgarone91}, \citet{Lauroesch00}, \citet{Price00},
\citet{Welty01b}, \citet{Crawford02}. Fractal cloud structures
\citep{Elmegreen02} arise from typically subsonic ($\simless
0.7$\,km/s) turbulent motions, e.g. \cite{Barlow95}, \citet{Welty01}.

Our project is tailored to study global dust characteristics in
clouds, and we do not consider local small-scale variations in the
detailed cloud morphology. We assume that a cloud has approximately
similar physical dust parameters in grain material, abundance, and
sizes, and that grain alignment might be triggered by the global
magnetic field structure. In contrast, these dust parameters are
assumed to vary on a large scale when clouds are well separated by
several 10s or 100s of pc. We use the term single-cloud sightline
when the observed interstellar line profiles show one dominant Doppler
component in high-resolution spectra at a resolving power of $\lambda
/\Delta \lambda \sim 75,000$ (full width at half maximum, FWHM $\sim ~
4$\,km/s) that accounts for more than half of the observed column
density. Of course some of those single-cloud sightlines may include
two or more fine-structure components with slightly different radial
velocities.  \citep{Welty14}. In fact many such objects are nearby and
so the Doppler components must be very close to each other. This
proximity is followed by similar spatial motions of the intervening
clouds which leads to similar radial velocities. Sightlines with
several clouds that are close to each other show similar extinction
properties as observed towards associations such as Sco OB2 and Per
OB2 \citep{FM07}. Therefore, the same properties of the dust and
presumably that of the global magnetic field can be assumed. Hence, we
apply a less strict nomenclature of single-cloud sightlines as
compared to the terminology used by \cite{Welty03}.

\citet{S18} have demonstrated that when sightlines are observed that
intersect different components (which likely do not resembling each
other), the pristine nature of the extinction and polarisation curve
of each individual cloud cannot be recovered. The possibility of
investigating the relation between the physical parameters of the dust
is lost, as are the observational characteristics of the extinction
and polarisation. The only way to investigate these relations is to
inspect sightlines that are dominated by single clouds.  The Large
Interstellar Polarisation Survey \citep[LIPS,][]{Bagnulo17} allowed us
to simultaneously fit the extinction and polarisation curves of 59
sightlines. \citet{S18} applied a dust model composed of silicate and
carbon grains, finding that large ($>$6\,nm) spheroidal silicate
particles, which are of prolate shape, account for the observed
polarisation curves. For 32 sightlines we complemented the LIPS data
set with UVES archive high-resolution spectra. This enabled to extract
a small number of 8 single-cloud-dominated sightlines. This study
confirmed several correlations between the observed extinction and
polarisation characteristics and the physical parameters of the dust,
and several previously unknown correlations were found that are
significant in single-cloud sightlines alone. This demonstrates the
validity of our approach. It was observed that interstellar
polarisation from multiple clouds is depolarised and therefore lower
than that from single-cloud sightlines, and large variations in the
dust characteristics from cloud to cloud is detected.  However, when a
number of clouds are averaged, a similar mean of the dust parameters
is retrieved that represents what is called the Milky Way mean
extinction curve; this is an ill-defined average.

In this paper we present a high-resolution spectroscopic search for
such single-cloud sightlines at a distance of a few kiloparsec in the
solar neighbourhood{} for which the far-UV extinction curve is known.


\begin{table*}[!htb]
\scriptsize
\begin{center}
  \caption {Study sample sorted by HD number\label{sample.tab}.
    Column~(1) gives the identification number, Col.~(2) the HD
    identifier, Col.~(3) if different, the main name of the star as
    used in {\sc Simbad}, Col.~(4) the V magnitude, Col.~(5) the
    reddening E(B-V), Col.~(6) the visual extinction $A_{\rm V}$,
    Col.~(7) the total-to-selective extinction $R_{\rm V}$, Col.~(8)
    the spectral type from previous literature, Col.~(9) the spectral
    type as derived in this work, and Col.~(10) the instrument used to
    obtain high-resolution spectroscopy.}
  \begin{tabular}{r l l c c c c l l l}
\hline\hline
1  & 2    & 3            & 4 &     5                   &     6                &            7           & 8     & 9           & 10 \\
\hline
ID & Name & {\sc Simbad} &$V$& $E_{\rm {(B-V)}}$          & $A_{\rm V}$       &  $R_{\rm V}$      &  Spectral type  & Spectral type  & Instrument \\
&      &              & mag  &                          &   mag                &                  &  literature & this work  &                 \\
\hline
1 & HD~023180  & * omi Per  & 3.83 &  	0.29$\pm$  0.03 &   0.91$\pm$  0.08 &   3.14$\pm$  0.31	&  B1IV$^{\rm{F}}$       &  B1III              & UVES$^{b}$     \\
  2 & HD~024263  & * 31 $\tau$& 5.69 &  	0.21$\pm$  0.06 &   0.72$\pm$  0.22 &   3.44$\pm$  0.65	&  B5V$^{\rm{V}}$	&  B3.5V + binary    & UVES$^{a}$     \\
  ... &            &          &      &                   &                   &                  &                    &         & \\
136 & HD~326364  &            & 9.60 &  	0.62$\pm$  0.03 &   1.80$\pm$  0.08 &   2.91$\pm$  0.25	&  B0IV$^{\rm{F}}$	&  {B1IV}    + cloud  & UVES$^{a}$     \\
\hline
\end{tabular}
\end{center}
    {\bf Notes:} Cols.~5 -- 7 extracted from \cite{Gordon09},
    \cite{FM07}, with uncertainties revised as explained in the text.
    Spectral types as derived by $^{\rm {F}}$ \cite{FM07}, $^{\rm
      {G}}$ \cite{Gordon09}, $^{\rm {H}}$ \cite{Houk75}, $^{\rm {HS}}$
    \cite{Houk99}, and $^{\rm {V}}$ \cite{Valencic}.  Spectroscopic
    data from: ${a}$: this work; ${b}$: \cite{Cox17}; ${c}$: UVES programme
    ID 096.D-0008(A) (unpublished ESO archive); ${d}$: UVES programme ID
    099.C-0637(A) (unpublished ESO archive); ${e}$: Bagnulo et
    al. (2013); ${f}$: \citet{Sembach}; ${g}$: \cite{Welty10}; ${h}$:
    \cite{Tolstoy03}; ${i}$: \citet{K10}; ${j}$: SAO 1.0\,m; ${k}$:
    SOPHIE archive; ${l}$: ELODIE archive, ${m}$: BOES
    \citep{Kim07}. For the complete list, see
    Table~\ref{appsample.tab}.
\end{table*}

\section{Observations}\label{observations.sec}
Our target selection is aimed at identifying a sample of single-cloud
sightlines that are suited for detailed studies of variations in dust
properties in the diffuse ISM. For these single-cloud sightlines we
require the wavelength dependence of the extinction and
polarisation. Therefore we started by considering all 544 sightlines
for which the far-UV extinction curves between $0.33 - 0.09\,\mu$m
were derived from the International Ultraviolet Explorer \citep[IUE,
][]{FM07}, \citet{Valencic} and/or the Far Ultraviolet Spectroscopic
Explorer \citep[FUSE, ][]{Gordon09} satellite data. For these stars we
performed an extensive search of available optical high-resolution
spectra ($\lambda /\Delta \lambda \ga 40,000$), which enables studying
the interstellar line profiles. We found archive spectra for 86 
sightlines and conducted new observations for 100 sightlines with the
UVES spectrometer, e.g. \citet{Dekker00} and \citet{Smoker09}.

There are 136 sightlines that have high-resolution spectra with a
broad wavelength coverage that is appropriate for a detailed
analysis. They are catalogued in Table~\ref{sample.tab}, where we
specify (1) the identification number (ID) used in this paper, the
target name for which we favour (2) the HD identifier, and if it is
different, provide the (3) main name as used in {\sc Simbad}, (4) the
$V$ -band magnitude, (5) the reddening $E(\rm{B - V})$, (6) the
extinction $A_{V}$, (7) the total-to-selective extinction $R_{V}$, (8)
the previous classification of the spectral type with its literature
reference, (9) our spectral classification based on the
high-resolution spectroscopy (Sect.~\ref{spectype.sec}), and (10) the
instrument used to obtain high-resolution spectroscopy. The
uncertainties for $A_{V}$, $E(\rm{B - V}),$ and $R_{V}$ given in
Table~\ref{sample.tab} take the scattering between different estimates
found in the literature, e.g. \citet{Wegner}, \citet{FM90},
\citet{Valencic}, \citet{Gordon09} for the same star into account. We
added a systematic error of 0.05, 0.02, and 0.2 magnitudes to the
uncertainties for $A_{V}$, $E(\rm{B - V}),$ and $R_{V}$,
respectively. We examined the mid-IR imaging for these stars.

We found accompanying UVES and FEROS spectra of {50} stars that
include the K\,{\sc i} line at 7699\,\AA \/.  The K\,{\sc i} line, as
we show below, is a good indicator for classifying a sightline that is
dominated by a single or multiple velocity components. These stars
(IDs 137 -- 186) are listed separately in Table~\ref{KI.tab} together
with the ESO programme ID as reference and fit parameters of the
K\,{\sc i} line profile.  The data were fitted using the {\sc vapid}
suite \citet{Howarth02} from which the velocities v$_{\odot}$, column
densities $N$, and broadening $b$ were estimated assuming instrumental
FWHM from the slit width for the UVES data or for 48,000 for the three
FEROS sightlines (HD\, 112244, HD\, 151804 and HD\, 166734). The line
broadening parameter is defined by \citet{Spitzer} being $b = 0.6 \
\times $ FWHM of the line. Fitting was redone with the instrumental
FWHM decreased by 15 percent as an estimate of the systematic error
which was added in quadrature to the errors derived by {\sc vapid}. We
also included in the error budget a systematic error of 0.3 km/s in
the UVES velocities.

\begin{table*}[!htb]
  \scriptsize
  \begin{center}
    \caption{UVES and FEROS archive spectra covering the K\,{\sc i}
      line. \label{KI.tab} Column~(1) gives the identification
      number, Col.~(2) the HD identifier, Col.~(3) the ESO programme
      ID, Cols.(4-6) the fit parameters of the K\,{\sc i} line
      profile, and Col.~(7) the instrumental FWHM. Single-cloud
      sight-lines are highlighted in boldface. }
\begin{tabular}{clcrrrc}
\hline  \hline
1    & 2    & 3         & 4             &  5   & 6 & 7 \\
ID   & Star & Prog. ID  & v$_{\bigodot}$  & $b$ &  log($N$) & FWHM$_{\rm{ins}}$ \\
            &                         &                       &      km/s        &   km/s         &   cm$^{-2}$  & km/s         \\
\hline
       137  &{\bf {\bf {\bf CPD\,592600}}}  &         071.C-0513(C)  &   -17.63$\pm$0.95  &   6.08$\pm$1.39  &  10.62$\pm$0.08  &  2.7 \\
            &                        &                        &    -4.73$\pm$0.32  &   1.70$\pm$0.31  &  10.99$\pm$0.02  &  2.7 \\
            &                        &                        &     1.81$\pm$0.30  &   1.50$\pm$0.20  &  11.74$\pm$0.03  &  2.7 \\
       138  &            CPD\,573509  &         094.D-0355(A)  &    -3.25$\pm$0.30  &   1.34$\pm$0.24  &  12.05$\pm$0.10  &  2.7 \\
            &                        &                        &     9.22$\pm$0.39  &   3.24$\pm$0.40  &  11.14$\pm$0.04  &  2.7 \\
 ...    &      &           &               &      &   &   \\
       186  &   {\bf {Walker\,67}}  &         092.C-0019(A)  &    26.82$\pm$0.31  &   0.67$\pm$0.67  &  12.30$\pm$0.49  &  2.7 \\
\hline
\end{tabular}
\end{center}
{\bf Notes: } For the complete list, see Table~\ref{appKI.tab}.
\end{table*}


\subsection{New UVES observations}
We performed UVES observations of 100 stars using the standard
390$+$760 setting, which led to a wavelength range of approximately
3270 -- 4450\,\AA \, in the blue arm, 5700 -- 7520\,\AA\, in the lower
red arm, and 7660 -- 9460\,\AA\, in the red upper arm, with some gaps
between the spectral orders. The setting was chosen to cover the
Na\,{\sc i} lines at 3302.4, 3303, 5890, and 5895.9\,\AA, the\,
Ca\,{\sc ii} H and K doublet at 3933.7 and 3968.5\,\AA, K\,{\sc i} at
7699\,\AA, the strong DIBs at 5780 and 5797\,\AA, as well as a host of
other interstellar lines. When our signal-to-noise ratio (S/N) was
sufficient, these other lines include Ti\,{\sc ii} 3383.8\,\AA, CH
4300.3\,\AA, CH$^+$ 4232.5\,\AA, and CN 3874.6\,\AA. The typical S/N
per pixel for the reduced data was 130 around Ca\,{\sc ii}
(3933.6\,\AA) and 200 around Na\,{\sc i} (5890\,\AA). The slit width
was set to 0.5$^{\prime\prime}$, which provided a spectral resolution
(as measured from telluric lines) of $\lambda / \Delta \lambda \sim
75,000$. We note that \cite{Welty94, Welty96} detected Ca\,{\sc ii}
and Na\,{\sc i} components separated by as little as 0.5\,km/s; these
components would remain unresolved in our data.

\begin{figure}
    \includegraphics[width=\columnwidth]{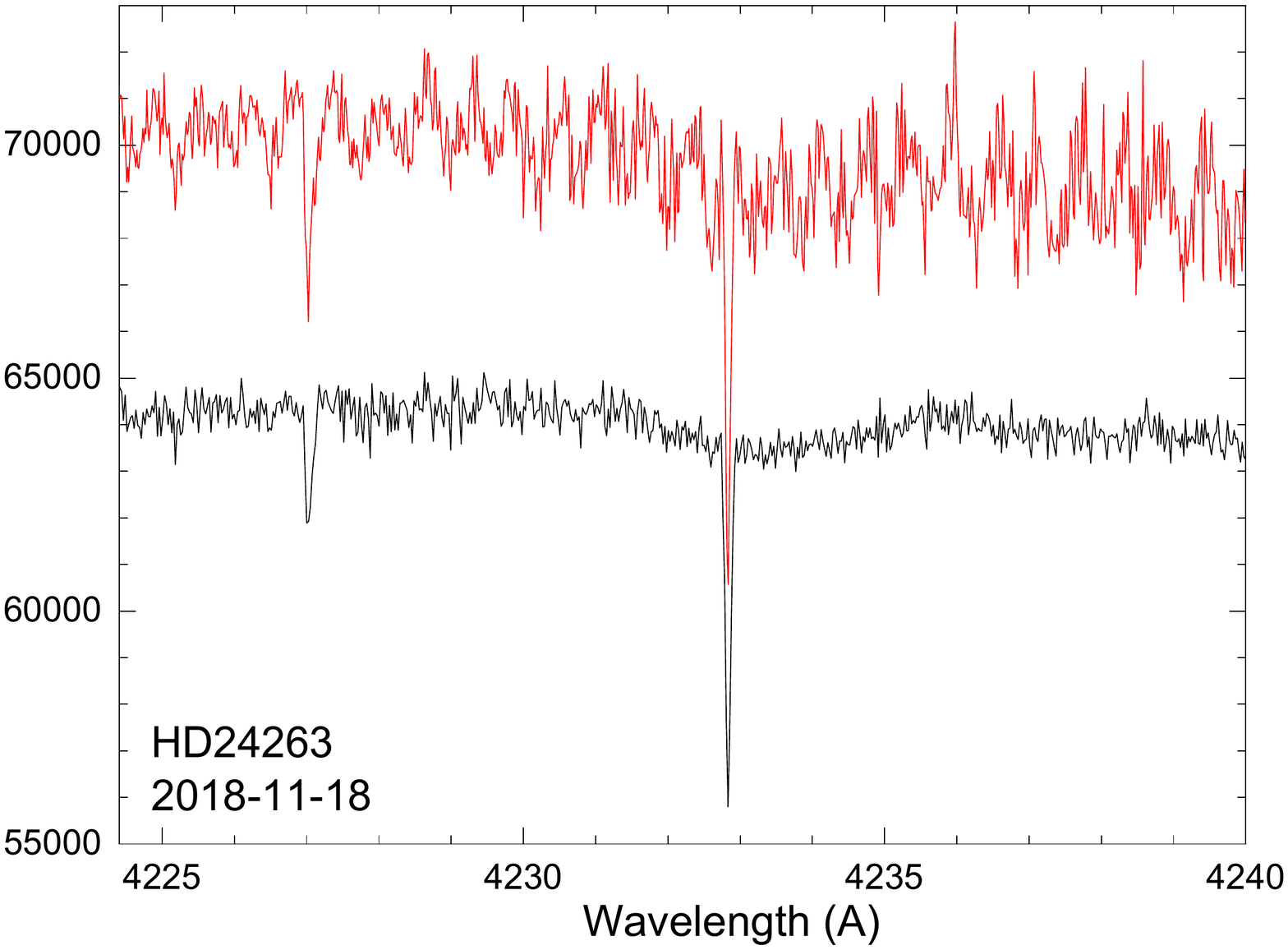}
    \caption{Small region of the spectrum of the HD~024263 sightline obtained with
      the aid of the UVES pipeline (upper line, in red) and with our own  interactive
      analysis system {\sc dech} (lower line, in black). The weak feature shortward of the
      CH$^+$ line is the feature of
      $^{13}$CH$^+$.  \label{dechvspipeline}}
\end{figure}

We processed UVES raw data and measured line profiles in the reduced
spectra with our interactive analysis software {\sc dech}\footnote{available
  upon request to G. Galazutdinov.}. The automatic data reduction of
the UVES spectra by the ESO pipeline may show several shortcomings
that might affect the quality of the extracted spectra. We note first
an imperfection of the algorithm that automatically finds the position
of the spectral orders; second,{\it } a sub-optimal setting of the
integration limits; and third, a robust but simplified algorithm that
averages different spectra of the same star in which low-quality data
are not excluded.  In 2 out of 126 UVES spectra, the unsupervised
automatic extraction with default parameters failed. In
Fig. \ref{dechvspipeline} we show the benefit of performing an
interactive data analysis as compared to an automatic reduction.

For the {\sc dech} data reduction, we averaged bias images for subsequent
correction of flat-field, wavelength calibration, and stellar spectra.
The scattered light was determined as a complex two-dimensional
surface function that was individually calculated for each stellar and
flat-field frame by a cubic-spline approximation over manually
selected areas of minima between the spectral orders.  The
pixel-to-pixel variations across the CCD were then corrected by
dividing all stellar frames by the averaged and normalised flat-field
frame. One-dimensional stellar spectra were extracted by simple
summation in the cross-dispersion direction along the width of each
spectral order. The extracted spectra of the same object observed in
the same night were averaged to achieve the highest S/N. Fiducial
continuum normalisation was based on a cubic-spline interpolation over
the interactively selected anchor points. The wavelength scale of the
spectra was calculated on the basis of a polynomial,

\begin{equation}
\lambda(x,m) = \sum_{i=0}^{k} \sum_{j=0}^{n} a_{ij} x^{i} m^{j} \/,
\end{equation}

where $a_{ij}$ are polynomial coefficients, $x^i$ is the pixel
position in dispersion direction, and $m^j$ is the order number.
Depending on the spectrograph arm, we typically used 700 and up to
1200 lines of the thorium lamp in the final wavelength solution. The
rms residual error between the fit and the position of the lines was
usually $\leq 0.003$\,\AA, that is, much lower than 1\,km/s. Because the
wavelength calibration was not taken directly after the observations,
but on the morning after, the absolute error is somewhat
larger. Radial velocities and equivalent width were measured using
Voigt multiple-component profile fits and the direct integration methods
available in the {\sc dech} code.


\begin{figure*} [h!tb]
  \includegraphics[width=18.5cm,clip=true,trim=0cm 3cm 0cm 0.cm]{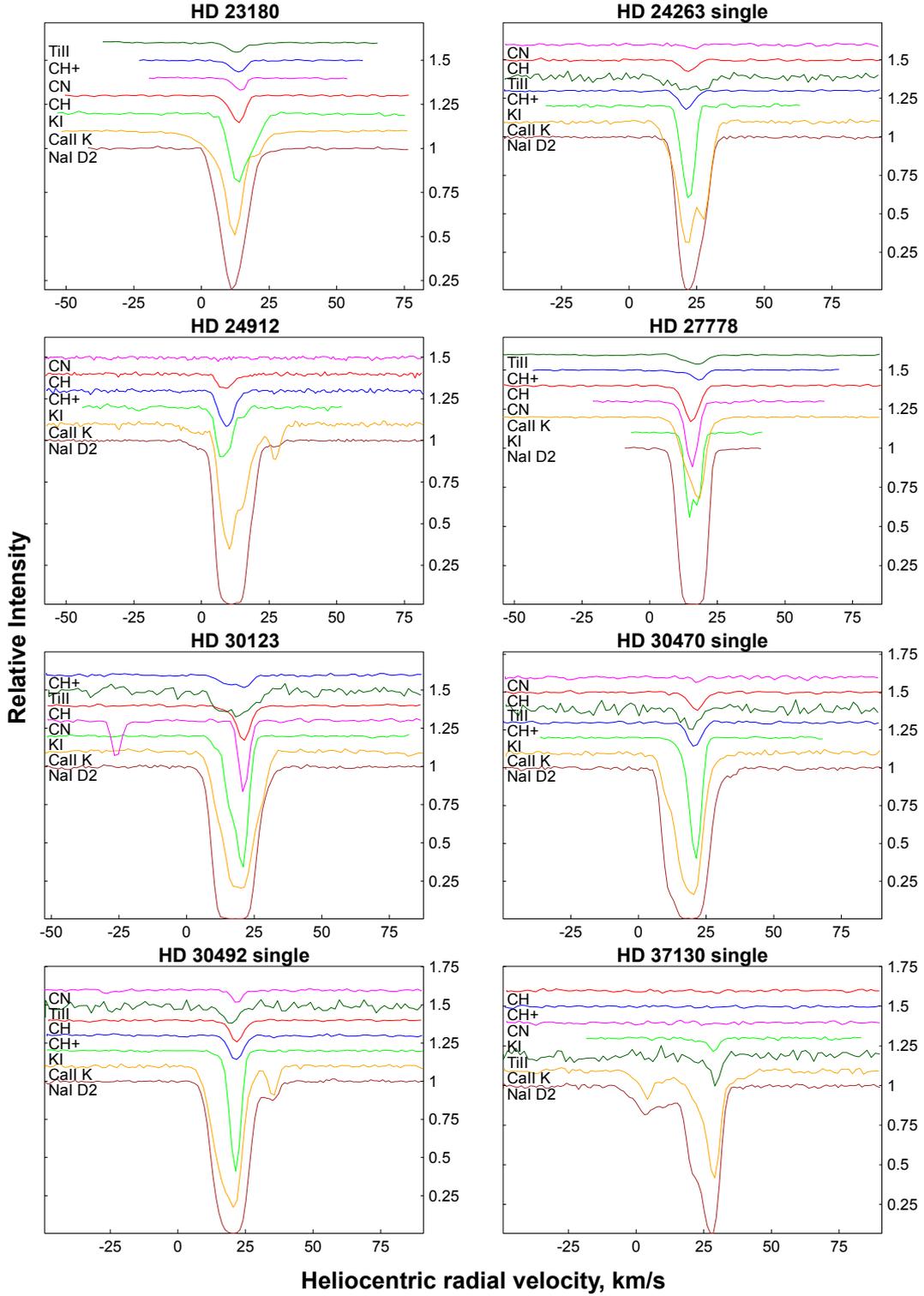}
  \caption{UVES-derived heliocentric velocity profiles in Ti\,{\sc ii}
    3383.8\,\AA\,(dark green), CN 3874.6\, \AA\, (magenta), CH
    4300.3\,\AA\, (red), CH$^+$ 4232.5\,\AA\, (blue), K\,{\sc i}
    7699\,\AA\ (green), Ca\,{\sc ii} K 3933.7\,\AA\, (orange), and
    Na\,{\sc i} D2 5889\,\AA\,(brown).  {Single-cloud sightlines are
      marked ``single''.} For the remaining spectra, see
    Fig.~\ref{appstart.fig} to
    Fig.~\ref{append.fig}. \label{spectra.fig}}
\end{figure*}


\begin{table*}[!htb]
\scriptsize
\begin{center}
 \caption {Result summary. \label{res.tab} Column~(1) lists the
   identification number as of Table~\ref{sample.tab}, Col.~(2) the HD
   identifier, the equivalent width in Col.~(3) of the Ca\,{\sc ii}~K
   and in Col.~(4) of the Ca\,{\sc ii}~H doublet used in Col.~(5) for
   the distance estimate \citep{Megier09}. Distances obtained from the
   GAIA parallax are listed in Col.~(6) and from the spectral
   type-luminosity ratio in Col.~(7). Appearance of single (S) or
   multiple (M) velocity components of profiles for the Ca\,{\sc ii}~H
   and K, the Na\,{\sc i}, and the K\,{\sc i} lines in Cols.~(8-10).
   The fit parameters of the K\,{\sc i} line profile are listed for
   the position, the column density, and the derived equivalent width
   in Cols.~(11-13). Column~(14) lists the classification of cloud
   environment as cold (c) or warm (w) with or without a CN\,{\sc i}
   detection. Column~(15) lists the infrared morphology of the WISE
   (W4) image. Single-cloud sightlines are highlighted in boldface. }
 \begin{tabular}{r | c |c  c | r r r | c c c | r c r | c c }
    \hline
    \hline
1  &  2     & 3       & 4         & 5           & 6                   & 7                               & 8      & 9    &10   & 11                   & 12         &  13      & 14 & 15 \\
\hline
ID & Target & \multicolumn{2}{c|}{Ca\,{\sc ii}} & \multicolumn{3}{c|}{Distance} & \multicolumn{3}{c|}{Components}  & \multicolumn{3}{c|}{K\,{\sc i}} & Environ. & WISE \\
   &        &   EW(K) &  EW(H)    & Ca~{\sc ii} & GAIA                & Sp/L        & Ca~{\sc ii}~K & Na~{\sc i} & K~{\sc i} & v$_{\bigodot}$  & log N  & EW  &class  & class \\
   &        &  (m\AA)  &  (m\AA)  & \multicolumn{3}{c|}{(pc)}         &  &  &   & (km/s) & ($10^9$\,/cm$^2$) & (m\AA) &  &  \\
\hline
  1 & {HD~023180} &  86$\pm$ 2&  47$\pm$ 1  & $340^{+20}_{-15}$     & $  256^{+72}_{- 47}$ &  320  &  S & S & {M}  &  13.08   &  464 $\pm$ 11   &   69.1 $\pm$  1.6  & c+CN & a \\                           
  2 & {\bf HD~024263} & 129$\pm$ 3&  84$\pm$ 2  & $670^{+65}_{-50}$     & $  222^{+ 8}_{-  8}$ &  210  &  M & S & S  &  21.93   &  706 $\pm$ 10   &   90.6 $\pm$  1.2  & - & a \\                              
  ... & & & & & & & &  & & & &  & & \\
136 &      HD~326364  &   378$\pm$23& 196$\pm$13 &{$1597^{+379}_{-233}$} & $ 1679^{+149}_{-127}$ & {1859} &  M & M & M  &  -8.67   &  437 $\pm$  7   &   64.3 $\pm$  1.2  & c+CN & a \\
    &  ''             &             &            &                     &                       &      &    &   &    &   1.69   &  945 $\pm$  8   &  106.9 $\pm$  0.9  & w-CN   &   \\
\hline
  \end{tabular}
\end{center}
    {\bf Notes: } $^{{LR}}${{low-resolution spectrum has insufficient resolving power for
        classification as single-cloud sightline,}} $^{\imath}$
    saturated line, $^{v}$: ignored because lines have different
    velocities, $^e$: WISE image shows artefact. For the complete list,
    see Table~\ref{appres.tab}.
\end{table*}


\subsection{Archive spectra}
In addition to the 100 stars observed by us with UVES, we have found
high-resolution spectra of 86 other sightlines. Seventeen UVES spectra
were taken in the context of the EDIBLES survey \citep{Cox17}, which
covers almost the entire optical wavelength range from 320 to 1000\,nm
with an S/N exceeding 500 per pixel. Two spectra were taken from
existing data (co-author Galazutdinov) of the 1.0\,m Zeiss Special
Astrophysical Observatory of the Russian Academy of Sciences and BOES
at the 1.8\,m Bohyunsan Optical Astronomy Observatory in Korea
\citep{Kim07}. Eighteen spectra were taken from ELODIE
\citep{Moultaka04}, FEROS \citep{Kaufer99},
SOPHIE\footnote{atlas.obs-hp.fr/sophie}, and ESO UVES archives,
alternatively, their single- or multiple-cloud status was determined
by a literature search using {\sc Simbad}. These 36 sightlines with
archive spectra are provided in Table~\ref{sample.tab}. We found
additional publicly available UVES and FEROS spectra for 50 stars that
include the K\,{\sc i} line. They are presented in Table~\ref{KI.tab}.


\subsection{WISE mid-IR imaging}
In addition to optical high-resolution spectroscopy, we searched for
mid-IR data, which provide dust emission signatures of the
clouds. Mid-IR imaging is obtained with the Wide-field Infrared Survey
Explorer \citep[WISE,][]{Wright10} for all sources of
Table~\ref{sample.tab}.  WISE\footnote{images available at:
  https://irsa.ipac.caltech.edu/applications/wise/} mapped the sky at
3.4, 4.6, 12, and 22\,$\mu$m (filters: W1, W2, W3, and W4) in 2010
with an angular resolution of 6.1", 6.4", 6.5", and 12.0" and a
5\,$\sigma$ point source sensitivities better than 0.08, 0.11, 1, and
6\,mJy in the four bands, respectively.

\section{Analysis of spectroscopic data and mid-IR imaging\label{result.sec}}
In this section we present our spectral classification
(Sect.~\ref{spectype.sec}) of the sample (Table~\ref{sample.tab}) and
measurements of radial velocities and equivalent widths
(Sect.~\ref{radew.sec}).  In Sect.~\ref{single.sec} we derive column
densities from the K\,{\sc i} profile and identify the single-cloud
sightlines. We then use the CH/CH$^+$ line ratio to classify the
environment of the clouds to be either cold or warm
(Sect.~\ref{env.sec}), and in Sect.~\ref{MIR.sec} we analyse WISE
mid-IR images of our sightlines to constrain the location of the dust
emission. We apply three different methods to estimate distances to
the targets (Sect.~\ref{distances.sec}).

HD~164747A and HD~164747B are separated by 3.4\arcsec\ and are not
resolved by IUE. Both stars are covered in our UVES image. They are of
same spectral type and show a different Doppler structure of Ca~{\sc
  ii} lines.  The DIB intensities are identical, and we therefore
assume that the extinction curves are also identical. We removed
HD~123335 (ID~59), HD~134591 (ID~62), HD~147331 (ID~70), and HD~151346
(ID~79) from our analysis because the lines are contaminated.  We
spectroscopically classified 136 stars of Table~\ref{sample.tab} and
carried out the remaining analysis on 132 stars. The results are
summarised in Table~\ref{res.tab}.

\subsection{Spectral classification \label{spectype.sec}}

Knowing the spectral type is important for estimating the reddening
and spectral type-luminosity distance of a star. Our high-resolution
spectra allow accurate measures of {line intensities and profiles. We
  applied a two-step approach by estimating the spectral class of the
  stars following the procedure described and exemplified by
  \cite{Krelowski18}. The first step is based on the measured ratios
  of the H~{\sc i}, He~{\sc i,} and Mg~{\sc ii} line strengths, as
  suggested by \cite{WF90} and by the Stony Brook
  catalogue\footnote{http://www.astro.sunysb.edu/fwalter/SMARTS/spstds\_f2.html}. The
  position of OB standard stars in the Mg~{\sc ii}/He~{\sc i} versus
  H~{\sc i}/He~{\sc i} plane was shown and tabulated by
  \cite{Krelowski18}. For example, in this scheme, a B6 star has
  Mg~{\sc ii}/He~{\sc} $\sim 0.9$ and H~{\sc i}/He~{\sc i} $\sim 20$,
  while a B9.5 star shows Mg~{\sc ii}/He~{\sc i} $\sim 2.6$ and H~{\sc
    i}/He~{\sc i} $\simgreat 70$.} The second step includes
confirmation or correction of our initial spectral type classification
through careful inspection of the high-resolution spectra of the
considered stars. This includes interpretation of possible
incompatibilities. In this second step, we inspect various line
profiles of H~{\sc i}, He~{\sc i}, He~{\sc ii}, C~{\sc iv}, Mg~{\sc
  ii}, Si~{\sc ii}, Si~{\sc iii}, and Si~{\sc iv}. The detection of
He~{\sc ii} and C~{\sc iv} lines points to O type stars, while the
Si~{\sc iv} line indicates an early-B star. That the strength ratio of
the He~{\sc i} 4471 is higher than that of Mg~{\sc ii} 4481 provides
another indication for early B-type stars. At mid-B, the intensity of
Si~{\sc iv} 4089 and Si~{\sc iii} 4552 lines relative to that of
Si~{\sc ii} 4128-30 is the defining characteristic, while for late-B
stars, the defining characteristic is the intensity of Mg~{\sc ii}
4481 relative to that of He~{\sc i} 4471, according to
\cite{Walborn08}. Hydrogen Balmer line profiles provide evidence of a
dwarf or a high-luminosity star. The similarity of lines such as the
series of He~{\sc i} at 3965, 4009, 4026, 4121, 4144, 4169, 4388,
4437, 4713, 4923, and 5876 (\AA) is characteristic for B-type stars
when He~{\sc ii} lines are absent.  We also searched for signatures of
multiple stellar components and binarity as apparent from the profile
shapes.  The intensities of weak stellar lines in fast rotators or in
binaries are hard to estimate and create uncertainties in
spectral-type determinations.

Our newly estimated spectral types and those previously given in the
literature are listed in Cols.~7 and 8 of Table~\ref{sample.tab},
respectively.  For 89 out of the 136 stars, the two estimates agree to
within one subclass. Fourteen stars were previously classified as B0-1
{I-V} while we identified them as O-type stars with 12 as O9 and the
other between spectral type O4-8. Twenty-two stars of our sample
appear to be spectroscopic binary stars, and 14 are associated with a
cloud in {\sc Simbad.}  In the following paragraphs we discuss the
discrepant cases and report the reason for our decision for a binary
classification for some objects.

The typical signatures of O-type stars are the C~{\sc iv} 5801.3 and
C~{\sc iv} 5811 lines, which are not seen in B-type stars.  In O7
stars, the He~{\sc i} line near 4471.5 is as deep as He~{\sc ii}
4541.6. In even hotter stars, the He~{\sc ii} feature becomes
stronger, while in late O-type stars, the He~{\sc i} line becomes
dominant.  For example, we classify Herschel~36 as O8V: the C~{\sc iv}
lines are detected, but Mg~{\sc ii} is barely visible, and He~{\sc i}
is stronger than He~{\sc ii}.

The luminosity class of a star can be estimated based on the H~{\sc i} line
profiles. Our classifications were made by comparing the spectra of our
targets with those given as standards by \cite{WF90}. Supergiants
and bright giants (class I and II) show narrow H~{\sc i} lines, such as
HD~046711 and HD~091943. Main-sequence stars (class IV and V) show broad
H lines, such as HD~037903, HD~108927, and HD~147196.

Spectral types of several objects may be uncertain when they belong to
a binary system, and the apparent binarity is revealed by double,
blended profiles that are difficult to separate in the two
components. We classify HD~315032 as an early B0 star because Mg~{\sc
  ii} is barely visible. HD~156247 appears as a binary in He~{\sc i},
Mg~{\sc ii,} and C~{\sc ii} lines.  The He~{\sc i} line profile
appears likely blended in HD~315033 and in HD~161653, for which broad
wings of H~{\sc i} lines and many O$^+$ lines are detected.  We
classify HD~161653 as B3~II, although it was classified as B0.5~III by
\cite{Valencic} and as B2~II by \cite{Houk75}. The classification is
uncertain because this star apparently is a binary, and the red
component of He~{\sc i} seems broader than in Mg~{\sc ii}. The
hydrogen lines have broad wings, which is characteristic for
dwarfs. However, if the star is of class IV, the spectral
type-luminosity distance becomes only $\sim 660$\,pc, while GAIA and
Ca~{\sc ii} measurements place it farther away than 1\,kpc. In our
class II, the spectral type-luminosity distance of HD~161653 becomes
1230\,pc, which is very consistent with the other distance estimates
(cf. Table~\ref{res.tab}).

\subsection{Radial velocity and equivalent width measurements}\label{radew.sec}
We converted the spectra of Table~\ref{sample.tab} into normalised
intensity versus heliocentric velocity for the profiles of the
following spectral lines: Ti\,{\sc ii} 3383.8\,\AA, CN 3874.6\,\AA,
Ca\,{\sc ii}~K 3933.7\,\AA, CH 4300.3\,\AA, CH$^+$ 4232.5\,\AA,
Na\,{\sc i} 5890\,\AA, and K\,{\sc i} 7699\,\AA\ (see
Fig.~\ref{spectra.fig} and
Figs.~\ref{appstart.fig}~--~\ref{append.fig}). The broadest Ca\,{\sc
  ii} component seen in the spectrum of HD~062542 at 15.5\,km/s is
probably of stellar origin.  We note that the profiles of Ti\,{\sc ii}
and Ca\,{\sc ii}~K lines often appear similar \citep{Hunter06} as are
those of Na\,{\sc i} and K\,{\sc i}. In addition to these seven lines,
we inspected other lines such as {Ca\,{\sc i} 4227\,\AA\,} and DIBs
covered in the observed wavelength range of our observations, finding
that they had often low S/N, or are frequently saturated, like
Na\,{\sc i} 5895.9\,\AA.  In addition to the heliocentric velocities,
we also measured the equivalent widths for the Ca\,{\sc ii} H and K
doublets for the seven lines, as given in Cols.~3 and 4 of
Table~\ref{res.tab}. Interstellar lines often do not show the same
radial velocities and appear in three distinct families. The radial
velocity of a) K\,{\sc i} and CH commonly coincides nearly perfectly,
and likely also with CN, while b) CH$+$ frequently shows a shift. c)
The Ca\,{\sc ii} lines quite frequently show very complex Doppler
structure, in contrast to other interstellar lines, while Ca\,{\sc ii}
and Ti\,{\sc ii} show similar profiles. DIBs seem to originate in
clouds following K\,{\sc i} \citep{Galazutdinov00}.

\begin{figure*}
  \includegraphics[angle=90,width=9.6cm,clip=true,trim=0cm 0cm 0cm 0.cm]{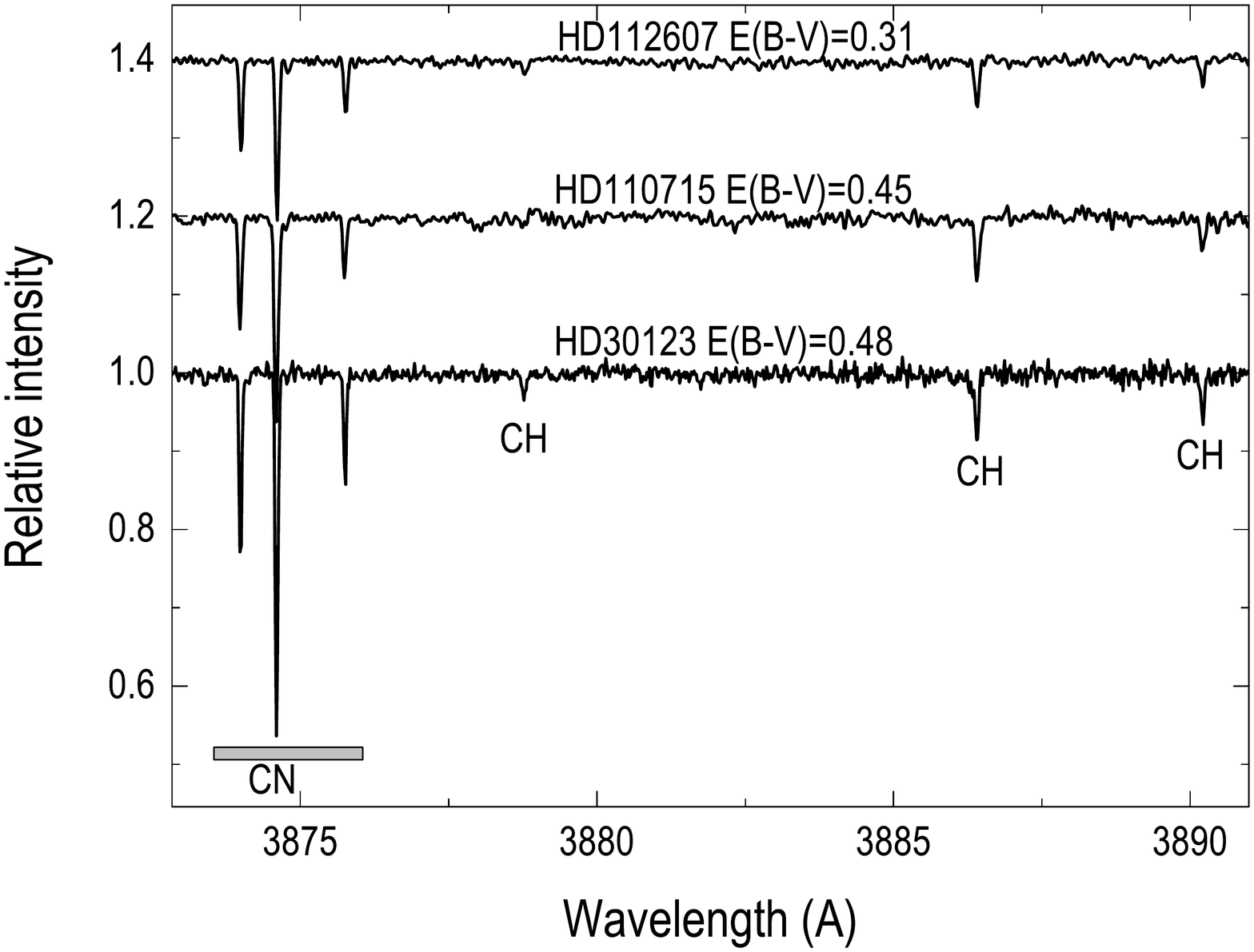}
 \includegraphics[angle=90,width=9.4cm,clip=true,trim=0cm 0cm 0cm 0.cm]{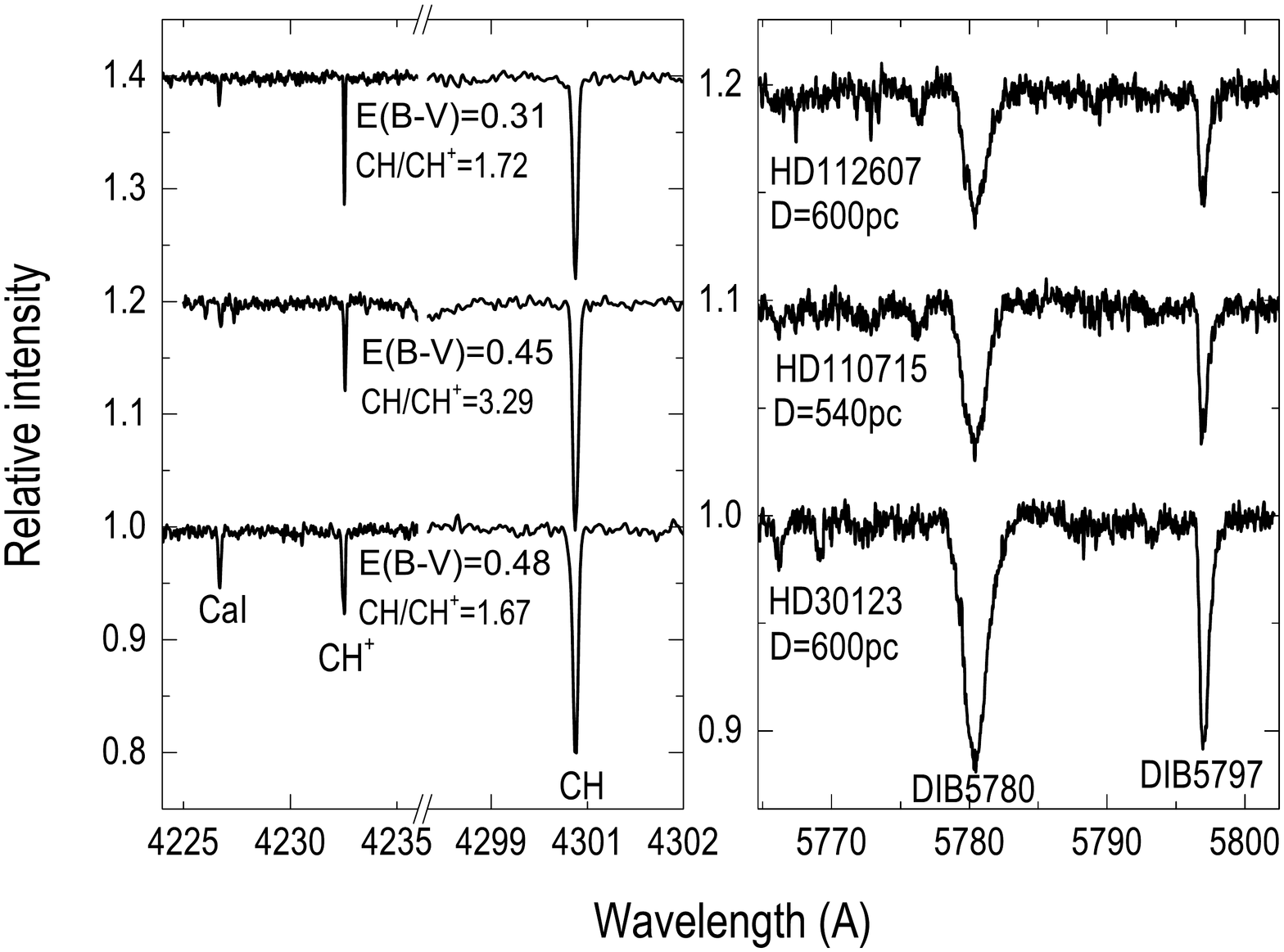}
 \caption{Spectral features of the rotational R1, R0, and P1
   transitions of cyanogen (CN) near 3875\AA\, (left), the CH$^+$
   4232.5\AA\ and CH A-X 4300.3\AA\, transition (middle), and DIBs at
   5780\AA\, and 5797\AA\, (right panel) for the three single-cloud
   sightlines HD~112607, HD~110715, and HD~030123. The CH/CH$^+$ line
   intensity ratio, $E_{\rm {(B-V)}}$, distance, and some other lines are
   labelled. \label{zetCN.fig}}
\end{figure*}
\begin{figure} 
  \includegraphics[angle=90,width=9.8cm,clip=true,trim=0cm 0cm 0cm 0.cm]{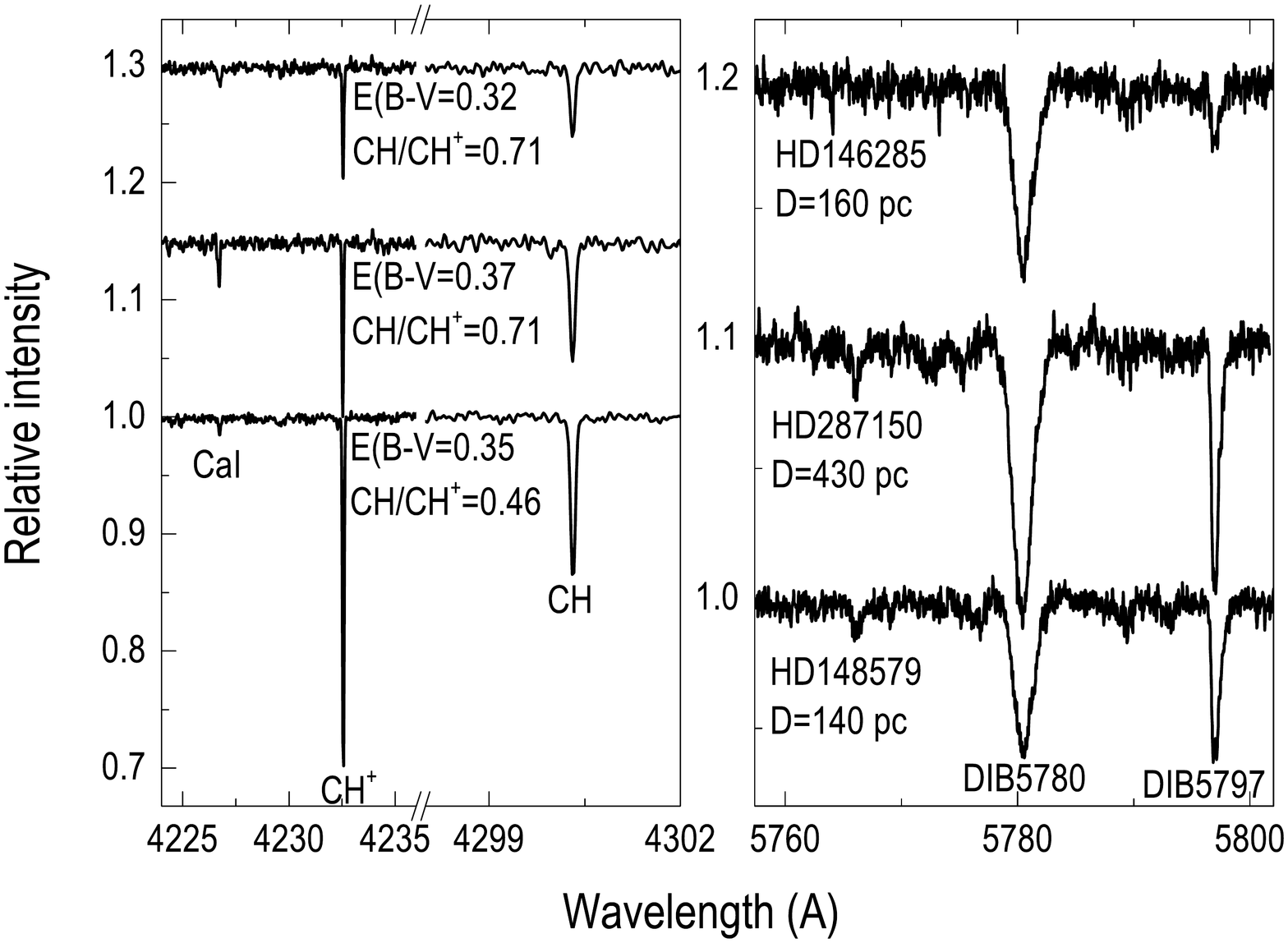}
  \caption{Spectra of the CH$^+$ 4232.5\AA\ and CH 4300.3\AA\, transition
    (left) and DIBs at 5780\AA\, and 5797\AA\, (right panel) for the three
    single-cloud sightlines HD~146285, HD~148579, and HD~287150. The
    CH/CH$^+$ ratio, $E_{\rm {(B-V)}}$ reddening, distance, and Ca~{\sc i} line are
    labelled.\label{sigmol.fig}}
\end{figure}

\subsection{Identification of single-cloud sightlines}\label{single.sec}
We classified the appearance of the radial velocity profiles of
Ca\,{\sc ii}~K, Na\,{\sc i}, and K\,{\sc i} as single (S) or multiple
(M) component sightlines. The results are given in Cols.~8 -- 10 of
Table~\ref{res.tab}. The saturated Na\,{\sc i} lines, for which a
classification was not possible, are marked by a hyphen. We find that
the K\,{\sc i} line is particularly suited for the S or M
classification. The K\,{\sc i} profile is fit by a multiple-component
Voigt function using the position and column density as free
parameters; in order to estimate the column density, we applied the
apparent optical depth method described by
\citeauthor{SavageSembach91} (\citeyear{SavageSembach91}) as
implemented in {\sc dech}. The best-fit parameters to the K\,{\sc i}
profiles are given in Col.~4 (position) and Col.~5 (column density) in
Table~\ref{KI.tab} and in Col.~11 and Col.~12 in Table~\ref{res.tab},
respectively. We compared our estimates of the column densities of the
K\,{\sc i} line as derived by {\sc vapid} and {\sc dech}.  Generally
both methods agree very well however for lines that are observed close
to saturation or resolution limits the systematic differences become
significant. We found that for 81 of the 132 sightlines of
Table~\ref{res.tab} that allowed us to perform this analysis, the fit
to the K\,{\sc i} profile required multiple components, while 51 of
these stars were fit by a single-component model. In addition, we
found that 14 of the 50 stars of Table~\ref{KI.tab} are dominated by a
single-component. In total there are 65 out of 186 stars that we
classify as single-cloud sightline.

A number of objects have been observed in the K\,{\sc i} and Ca\,{\sc
  I} surveys by \cite{Welty01} and \cite{Welty03}, who used the UHRF
or McDonald spectrographs offering resolution of 0.5\,km/s. These
observations are consistent with our classifications of
multiple-component sightlines. For example, the multiple-component
sightline HD~023180 shows a very asymmetric K\,{\sc i} profile in UVES
and reveals four components between 10.5 - 14.9\,km/s in the spectra
by \cite{Welty03}. HD~027778 shows a double peak in K\,{\sc i} in
UVES, which is confirmed in the spectrum by \cite{Welty01}, which
shows four about equally strong components between 14.1 -
18.6\,km/s. More interestingly, HD~207198 and HD~209975 show multiple
components in the \cite{Welty01} and \cite{Welty03} spectra, but these
lines appear in ELODIE and SOPHIE spectra as a
single-component. ELODIE and SOPHIE have about half the resolving
power provided by UVES, which obviously is insufficient to
characterise single-cloud sightlines. Nine spectra in our sample
observed at similarly low resolution therefore cannot be
classified. These stars are marked LR in Table~\ref{res.tab}.  Three
stars have decent S/N spectra in \cite{Welty01} and \cite{Welty03} and
are in common to our single-cloud sightlines: HD~147933 and HD~149757
show at 0.5\,km/s resolution significant fine structure in the K~{\sc
  i} line, while towards HD~147165 the second strongest component is a
factor two fainter. Recently, \cite{Welty20} finds that most of the
material towards HD~062542 resides indeed in a single narrow velocity
component. The ultra-high resolution spectra often exhibit fine
structures in lines that appear to be single when observed at
resolution of 1.5\,km/s \citep{Welty14}. In total, we found that in
our terminology 65 of the 186 stars of this investigation appear as
single-cloud sightlines. Of these, about half have previously been
identified as such by \cite{Sembach}, \cite{Ensor}, \cite{S18}.

\begin{figure*} 
 \includegraphics[width=4.5cm,clip=true,trim=3.5cm 6.5cm 3.5cm 5.5cm]{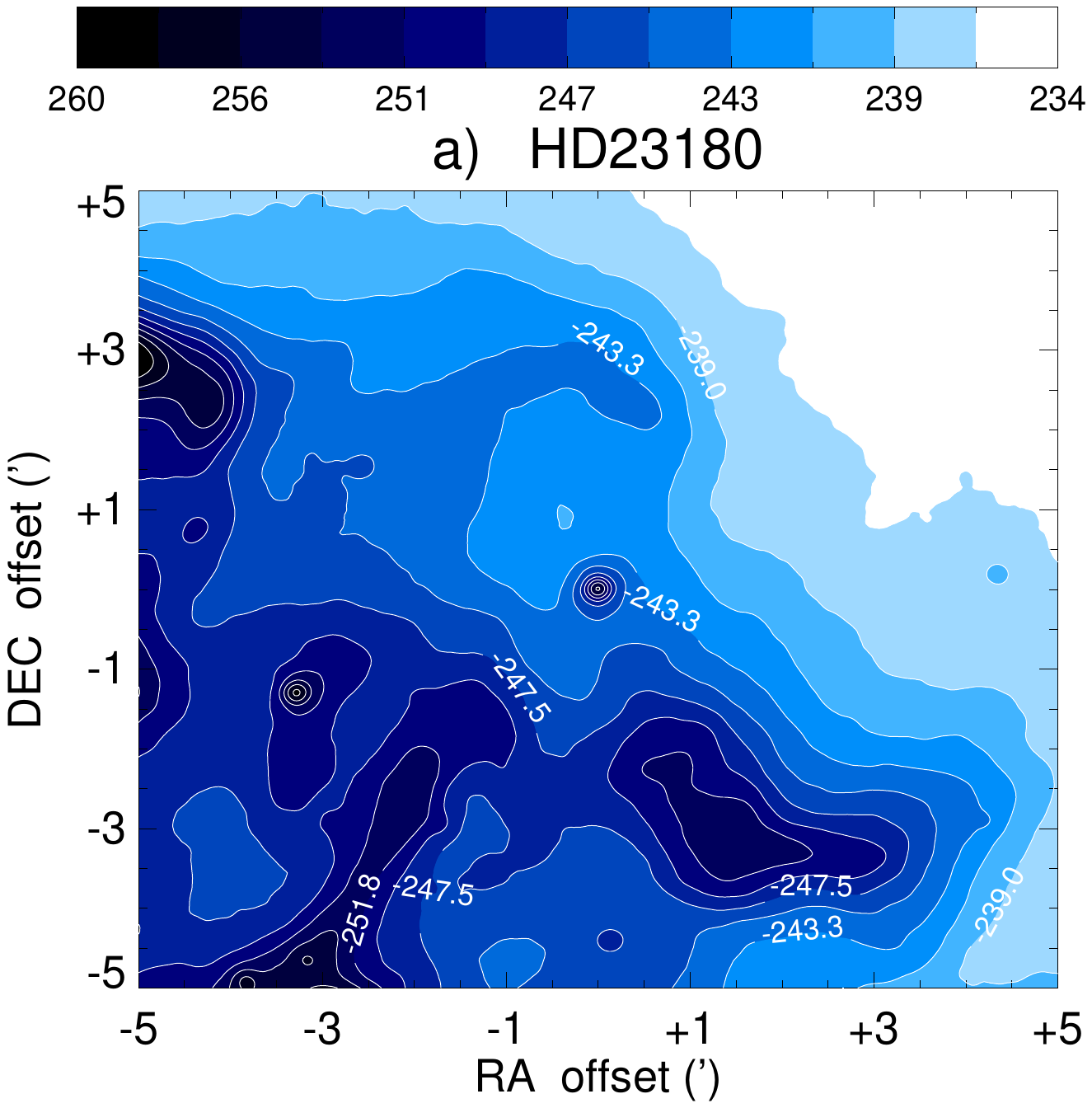}
 \includegraphics[width=4.5cm,clip=true,trim=3.5cm 6.5cm 3.5cm 5.5cm]{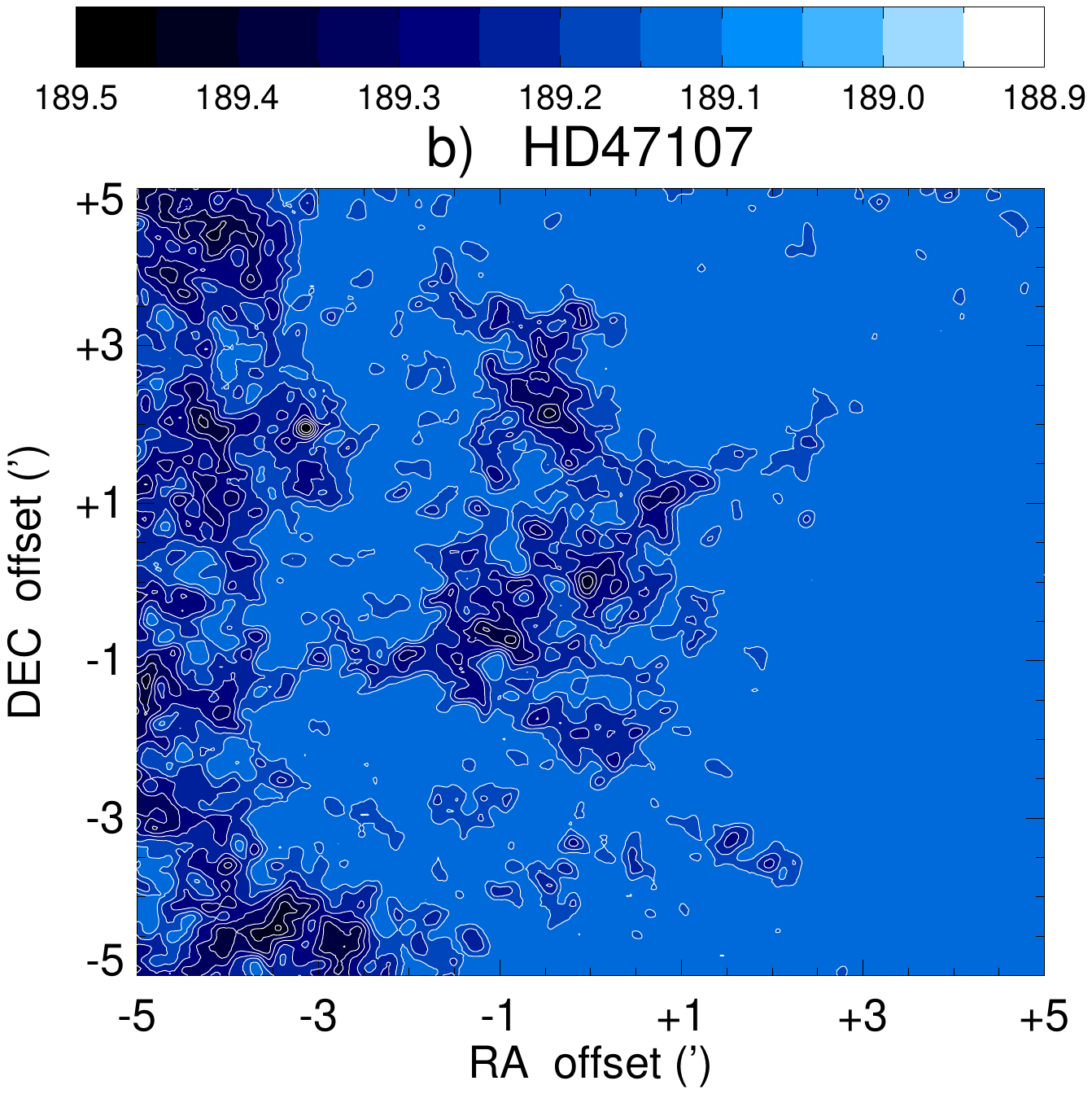}
 \includegraphics[width=4.5cm,clip=true,trim=3.5cm 6.5cm 3.5cm 5.5cm]{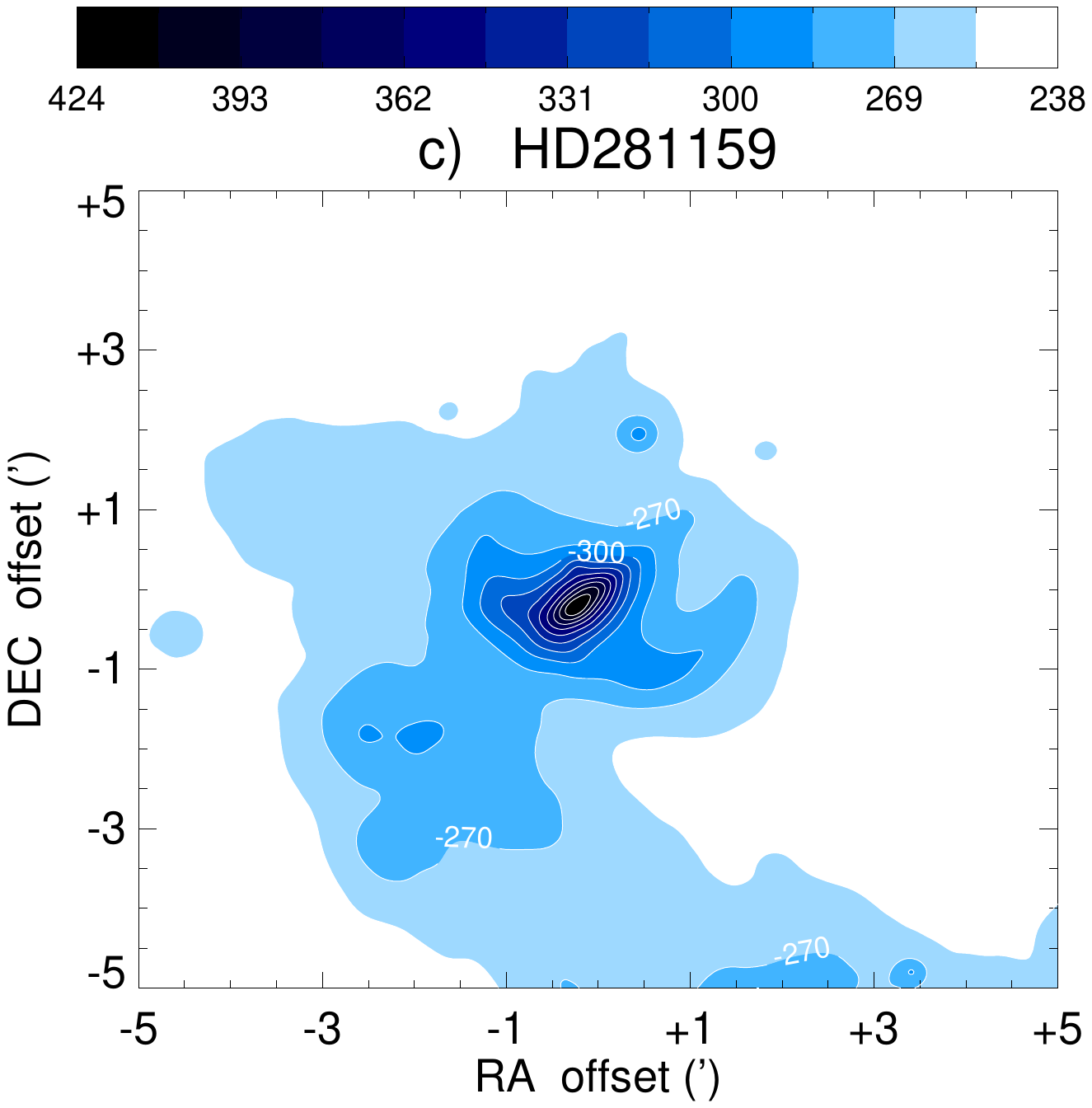}
 \includegraphics[width=4.5cm,clip=true,trim=3.5cm 6.5cm 3.5cm 5.5cm]{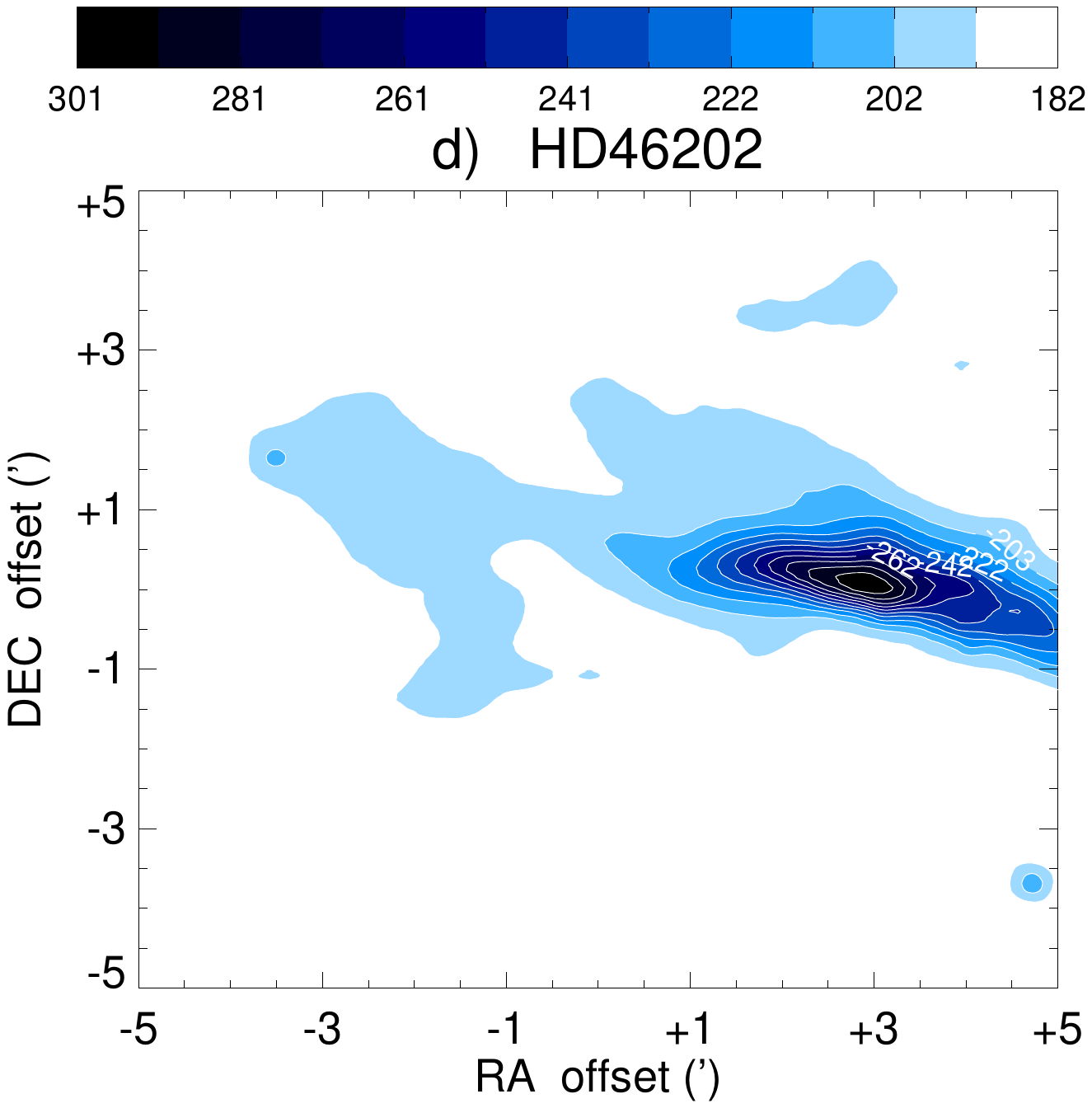}
 \caption{WISE images in filter W4 ($\sim 22$\,$\mu$m) centred on
   stars of different morphological types. From left to right:
   HD~023180 (type a), HD~047107 (type b), HD~281159 (type c), and
   HD~046202 (type d).  Fluxes levels are measured in
   mJy. \label{wise.fig}}
\end{figure*}

\subsection{Cloud environment \label{env.sec}}
Diffuse clouds show significant variations in their physical
conditions depending on their density, the strength of the radiation
field, the cosmic-ray intensity, and the presence or absence of
shocks. The strength of simple radicals such as CN, CH, and CH$^+$ and
of DIBs are known to vary from cloud to cloud \citep{K87}.  The
presence of CH$^+$ gives particular insights into the environmental
condition of the clouds.  CH$^+$ is produced in media that are heated
by a warm shock wave or a strong UV field, and it is rapidly destroyed
by collisions with H, H$_2$, and e$^-$ \citep{Morris16}.  The carriers
of broad DIBs such as the 5780\,\AA\ band are more abundant when
simple radicals are destroyed.  This seems to occur in strongly
irradiated warm clouds where CN is rarely seen. Regions showing weak
ionisation with line intensity ratios CH/CH$^+ \simgreat 1$ appear to
be connected to low-irradiation cold environments \citep{K19b}, and
the CN line is detected there.  We show this result in
Fig.~\ref{zetCN.fig} for HD~112607, HD~110715, and HD~030123, in which
CN is detected and the cloud is cold (CH/CH$^+\simgreat 1$), and in
Fig.~\ref{sigmol.fig} for HD~146285, HD~148579, and HD~287150, in
which CN is not detected and the cloud environment is warm (CH/CH$^+ <
1$). This finding has been reported for five of our
stars\footnote{HD~147888, HD~147933, HD~204827, HD~206267, and
  HD~207198} by \cite{Pan}, where CH and CN column densities for
moderately high-density gas ($n > 30$\,cm$^{-3}$) correlate and no CN
is detected in low-density regions showing CH$^+$.

The strength of the lines and DIBs is further discussed by
\cite{Weselak} and \cite{K19b}. We have inspected the CH A-X
4300.3\AA, CH$^+$ 4232.5\AA,\ and the CN 3874.6 \AA\ line in spectra
of our sample (see Fig.~\ref{spectra.fig} and
Figs.~\ref{appstart.fig}~--~\ref{append.fig}).  The wavelength ranges
of these lines are included in the spectra of 109 of our stars.
Thirteen of them have a low S/N in CH and CN.  In five stars, the CH
lines were not detected. Three multiple-component sightlines show
significant velocity shifts between the lines and were excluded from
the analysis. This leaves us with 88 spectra.  In six of them, CH and
CN are split into two clearly separated velocity components, and
because of the multiple-component sightlines, we were able to measure
97 ratios of CH/CH$^+$ with or without a detection of CN lines. For
these, we report in Col.~14 of Table~\ref{res.tab} thirty-nine warm
clouds (CH/CH$^+ < 1$) in which the CN line is not detected, labelled
w-CN; 54 cold clouds (CH/CH$^+ \simgreat 1$) in which the CN line is
detected, labelled c+CN; and only four warm clouds in which CN is
detected, labelled w+CN.  The CN line is detected in all clouds
labelled ``cold''.

Data in CH and CN bands are available in {39 out of 51} single-cloud
sightlines with spectra covering these features (Table~\ref{res.tab}).
For these, CN is detected in 23 out of {37} single clouds that are classified as
cold. {Except in HD~156247,} CN is missing in all other {14}
single clouds that are classified as warm.  When the cloud is close to
the OB star, it should be warm. Clouds of single-cloud sightlines are
cold and are predominantly located far away from the star in the cold
ISM.

\subsection{Analysis of mid-IR imaging \label{MIR.sec}}
We complemented our spectroscopic analysis  using archive mid-IR
imaging to assess the nature of the cloud environment. Dust is heated
by UV and optical stellar light and re-emits the absorbed energy at
infrared (IR) wavelengths.  Inspecting the appearance of our sources
in the mid-IR may allow us to constrain the location of the dust
cloud(s) along the sightline that is either nearby the star in a
circumstellar envelope or farther away in the diffuse ISM. We studied
the connection of star and dust emission along the sightlines by
inspecting the morphology of mid-IR images provided by WISE in a $5'
\times 5'$ field centred on our stars.

The observed mid-IR fluxes are dominated either by star or dust
emission.  When the flux is dominated by dust emission, it is likely
resolved in the WISE W3 and W4 bands. This can be demonstrated by
computing the dust emission of a circumstellar dust envelope of a
mixture of large 600\,nm sized carbon and silicate grains that are
heated by a typical B-type star of our target list.  The star has a
luminosity of $10^4$\,\Lsun \, and temperature of 20,000\,K. The shell
has an inner boundary set by dust evaporation at temperature $\sim
1500$\,K and an outer radius of 1\,pc. For a dust envelope with a
constant density and dust mass of 0.8\, \Msun \, the total optical
depth $\tau_{\rm V} = 1$. In this shell dust becomes warmer than 80\,K
up to distances of a few $10^{17}$\,cm. This envelope may be detected
with the spatial resolution of WISE\footnote{A 6$''$ beam has a linear
  scale of $10^{17}$\,cm at 1115 pc.} up to 1\,kpc distance, and more
so when the emission, for example in W3, is dominated by very small
(nanometre) particles such as polycyclic aromatic hydrocarbons,
e.g. \citet{S93}, \citet{K08}. It may be considered that dust clouds
have much higher optical depth than estimated from adopting the
$A_{\rm V}$ of the sightline to the star (Table~\ref{sample.tab}). In
this case, the clouds could be heated by embedded stars.  Because
stars like this are unseen in the near-IR but the optical depth must
be very high ($A_{\rm V} \geq 10$\,mag) in the mid-IR, a scenario for
the given $A_{\rm V}$ of our sightlines is unlikely.

We classified the morphology of the WISE images and distinguished the
following cases: \/ {\bf a) }Images without dust emission
signatures. The WISE image at the position of the star appears
point-like, and is brighter in W1 and W2 than in W3 and W4, where the
star is sometimes not even detected. \/ {\bf b) }Image background is
enhanced. The emission near the star shows a somewhat enhanced diffuse
background component in W3 and even more so in W4. \/ {\bf c) }A cloud
is detected. The star is located in or seems within less than
1\arcmin\ associated with a bright extended and often ellipsoidal
dust cloud. The clouds are often brighter in the W3 and more so in the
W4 band than in the W1 and W2 bands. \/ {\bf d) }A distant dust lane
or cloud is detected more than 1\arcmin\ away from the star, and is
likely not associated with the star. The dust lane often extends
across the full WISE image. The dust lane or cloud is often brighter
in W3 and more so in W4 than in W1 and W2.

Typical examples of our classification of the emission morphology as
seen by WISE in filter W4 are shown in Fig.~\ref{wise.fig} for
HD~023180 (case a), HD~047107 (case b), HD~281159 (case c), and HD~046202
(case d).  In the W4 (22\,$\mu$m) image of HD~023180, two point sources
are detected, one lies at the position of the star.  In HD~47107 the
star is detected and surrounded by some diffuse enhanced background.
HD~281159 is detected near the centre of a cloud and surrounded by a
structure that is bright in W4, visible in W1 and W2, but not detected
in W3. The star HD~046202 is not detected, but a cloud 2\arcmin\ W of
it is visible. Other remarkable structures are seen for HD~172028,
which is inside a cloud that is bright in all bands. HD~155756 shows a
nearby cloud $1'$ NE that is likely not associated with the
star. HD~200775 shows two bright dust clouds and one around the
star. Of the 132 objects with a WISE classification in Col.~15 of
Table~\ref{res.tab}, 83 ({63\,\%) fall in category a, 20 (15\,\%) in
  category b, 14 (11\,\%) in category c, and 14 (11\,\%)} in category
d. The distribution is similar for the {51} single-cloud sightlines:
WISE images exist for {all} these stars, and {32 (63\,\%) of them are
  in category a, 9 (15\,\%) in category b, 2 (4\,\%) in category c,
  and 9 (17\,\%)} in category d.

An extended dust halo or circumstellar shell that might contribute
significantly to the total dust column density along the sightline is
detected in a minority of our sample. This is category c, which
includes only $10$\,\% of the stars. Therefore dust along our
sightlines seems predominantly (90\,\%) located in the diffuse ISM, at
least the clouds are not detected by dust emission within a $ 1'$
neighbourhood to the stars.

\section{Distance estimates \label{distances.sec}}
We have derived the distances of our sample of early-type stars using
three different methods: {\it i)} GAIA parallax ($D_{\rm GAIA}$), {\it
  ii)} spectral type-luminosity ($D_{\rm SpL}$), and {\it iii)} the
amount of interstellar matter derived from the Ca~{\sc ii}~(H, K)
doublet ($D_{\rm {Ca}~II}$), as proposed by \cite{Megier05}.

{\it \textup{For the first method,} } the geometric distances $D_{\rm
  {GAIA}}$ were taken from \cite{Bailer18}. They are based on the
second GAIA data release, where an inference procedure is used, which
does not consider the physical properties of the stars. Furthermore,
the uncertainty of parallax measurements increases with increasing
distance to the star.

{\it } In the second method, the spectral type-luminosity distances
$D_{\rm SpL}$ were obtained from the spectral type and luminosity
using the spectral classification of Sect.~\ref{spectype.sec} and dust
extinction given in Table~\ref{sample.tab}. Spectral type-luminosity
distances depend on the correct identification of spectral type and
luminosity class of a star. A difference of one spectral subclass in
OB stars changes the derived stellar temperature already by 10\,\%,
implying an error in the distance of 40\,\%.  This method also
requires a correction for dust extinction with its own uncertainties,
e.g.  \citet{K09}, \citet{Scicluna15}. Unresolved binaries, which are
common among early-type stars, result in errors in the spectral type
and affect the estimate of the spectral type-luminosity distance.

The distance estimates $D_{\rm {Ca}~II}$ were obtained in the third
method using the relation between interstellar absorption to distance
that has been studied by \cite{Struve} and more recently by
\cite{Smoker06}. This method works well when the absorbing material is
relatively uniformly distributed, and within the Galactic
plane. \cite{Gazinur05} showed that the Ca~{\sc ii} is a better
distance indicator than K~{\sc i}, Na~{\sc i} or the reddening $E_{\rm
  {(B-V)}}$.  \cite{Megier09} reported a simple relation between the
equivalent width of the Ca~{\sc ii}~(H, K) doublet and the distance
$D_{\rm {Ca~II}}$. Large-scale structures and dependences on galactic
latitude, local enhancements in the Ca~{\sc ii} column density, likely
related to stellar clusters and associations, and a frequent
saturation of the Ca~{\sc ii} line limit the accuracy of the
method. Therefore we used our equivalent width measurements of the
Ca~{\sc ii}~(H, K) doublet (Cols. 3-4 of Table~\ref{res.tab}) and used
the formulae of \cite{Megier09} to compute the distance $D_{\rm
  {Ca~II}}$. Distance estimates derived from the three methods for the
132 stars of our sample are given in Cols. 5-7 of Table~\ref{res.tab}.

\begin{figure}
 \includegraphics[width=8cm,clip=true,trim=0.7cm 4.2cm 4.cm 6.3cm]{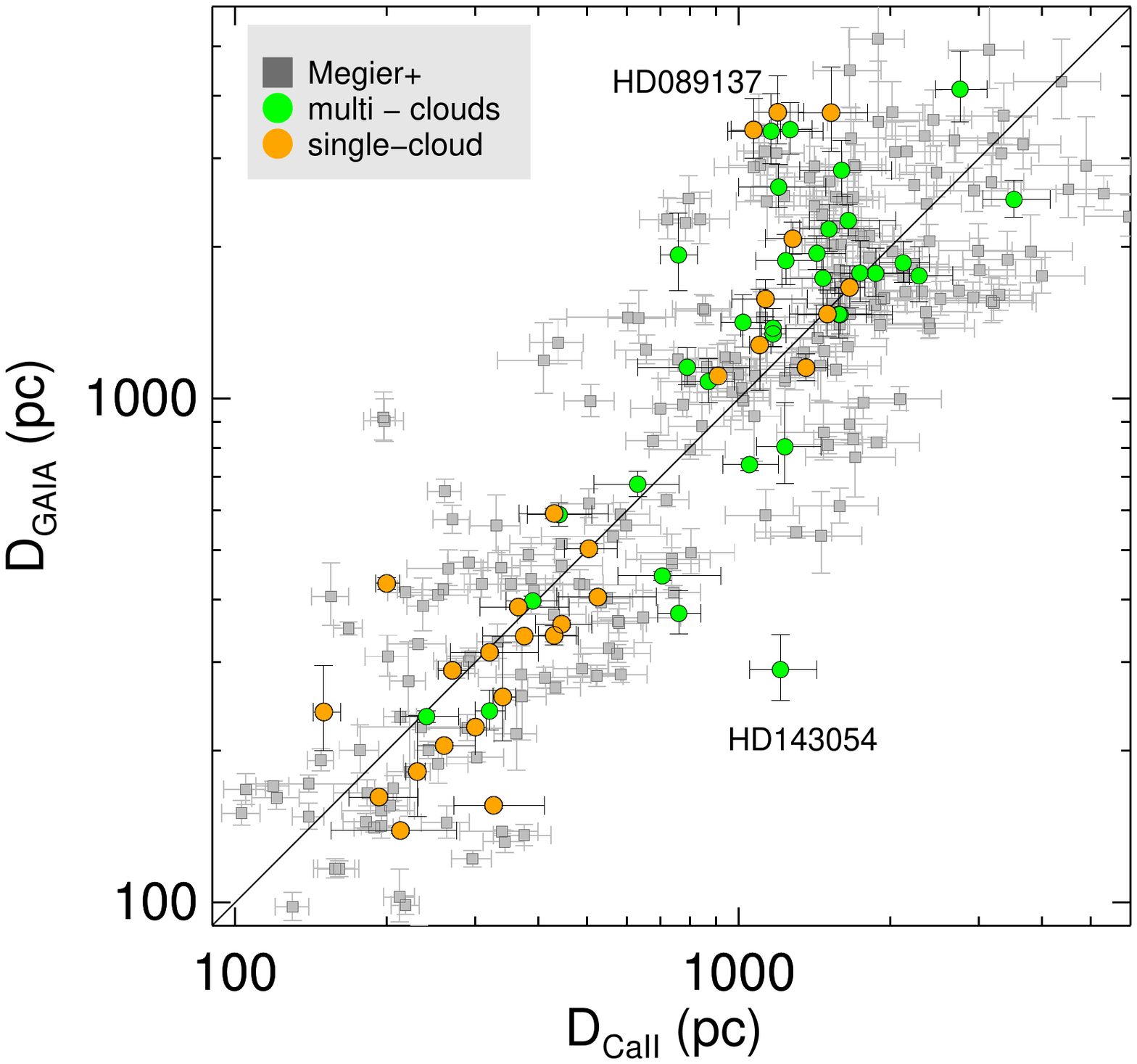}  
  \includegraphics[width=8cm,clip=true,trim=0.7cm 7.02cm 4.cm 7.cm]{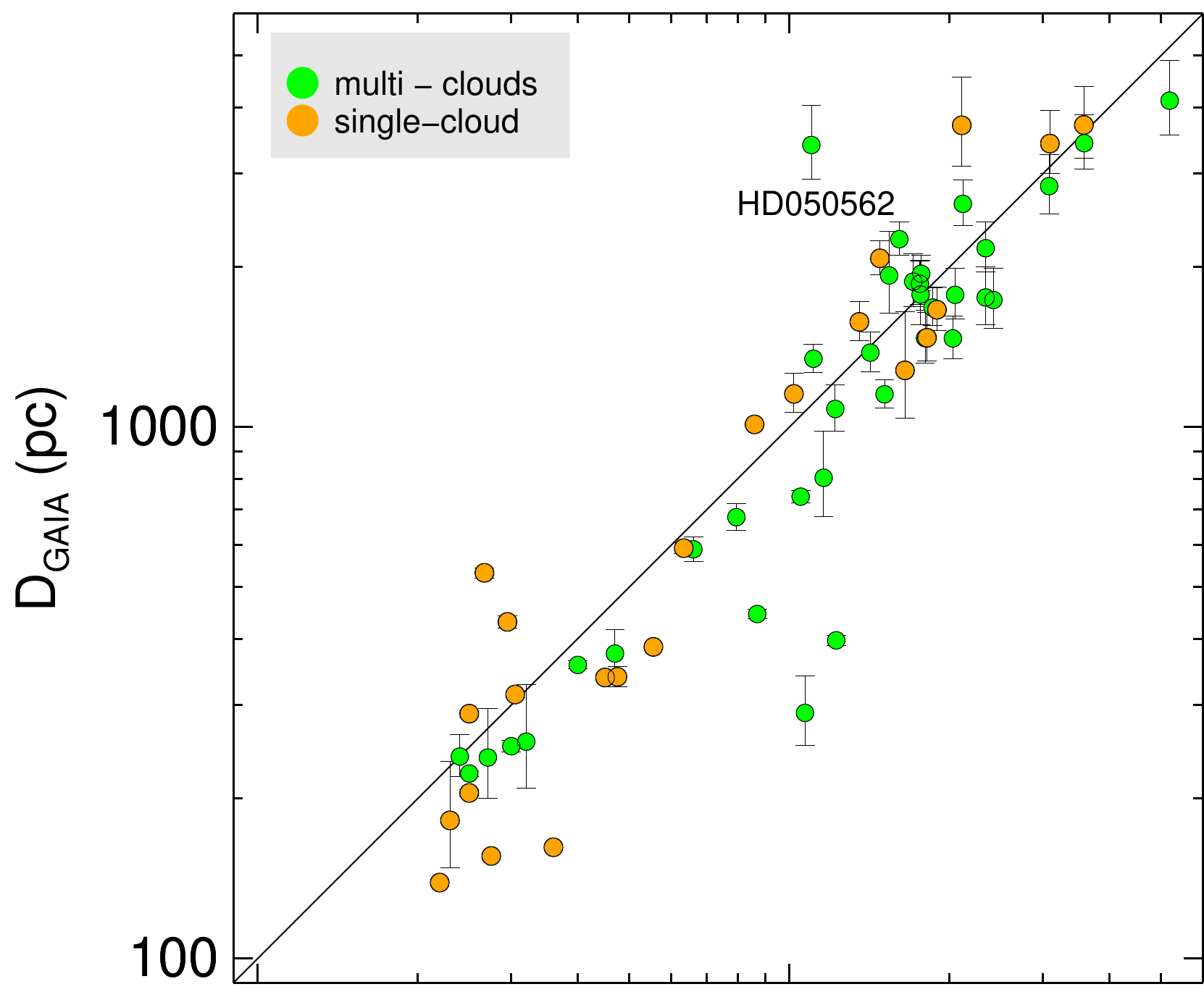}
   \includegraphics[width=8cm,clip=true,trim=0.7cm 4.5cm 4.cm 7.2cm]{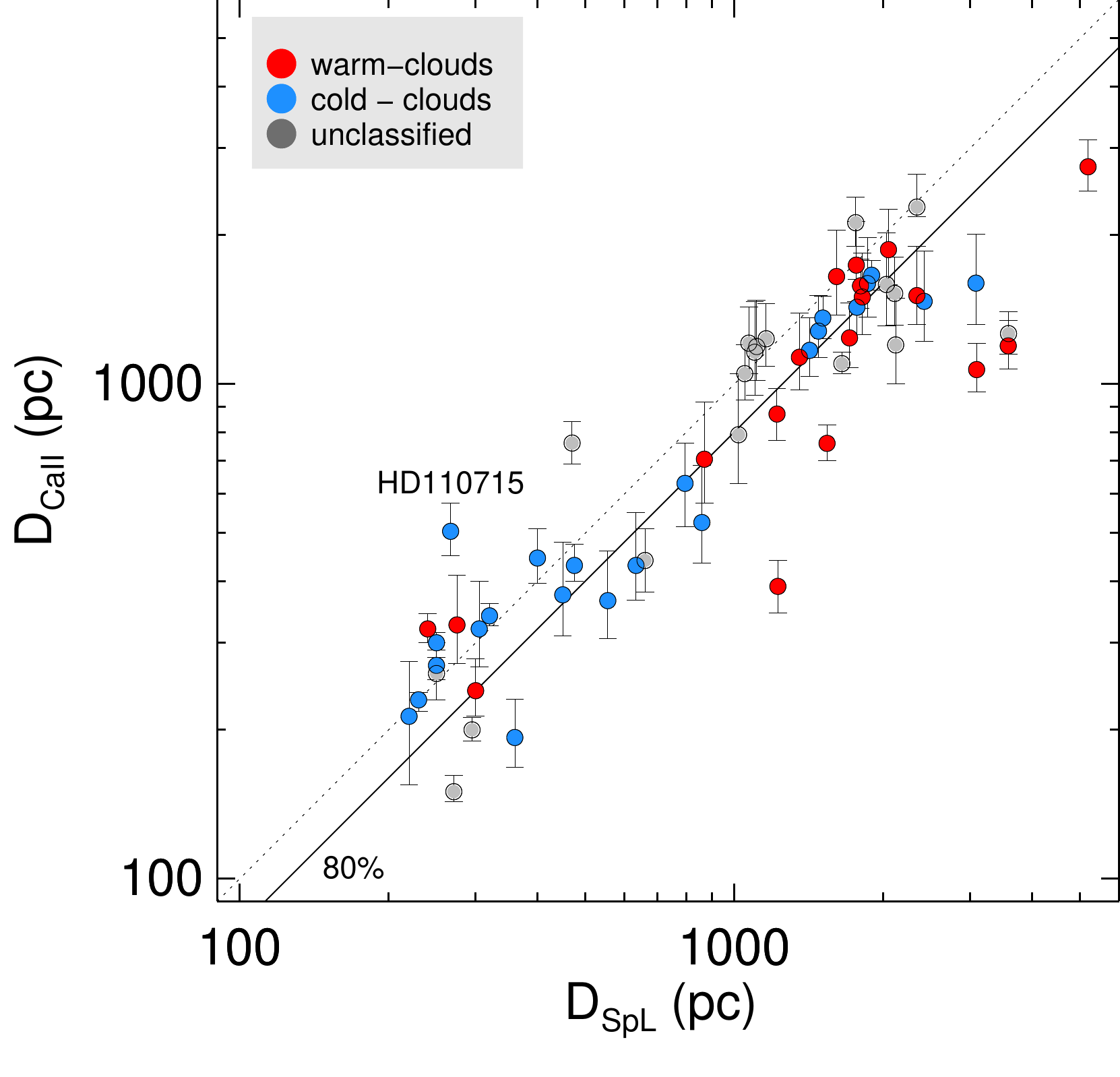}
   \caption{Relation between distance estimates using GAIA parallax
     $D_{\rm GAIA}$ \citep{Bailer18}, spectral type-luminosity
     $D_{\rm SpL}$, and the Ca~{\sc ii} H, K doublet, $D_{\rm
       {Ca}~II}$. Squares (grey) represent the Ca~{\sc ii} distances
     by \cite{Megier09}, and circles indicate the distances as listed
     in Table~\ref{res.tab}, with sightlines dominated by single-cloud
     (orange), multiple clouds (green), cold (blue), and warm (red)
     clouds, as labelled.
\label{dist.fig}}
\end{figure}

The relationships between the distance estimates $D_{\rm {GAIA}}$,
$D_{\rm {Ca~II}}$, and $D_{\rm {SpL}}$ derived in
Sect.~\ref{distances.sec} are shown in Fig.~\ref{dist.fig}.  Binaries
and stars associated with clouds are ignored.  For a reasonable
$D_{\rm {Ca~II}}$ estimate the ratio of the equivalent width of the
Ca~{\sc ii} H and K doublet is $EW(H)/EW(K) > 1.32$. We considered
stars with all three estimates measured at $>3\sigma$ confidence and
excluded binaries. This returned samples that included 59 pairs of
distance estimates that correlate with Pearson’s coefficient $\rho
(D_{\rm {GAIA}},D_{\rm {Ca~II}}) \sim 0.74$, $\rho(D_{\rm
  {GAIA}},D_{\rm {SpL}}) \sim 0.89$, and $\rho(D_{\rm {Ca~II}},D_{\rm
  {SpL}}) \sim 0.82$. At short distances, $D_{\rm {GAIA}} \leq
2$\,kpc, there are 48 pairs that are correlated more strongly ($\rho
\simgreat 0.87$), while at larger distance, $D_{\rm {GAIA}} > 2$\,kpc,
11 pairs remain and the correlation with GAIA distances breaks down.

In Fig.~\ref{dist.fig} we also show 232 distance estimates derived
from the Ca~{\sc ii} H, K doublet measurements at $>3\sigma$
confidence and $EW(H)/EW(K) > 1.32$ by \cite{Megier09}. They show
similar scatter, and their $D_{\rm {Ca~II}}$ estimates are correlated
at $D_{\rm {GAIA}} < 6$\,kpc with $\rho = 0.7$, and a subsample at
$D_{\rm {GAIA}} \leq 2$\,kpc is correlated with $\rho = 0.83$. The
sample of \cite{Megier09} and our sample have 17 stars in common. The
equivalent width of the Ca~{\sc ii} H, K generally agrees to better
than 4\%, with three exceptions: HD~027778, HD~099872, and
HD~141318. For these stars the equivalent width shows larger
variations, so that the derived Ca~{\sc ii} distances differ by more
than 20\%.  When it was available, we used the UVES-derived $D_{\rm
  {Ca~II}}$ estimates because this instrument provides the highest
resolving power. \cite{Megier09} applied HIPPARCOS \citep{HIPPARCOS}
data to derive their $D_{\rm {Ca~II}}$ formulae.  We show the two data
samples in top panel of Fig.~\ref{dist.fig}. No significant trend in
the equivalent width of the Ca~{\sc ii}~(H, K) doublet with the
parallax distance estimates is visible that would reduce the observed
scatter.

Overall, we note a large variation of about a factor of two when
different distance estimates of the same star are compared.  Within
this scatter, $D_{\rm {Ca~II}} \sim D_{\rm {GAIA}}$ while $D_{\rm
  {Ca~II}} \sim 0.8 \times D_{\rm {SpL}}$, but the sparse statistics
prevents firm conclusions.  Fig.~\ref{dist.fig} shows that the
variation in the distances remains when subsamples of sightlines are
built that are dominated by single or multiple clouds and warm or cold
cloud environments.  The reddening of our sample is not correlated
with the GAIA distance either ($\rho(E_{\rm {(B-V)}}, D_{\rm
  {GAIA}})=0.2$).

We mark in Fig.~\ref{dist.fig} cases where the spectral
type-luminosity and Ca~{\sc ii} distances agree but the GAIA distance
at low and high value of the parallax differs by more than a factor of
three. For example, HD~050562 is in GAIA located far away at 3.4\,kpc,
while $D_{\rm {Ca~II}} \sim D_{\rm {SpL}} \sim 1.1$\,kpc are closer by
a factor three. On the other hand, HD~143054 is predicted to be at
300\,pc nearby when GAIA is used, while $D_{\rm {Ca~II}} \sim D_{\rm
  {SpL}} \simgreat 1.1$\,kpc, which is a factor four farther away. The
distances of HD~024263 derived from GAIA and spectral type-luminosity
ratio are at $\sim 220$\,pc, while $D_{\rm {Ca~II}}$ estimates are at
670\,pc, or finally, HD~110715 shows $D_{\rm {Ca~II}} \sim D_{\rm
  {SpL}} \sim {530}$\,pc, while $D_{\rm {SpL}} \sim 270$\,pc. Further
discrepancies can be identified in Table~\ref{res.tab}. No method can
be favoured for individual stars over another method for deriving
distances.

\section{Dark dust \label{dark.sec}}

The distance $D$ of a star, its apparent $m_{\rm V}$ and absolute
magnitude $M_{\rm V}$, and the dust extinction along the sightline is
connected by the photometric equation. Dust absorbs relatively more
blue than red photons, so that the interstellar extinction is often
exchanged by the term “reddening” and simplified to be given by
$A_{\rm V} = R_{\rm V} \times E_{\rm {B-V}}$. However, it was never
proved that it is allowed to neglect any additional constant
extinction that represents some neutral, grey, or as we call it, {\it
  \textup{dark dust}}. The original form of the photometric equation
by \cite{Trumpler} includes in addition to the wavelength-dependent selective
interstellar extinction also a constant $C_{\rm {Dark}}$
that represents dark dust,

\begin{equation}
m_{\rm V} - M_{\rm V} = 5 \log{D} -5 + R_{\rm V}
\times E_{\rm {B-V}} + C_{\rm {Dark}}
.\end{equation}

\noindent \cite{Trumpler} estimated $C_{\rm {Dark}} \sim
0.2$\,mag/kpc, which today appears a crude underestimate. He proposed
``meteoric bodies'', and we claim grains of sizes larger than the
wavelength of the obscured light to explain dark dust
extinction. Several studies have claimed the detection of a dark dust
component populated by $\simgreat 1\mu$m large grains. Dark dust is
frequently found around circumstellar shells, e.g. \citet{Strom},
\citet{Sitko94}, \citet{Lanz95} or evolved stars,
e.g. \citet{Apruzese74}, \citet{Andriesse78}, \citet{Jura01},
\citet{Scicluna15b}. This dark dust often occurs together with a
total-to-selective extinction that is higher than the canonical value
of Milky Way dust of $R_{\rm V} \simgreat 3.1$
\citep{Fitzpatrick19}. More recently, \cite{Krelowski16} reported that
the Ca~{\sc ii} as well as VLBI parallax distances of the Orion
Trapezium star HD~037020 differ from the spectral type-luminosity
distance by a factor 2.5. They interpreted this as being due to
$A_{\rm {Dark}} = 1.8$\,mag dark dust extinction in front of the
star. Dark dust has also been detected in the general field by
\cite{Skorzynski03}.  They reported that towards some OB stars that
are located within 400\,pc of the solar neighbourhood, the intrinsic
absolute magnitudes are too faint so that the derived spectral
type-luminosity distances are significantly overestimated when
compared to HIPPARCOS distances.  This discrepancy is explained by a
few magnitudes of dark dust extinction in the ISM.

We considered stars of Table~\ref{res.tab} at GAIA distances
{$\simless 2$\,kpc}, measured at $>{4}\sigma$ confidence, and excluded
binaries, as reported in Table~\ref{sample.tab}. For 9 out of 77 of
such stars the spectral type-luminosity distance differs significantly
by more than $\pm 50$\,\% from the GAIA distance.  This interval
includes eight stars with $D_{\rm {SpL}} \simgreat 1.5 \times D_{\rm
  {GAIA}}$ (Table~\ref{dark.tab}) and one peculiar star, HD~112954,
for which $D_{\rm {GAIA}} \sim 2.25 \ D_{\rm {SpL}}$. When these eight
stars are excluded, the scatter in $D_{\rm {SpL}} / D_{\rm {GAIA}}$ is
reduced to {$\sim 22$}\%. We find first that dark dust sightlines
appear most frequently in the field. Our sample includes only one
star, HD~037903, with which a dust cloud can be associated in WISE
imaging (Sect.~\ref{MIR.sec}). This star illuminates the bright
reflection nebula NGC~2023 in Orion and shows vibrationally excited
interstellar H$_2$, which is indicative of a photodissociation region
around the star \citep{Meyer01,Gnacinski}.

We also find that the extinction of light by dark dust occurs in the
cold ISM. At least all dark dust sightlines detected by us are
associated with cold cloud environments (Sect.~\ref{env.sec}).

Finally, dark dust sightlines show predominantly flat extinction curves.
Six such stars lie at $R_{\rm V} \simgreat 3.8$. Flat extinction
curves are associated with large grains.

We estimated the amount of dark extinction under the assumption that the GAIA
distance is precise and the luminosity $L$ of the star is correctly
derived by our spectral classification.  The observed flux is given by

\begin{equation}
  {L \ e^{-\tau} \over 4 \pi \ D^2_{\rm {SpL}}} =
  {L \ e^{-(\tau + \tau_{\rm {dark}})} \over 4 \pi \ D^2_{\rm {GAIA}}} 
,\end{equation}

\noindent hence $\tau_{\rm {dark}} = \ln {({D^2_{\rm {SpL}} / D^2_{\rm
      {GAIA}}}),}$ which amounts to typically $1 - 3$\, mag of dark
dust extinction (Table~\ref{dark.tab}). High-precision extinction
measurements in the near-IR or at even longer wavelengths are required
for a clear identification of large particles that remain hidden in
the dark dust component of the ISM.

\begin{table}[!htb]
\scriptsize
\begin{center}
  \caption {Dark dust detections. \label{dark.tab} Column (1) lists
    the counter, Col. (2) the HD identifier, Col. (3) the
    total-to-selective extinction $R_{\rm V}$, Col. (4) the spectral
    type-luminosity distance, Col. (5) the GAIA distance, Col. (6)
    the visual extinction, and Col. (7) the dark dust extinction
    $A_{\rm {Dark}}$\,.}
    \begin{tabular}{r c c r r c c}
\hline\hline
1  & 2    & 3            & 4 &     5                   &     6   & 7    \\
\hline
& Name & $R_{\rm V}^{\it a}$   & $D_{\rm {SpL}}$ & $D_{\rm {GAIA}}^{\it b}$ &  $A_{\rm V}$ &  $A_{\rm {Dark}}$  \\
    &             &         & pc      & pc      & mag     & mag \\
\hline
 1  &   HD~037130  &    5.5  &    705  &    400  &    1.3  &    1.2 \\ 
 2  &   HD~037903  &    4.1  &   1225  &    397  &    1.5  &    2.4 \\ 
 3  &   HD~096675  &    3.8  &    360  &    162  &    1.1  &    1.7 \\ 
 4  &   HD~143054  &    2.8  &   1070  &    290  &    1.5  &    2.8 \\ 
 5  &   HD~146285  &    3.8  &    275  &    156  &    1.2  &    1.2 \\ 
 6  &   HD~147888  &    4.1  &    185  &     92  &    2.0  &    1.5 \\ 
 7  &   HD~169582  &    3.0  &   3500  &   1681  &    2.6  &    1.6 \\ 
 8  &   HD~294264  &    5.5  &    755  &    445  &    2.8  &    1.1 \\
 \hline
 \end{tabular}
\end{center}
    {\bf Notes:} $^{\it a}$ Table~\ref{appsample.tab}, $^{\it a}$  \cite{Bailer18}
\end{table}

\section{Summary and conclusions \label{concl.sec}}
We have argued here that our interpretation of the extinction and
polarisation of the interstellar medium is strongly biased by the fact
that multiple interstellar clouds may be present along any line of
sight. Recently, we have shown that when the analysis is limited to
{what we call} single-cloud sightlines, new relationships appear
between the observing characteristics of extinction and polarisation
and the physical properties of the dust. These relations are hidden in
multiple-cloud sightlines where interstellar lines are Doppler-split,
while single-cloud sightlines are suitable to investigate the pristine
nature of interstellar material. Unfortunately, while hundreds of
extinction and polarisation curves have been measured for the ISM, the
number of single-cloud sightlines that have been investigated is still
very limited.  \citet{S18} argued that data for detailed dust
modelling are available for only eight sightlines.  To overcome this
observational bias, we performed an extensive survey of
high-resolution spectra of reddened OB stars for which the far-UV
extinction curve has been measured by IUE or FUSE.  We compiled a
sample of {186} high-resolution spectra of such stars, 100 of which
were observed by us using the UVES instrument of the ESO Very Large
Telescope. Archive spectra with a broad wavelength coverage for
another 36 sightlines where retrieved from various archives, as well
as {50} accompanying UVES narrow-band archive spectra that include the
K~{\sc i} line. All sightlines for which UVES raw data were available
were reprocessed by us with our interactive analysis package {\sc
  dech}.

Dust properties are assumed to vary on a large scale, typically when
dust clouds are separated by several 10s or 100s of pc, whereas within
a single-cloud, the mean characteristics of dust remain. This paper
was tailored to detect single-cloud dominated sightlines that can be
used for further investigations of the dust characteristics in the
diffuse ISM.  In this context, we assign the term single-cloud
sightline when the observed interstellar lines observed at $\sim
4$\ km/s (FWHM) resolution showed one dominating Doppler component. In
this way, we realised that interstellar lines come in families of
distinct radial velocities with K\,{\sc i} coinciding with CH, while
CH$+$ is often shifted, and Ca\,{\sc ii} and Ti\,{\sc ii} show similar
and more complex structures. These offsets in the velocity profiles,
as well as local small-scale variations and details in the cloud
morphology, were not considered. Comparisons of our stars that are in
common with previous ultra-high resolution spectra by \citet{Welty01}
and \citet{Welty03}, for example, confirmed our assignment of
multiple-component sightlines.  Such spectra often exhibit fine
structures in lines that appear to be single when observed at
resolution of 1.5\,km/s \citep{Welty14}.  Nevertheless, the comparison
also demonstrated that lower resolution spectra at $> 4$\,km/s (FWHM)
are insufficient for this classification. We identified a sample of 65
single-cloud sightlines in the 186 high-resolution spectra, the
majority of which were previously unknown.
 
In 97 sightlines we detected CH and CH$^+$, and we used their strength
ratio for the classification into warm (line ratio CH/CH$^+ < 1$) or
cold (CH/CH$^+ > 1$) clouds. In all {52} cold clouds we did detect CN,
while we did not detect it in {32 out of 36} warm clouds. Most of the
clouds are cold and located sufficiently far away from the heating
source in the cold diffuse ISM. We inspected the WISE $3 - 22 \, \mu$m
emission morphology of our clouds. They appear predominantly stellar,
while dust emission that is associated with the star is detected in
only 10\% of the cases. Therefore we exclude a circumstellar nature of
the observed reddening for most cases, and the extinction towards our
stars is due to dust located in the diffuse ISM.

The high resolving power of our data has allowed us to accurately
measure the Mg~{\sc ii}/He~{\sc i} and He~{\sc i}/H~{\sc i} line
intensity ratios, which {were used in addition to other features} to
revisit the spectral classification of our targets.  In the majority
of cases (89 of 136), we confirmed previous assignments found in the
literature, but for 47 of 136 stars, our revision of spectral type was
substantial. We identified binary systems in 22 sightlines. We used
our revised classification to compute the spectral type-luminosity
distance of the stars. We also measured the equivalent width of the
Ca\,{\sc ii} H and K doublet and estimated the spectroscopic distance
of the stars. We compared the two distance estimates to the GAIA
parallaxes. Overall, the scatter in the distance estimates is large at
$\sim 40$\,\%. In some cases, only one of the three distances differs
heavily from the others. Sometimes, to agree with Gaia, spectroscopic
estimates require an unacceptably low luminosity of the stars.  We
cannot favour one method over others to derive the distances.

Finally, we detected a hidden dust population that amounts to a 1 -- 3
mag extinction in the ISM. We called this {\it {dark dust}}. For these
stars, the spectral type-luminosity distances are significantly
larger than derived by GAIA.  Dark dust predominantly appears in the
cold ISM with flat extinction curves. Dark dust does not show a
wavelength-dependent absorption of stellar light in the optical. It is
presumably made of very large $\simgreat 1\mu$m sized grains.  For a
direct identification of such dark dust particles, high-precision
extinction measurements in the near-IR or at even longer wavelength
are required.

In a forthcoming paper we will detail our stellar classifications and
we will use the sample of {65} single-cloud sightlines to investigate
the polarisation properties of the interstellar medium, and we will
search for correlations between diffuse interstellar bands, atomic and
molecular lines, and other properties of ISM dust such as the reported
dark dust component.

\begin{acknowledgements}
  We are greatful to the referee for pointing us to the UHRF data and
  valuable comments by Daniel Welty. This research is based on data
  obtained from the ESO Science Archive Facility and in particular on
  observations collected under ESO observing programme ID
  0102.C-0040. Additionally, data from the following ESO programme IDs
  was used in this paper: 065.I-0526(A), 065.N-0378(A), 066.B-0320(A),
  071.C-0513(C), 073.C-0337(A), 073.D-0024(A), 076.C-0431(A),
  096.D-0008(A), 099.C-0637(A), 194.C-0833(A). We also used spectra
  retrieved from the ELODIE and SOPHIE archives at Observatoire de
  Haute-Provence (OHP)\footnote{available at atlas.obs-hp.fr/elodie
    and atlas.obs-hp.fr/sophie}.  This research has also made use of
  the SIMBAD database, operated at the CDS, Strasbourg, France.  GAG
  and JK acknowledge the financial support of the Chilean fund CONICYT
  grant REDES 180136.  JK acknowledge the financial support of the
  Polish National Science Centre, Poland (2017/25/B/ST9/01524) for the
  period 2018 - 2021.
\end{acknowledgements}


\bibliographystyle{aa}
\bibliography{References}

\newcommand{\noop}[1]{}
\begin{thebibliography}{85}
\expandafter\ifx\csname natexlab\endcsname\relax\def\natexlab#1{#1}\fi

\bibitem[{{Andriesse} {et~al.}(1978){Andriesse}, {Donn}, \&
  {Viotti}}]{Andriesse78}
{Andriesse}, C.~D., {Donn}, B.~D., \& {Viotti}, R. 1978, \mnras, 185, 771

\bibitem[{{Apruzese}(1974)}]{Apruzese74}
{Apruzese}, J.~P. 1974, \apj, 188, 539

\bibitem[{{Bagnulo} {et~al.}(2017){Bagnulo}, {Cox}, {Cikota}, {Siebenmorgen},
  {Voshchinnikov}, {Patat}, {Smith}, {Smoker}, {Taubenberger}, {Kaper}, {Cami},
  \& {LIPS Collaboration}}]{Bagnulo17}
{Bagnulo}, S., {Cox}, N. L.~J., {Cikota}, A., {et~al.} 2017, \aap, 608, A146

\bibitem[{{Bailer-Jones} {et~al.}(2018){Bailer-Jones}, {Rybizki}, {Fouesneau},
  {Mantelet}, \& {Andrae}}]{Bailer18}
{Bailer-Jones}, C.~A.~L., {Rybizki}, J., {Fouesneau}, M., {Mantelet}, G., \&
  {Andrae}, R. 2018, \aj, 156, 58

\bibitem[{{Barlow} {et~al.}(1995){Barlow}, {Crawford}, {Diego}, {Dryburgh},
  {Fish}, {Howarth}, {Spyromilio}, \& {Walker}}]{Barlow95}
{Barlow}, M.~J., {Crawford}, I.~A., {Diego}, F., {et~al.} 1995, \mnras, 272,
  333

\bibitem[{{Chini} \& {Kruegel}(1983)}]{CK}
{Chini}, R. \& {Kruegel}, E. 1983, \aap, 117, 289

\bibitem[{{Cox} {et~al.}(2017){Cox}, {Cami}, {Farhang}, {Smoker},
  {Monreal-Ibero}, {Lallement}, {Sarre}, {Marshall}, {Smith}, {Evans}, {Royer},
  {Linnartz}, {Cordiner}, {Joblin}, {van Loon}, {Foing}, {Bhatt}, {Bron},
  {Elyajouri}, {de Koter}, {Ehrenfreund}, {Javadi}, {Kaper}, {Khosroshadi},
  {Laverick}, {Le Petit}, {Mulas}, {Roueff}, {Salama}, \& {Spaans}}]{Cox17}
{Cox}, N. L.~J., {Cami}, J., {Farhang}, A., {et~al.} 2017, \aap, 606, A76

\bibitem[{{Crane} {et~al.}(1995){Crane}, {Lambert}, \& {Sheffer}}]{Crane95}
{Crane}, P., {Lambert}, D.~L., \& {Sheffer}, Y. 1995, \apjs, 99, 107

\bibitem[{{Crawford}(2002)}]{Crawford02}
{Crawford}, I.~A. 2002, \mnras, 334, L33

\bibitem[{{Dekker} {et~al.}(2000){Dekker}, {D'Odorico}, {Kaufer}, {Delabre}, \&
  {Kotzlowski}}]{Dekker00}
{Dekker}, H., {D'Odorico}, S., {Kaufer}, A., {Delabre}, B., \& {Kotzlowski}, H.
  2000, in Society of Photo-Optical Instrumentation Engineers (SPIE) Conference
  Series, Vol. 4008, \procspie, ed. M.~{Iye} \& A.~F. {Moorwood}, 534--545

\bibitem[{Diego {et~al.}(1995)Diego, Fish, Barlow, Crawford, Spyromilio,
  Dryburgh, Brooks, Howarth, \& Walker}]{Diego95}
Diego, F., Fish, A.~C., Barlow, M.~J., {et~al.} 1995, Monthly Notices of the
  Royal Astronomical Society, 272, 323

\bibitem[{{Elmegreen}(2002)}]{Elmegreen02}
{Elmegreen}, B.~G. 2002, \apj, 564, 773

\bibitem[{{Ensor} {et~al.}(2017){Ensor}, {Cami}, {Bhatt}, \& {Soddu}}]{Ensor}
{Ensor}, T., {Cami}, J., {Bhatt}, N.~H., \& {Soddu}, A. 2017, \apj, 836, 162

\bibitem[{ESA(1997)}]{HIPPARCOS}
ESA. 1997, {The HIPPARCOS and TYCHO catalogues. Astrometric and photometric
  star catalogues derived from the ESA HIPPARCOS Space Astrometry Mission},
  Vol. 1200 (https://ui.adsabs.harvard.edu/abs/1997ESASP1200.....E)

\bibitem[{{Falgarone} {et~al.}(1991){Falgarone}, {Phillips}, \&
  {Walker}}]{Falgarone91}
{Falgarone}, E., {Phillips}, T.~G., \& {Walker}, C.~K. 1991, \apj, 378, 186

\bibitem[{{Falle} \& {Hartquist}(2002)}]{FalleHartquist02}
{Falle}, S.~A.~E.~G. \& {Hartquist}, T.~W. 2002, \mnras, 329, 195

\bibitem[{{Fan} {et~al.}(2019){Fan}, {Hobbs}, {Dahlstrom}, {Welty}, {York},
  {Rachford}, {Snow}, {Sonnentrucker}, {Baskes}, \& {Zhao}}]{Fan19}
{Fan}, H., {Hobbs}, L.~M., {Dahlstrom}, J.~A., {et~al.} 2019, \apj, 878, 151

\bibitem[{{Field}(1974)}]{Field}
{Field}, G.~B. 1974, \apj, 187, 453

\bibitem[{{Fitzpatrick} \& {Massa}(1990)}]{FM90}
{Fitzpatrick}, E.~L. \& {Massa}, D. 1990, \apjs, 72, 163

\bibitem[{{Fitzpatrick} \& {Massa}(2007)}]{FM07}
{Fitzpatrick}, E.~L. \& {Massa}, D. 2007, \apj, 663, 320

\bibitem[{{Fitzpatrick} {et~al.}(2019){Fitzpatrick}, {Massa}, {Gordon},
  {Bohlin}, \& {Clayton}}]{Fitzpatrick19}
{Fitzpatrick}, E.~L., {Massa}, D., {Gordon}, K.~D., {Bohlin}, R., \& {Clayton},
  G.~C. 2019, \apj, 886, 108

\bibitem[{{Galazutdinov}(2005)}]{Gazinur05}
{Galazutdinov}, G. 2005, Journal of Korean Astronomical Society, 38, 215

\bibitem[{{Galazutdinov} {et~al.}(2000){Galazutdinov}, {Kre{\l}owski}, \&
  {Musaev}}]{Galazutdinov00}
{Galazutdinov}, G.~A., {Kre{\l}owski}, J., \& {Musaev}, F.~A. 2000, \mnras,
  315, 703

\bibitem[{{Gnaci{\'n}ski}(2011)}]{Gnacinski}
{Gnaci{\'n}ski}, P. 2011, \aap, 532, A122

\bibitem[{{Gordon} {et~al.}(2009){Gordon}, {Cartledge}, \&
  {Clayton}}]{Gordon09}
{Gordon}, K.~D., {Cartledge}, S., \& {Clayton}, G.~C. 2009, \apj, 705, 1320

\bibitem[{{Heger}(1922)}]{Heger22}
{Heger}, M.~L. 1922, Lick Observatory Bulletin, 10, 141

\bibitem[{{Hong} \& {Greenberg}(1980)}]{HG80}
{Hong}, S.~S. \& {Greenberg}, J.~M. 1980, \aap, 88, 194

\bibitem[{{Houk}(1982)}]{Houk75}
{Houk}, N. 1982, {Michigan Catalogue of Two-dimensional Spectral Types for the
  HD stars} (Havard 1975 MCTS Book)

\bibitem[{{Houk} \& {Swift}(1999)}]{Houk99}
{Houk}, N. \& {Swift}, C. 1999, Michigan Spectral Survey, 5, 0

\bibitem[{{Howarth} {et~al.}(2002){Howarth}, {Price}, {Crawford}, \&
  {Hawkins}}]{Howarth02}
{Howarth}, I.~D., {Price}, R.~J., {Crawford}, I.~A., \& {Hawkins}, I. 2002,
  \mnras, 335, 267

\bibitem[{{Hunter} {et~al.}(2006){Hunter}, {Smoker}, {Keenan}, {Ledoux},
  {Jehin}, {Cabanac}, {Melo}, \& {Bagnulo}}]{Hunter06}
{Hunter}, I., {Smoker}, J.~V., {Keenan}, F.~P., {et~al.} 2006, \mnras, 367,
  1478

\bibitem[{{Jura} {et~al.}(2001){Jura}, {Webb}, \& {Kahane}}]{Jura01}
{Jura}, M., {Webb}, R.~A., \& {Kahane}, C. 2001, \apjl, 550, L71

\bibitem[{{Kaufer} {et~al.}(1999){Kaufer}, {Stahl}, {Tubbesing},
  {N{\o}rregaard}, {Avila}, {Francois}, {Pasquini}, \& {Pizzella}}]{Kaufer99}
{Kaufer}, A., {Stahl}, O., {Tubbesing}, S., {et~al.} 1999, The Messenger, 95, 8

\bibitem[{{Kim} {et~al.}(2007){Kim}, {Han}, {Valyavin}, {Plachinda}, {Jang},
  {Jang}, {Seong}, {Lee}, {Kang}, {Park}, {Yoon}, \& {Vogt}}]{Kim07}
{Kim}, K.-M., {Han}, I., {Valyavin}, G.~G., {et~al.} 2007, \pasp, 119, 1052

\bibitem[{{Klein} {et~al.}(1994){Klein}, {McKee}, \& {Colella}}]{Klein94}
{Klein}, R.~I., {McKee}, C.~F., \& {Colella}, P. 1994, \apj, 420, 213

\bibitem[{{Kre{\l}owski} {et~al.}(2010){Kre{\l}owski}, {Beletsky},
  {Galazutdinov}, {Ko{\l}os}, {Gronowski}, \& {LoCurto}}]{K10}
{Kre{\l}owski}, J., {Beletsky}, Y., {Galazutdinov}, G.~A., {et~al.} 2010,
  \apjl, 714, L64

\bibitem[{{Kre{\l}owski} {et~al.}(2019{\natexlab{a}}){Kre{\l}owski},
  {Galazutdinov}, \& {Bondar}}]{K19b}
{Kre{\l}owski}, J., {Galazutdinov}, G., \& {Bondar}, A. 2019{\natexlab{a}},
  \mnras, 486, 3537

\bibitem[{{Kre{\l}owski} {et~al.}(2016){Kre{\l}owski}, {Galazutdinov},
  {Bondar}, \& {Beletsky}}]{Krelowski16}
{Kre{\l}owski}, J., {Galazutdinov}, G.~A., {Bondar}, A., \& {Beletsky}, Y.
  2016, \mnras, 460, 2706

\bibitem[{{Krelowski} {et~al.}(1992){Krelowski}, {Snow}, {Seab}, \&
  {Papaj}}]{Krelowski92}
{Krelowski}, J., {Snow}, T.~P., {Seab}, C.~G., \& {Papaj}, J. 1992, \mnras,
  258, 693

\bibitem[{{Kre{\l}owski} {et~al.}(2019{\natexlab{b}}){Kre{\l}owski}, {Strobel},
  {Galazutdinov}, {Bondar}, \& {Valyavin}}]{K19}
{Kre{\l}owski}, J., {Strobel}, A., {Galazutdinov}, G.~A., {Bondar}, A., \&
  {Valyavin}, G. 2019{\natexlab{b}}, \mnras, 486, 112

\bibitem[{{Kre{\l}owski} {et~al.}(2018){Kre{\l}owski}, {Strobel},
  {Galazutdinov}, {Musaev}, \& {Bondar}}]{Krelowski18}
{Kre{\l}owski}, J., {Strobel}, A., {Galazutdinov}, G.~A., {Musaev}, F., \&
  {Bondar}, A. 2018, \actaa, 68, 285

\bibitem[{{Krelowski} \& {Walker}(1987)}]{K87}
{Krelowski}, J. \& {Walker}, G.~A.~H. 1987, \apj, 312, 860

\bibitem[{{Kr{\"u}gel}(2008)}]{K08}
{Kr{\"u}gel}, E. 2008, {An introduction to the physics of interstellar dust}
  (IOP)

\bibitem[{{Kr{\"u}gel}(2009)}]{K09}
{Kr{\"u}gel}, E. 2009, \aap, 493, 385

\bibitem[{{Lanz} {et~al.}(1995){Lanz}, {Heap}, \& {Hubeny}}]{Lanz95}
{Lanz}, T., {Heap}, S.~R., \& {Hubeny}, I. 1995, \apjl, 447, L41

\bibitem[{{Lauroesch} {et~al.}(2000){Lauroesch}, {Meyer}, \&
  {Blades}}]{Lauroesch00}
{Lauroesch}, J.~T., {Meyer}, D.~M., \& {Blades}, J.~C. 2000, \apjl, 543, L43

\bibitem[{{McGuire}(2018)}]{McGuire18}
{McGuire}, B.~A. 2018, \apjs, 239, 17

\bibitem[{{Megier} {et~al.}(2005){Megier}, {Strobel}, {Bondar}, {Musaev},
  {Han}, {Kre{\L}owski}, \& {Galazutdinov}}]{Megier05}
{Megier}, A., {Strobel}, A., {Bondar}, A., {et~al.} 2005, \apj, 634, 451

\bibitem[{{Megier} {et~al.}(2009){Megier}, {Strobel}, {Galazutdinov}, \&
  {Kre{\l}owski}}]{Megier09}
{Megier}, A., {Strobel}, A., {Galazutdinov}, G.~A., \& {Kre{\l}owski}, J. 2009,
  \aap, 507, 833

\bibitem[{{Meyer} {et~al.}(2001){Meyer}, {Lauroesch}, {Sofia}, {Draine}, \&
  {Bertoldi}}]{Meyer01}
{Meyer}, D.~M., {Lauroesch}, J.~T., {Sofia}, U.~J., {Draine}, B.~T., \&
  {Bertoldi}, F. 2001, \apjl, 553, L59

\bibitem[{{Morris} {et~al.}(2016){Morris}, {Gupta}, {Nagy}, {Pearson},
  {Ossenkopf-Okada}, {Falgarone}, {Lis}, {Gerin}, {Melnick}, {Neufeld}, \&
  {Bergin}}]{Morris16}
{Morris}, P.~W., {Gupta}, H., {Nagy}, Z., {et~al.} 2016, \apj, 829, 15

\bibitem[{{Moultaka} {et~al.}(2004){Moultaka}, {Ilovaisky}, {Prugniel}, \&
  {Soubiran}}]{Moultaka04}
{Moultaka}, J., {Ilovaisky}, S.~A., {Prugniel}, P., \& {Soubiran}, C. 2004,
  \pasp, 116, 693

\bibitem[{{Pan} {et~al.}(2005){Pan}, {Federman}, {Sheffer}, \&
  {Andersson}}]{Pan}
{Pan}, K., {Federman}, S.~R., {Sheffer}, Y., \& {Andersson}, B.~G. 2005, \apj,
  633, 986

\bibitem[{{Price} {et~al.}(2000){Price}, {Crawford}, \& {Barlow}}]{Price00}
{Price}, R.~J., {Crawford}, I.~A., \& {Barlow}, M.~J. 2000, \mnras, 312, L43

\bibitem[{{Savage} \& {Sembach}(1991)}]{SavageSembach91}
{Savage}, B.~D. \& {Sembach}, K.~R. 1991, \apj, 379, 245

\bibitem[{{Scicluna} \& {Siebenmorgen}(2015)}]{Scicluna15}
{Scicluna}, P. \& {Siebenmorgen}, R. 2015, \aap, 584, A108

\bibitem[{{Scicluna} {et~al.}(2015){Scicluna}, {Siebenmorgen}, {Wesson},
  {Blommaert}, {Kasper}, {Voshchinnikov}, \& {Wolf}}]{Scicluna15b}
{Scicluna}, P., {Siebenmorgen}, R., {Wesson}, R., {et~al.} 2015, \aap, 584, L10

\bibitem[{{Sembach} {et~al.}(1993){Sembach}, {Danks}, \& {Savage}}]{Sembach}
{Sembach}, K.~R., {Danks}, A.~C., \& {Savage}, B.~D. 1993, \aaps, 100, 107

\bibitem[{{Siebenmorgen}(1993)}]{S93}
{Siebenmorgen}, R. 1993, \apj, 408, 218

\bibitem[{{Siebenmorgen} {et~al.}(2018){Siebenmorgen}, {Voshchinnikov},
  {Bagnulo}, {Cox}, {Cami}, \& {Peest}}]{S18}
{Siebenmorgen}, R., {Voshchinnikov}, N.~V., {Bagnulo}, S., {et~al.} 2018, \aap,
  611, A5

\bibitem[{{Sitko} {et~al.}(1994){Sitko}, {Halbedel}, {Lawrence}, {Smith}, \&
  {Yanow}}]{Sitko94}
{Sitko}, M.~L., {Halbedel}, E.~M., {Lawrence}, G.~F., {Smith}, J.~A., \&
  {Yanow}, K. 1994, \apj, 432, 753

\bibitem[{{Sk{\'o}rzy{\'n}ski} {et~al.}(2003){Sk{\'o}rzy{\'n}ski}, {Strobel},
  \& {Galazutdinov}}]{Skorzynski03}
{Sk{\'o}rzy{\'n}ski}, W., {Strobel}, A., \& {Galazutdinov}, G.~A. 2003, \aap,
  408, 297

\bibitem[{{Smoker} {et~al.}(2009){Smoker}, {Haddad}, {Iwert}, {Deiries},
  {Modigliani}, {Randall}, {D'Odorico}, {James}, {Lo Curto}, {Robert},
  {Pasquini}, {Downing}, {Ledoux}, {Martayan}, {Dall}, {Vinther}, {Melo},
  {Fox}, {Pritchard}, {Baade}, \& {Dekker}}]{Smoker09}
{Smoker}, J., {Haddad}, N., {Iwert}, O., {et~al.} 2009, The Messenger, 138, 8

\bibitem[{{Smoker} {et~al.}(2006){Smoker}, {Lynn}, {Christian}, \&
  {Keenan}}]{Smoker06}
{Smoker}, J.~V., {Lynn}, B.~B., {Christian}, D.~J., \& {Keenan}, F.~P. 2006,
  \mnras, 370, 151

\bibitem[{{Spitzer}(1978)}]{Spitzer}
{Spitzer}, L. 1978, {Physical Processes in the Interstellar Medium}
  (WILEY‐VCH Verlag GmbH \& Co. KGaA)

\bibitem[{{Strom} {et~al.}(1971){Strom}, {Strom}, \& {Yost}}]{Strom}
{Strom}, K.~M., {Strom}, S.~E., \& {Yost}, J. 1971, \apj, 165, 479

\bibitem[{{Struve}(1928)}]{Struve}
{Struve}, O. 1928, \apj, 67, 353

\bibitem[{{Tolstoy} {et~al.}(2003){Tolstoy}, {Venn}, {Shetrone}, {Primas},
  {Hill}, {Kaufer}, \& {Szeifert}}]{Tolstoy03}
{Tolstoy}, E., {Venn}, K.~A., {Shetrone}, M., {et~al.} 2003, \aj, 125, 707

\bibitem[{{Trumpler}(1930)}]{Trumpler}
{Trumpler}, R.~J. 1930, \pasp, 42, 214

\bibitem[{Tull {et~al.}(1995)Tull, MacQueen, Sneden, \& Lambert}]{Tull95}
Tull, R.~G., MacQueen, P.~J., Sneden, C., \& Lambert, D.~L. 1995, PASP, 107,
  251

\bibitem[{{Valencic} {et~al.}(2004){Valencic}, {Clayton}, \&
  {Gordon}}]{Valencic}
{Valencic}, L.~A., {Clayton}, G.~C., \& {Gordon}, K.~D. 2004, \apj, 616, 912

\bibitem[{{Wada}(2008)}]{Wada08}
{Wada}, K. 2008, \apj, 675, 188

\bibitem[{{Walborn}(2008)}]{Walborn08}
{Walborn}, N.~R. 2008, in Revista Mexicana de Astronomia y Astrofisica
  Conference Series, Vol.~33, 5--14

\bibitem[{{Walborn} \& {Fitzpatrick}(1990)}]{WF90}
{Walborn}, N.~R. \& {Fitzpatrick}, E.~L. 1990, \pasp, 102, 379

\bibitem[{{Wegner}(2003)}]{Wegner}
{Wegner}, W. 2003, Astronomische Nachrichten, 324, 219

\bibitem[{{Welty}(2014)}]{Welty14}
{Welty}, D.~E. 2014, in IAU Symposium, Vol. 297, The Diffuse Interstellar
  Bands, ed. J.~{Cami} \& N.~L.~J. {Cox}, 153--162

\bibitem[{{Welty} \& {Crowther}(2010)}]{Welty10}
{Welty}, D.~E. \& {Crowther}, P.~A. 2010, \mnras, 404, 1321

\bibitem[{{Welty} \& {Fitzpatrick}(2001)}]{Welty01b}
{Welty}, D.~E. \& {Fitzpatrick}, E.~L. 2001, \apjl, 551, L175

\bibitem[{{Welty} \& {Hobbs}(2001)}]{Welty01}
{Welty}, D.~E. \& {Hobbs}, L.~M. 2001, \apjs, 133, 345

\bibitem[{{Welty} {et~al.}(1994){Welty}, {Hobbs}, \& {Kulkarni}}]{Welty94}
{Welty}, D.~E., {Hobbs}, L.~M., \& {Kulkarni}, V.~P. 1994, \apj, 436, 152

\bibitem[{{Welty} {et~al.}(2003){Welty}, {Hobbs}, \& {Morton}}]{Welty03}
{Welty}, D.~E., {Hobbs}, L.~M., \& {Morton}, D.~C. 2003, \apjs, 147, 61

\bibitem[{{Welty} {et~al.}(1996){Welty}, {Morton}, \& {Hobbs}}]{Welty96}
{Welty}, D.~E., {Morton}, D.~C., \& {Hobbs}, L.~M. 1996, \apjs, 106, 533

\bibitem[{{Welty} {et~al.}(2020){Welty}, {Sonnentrucker}, {Snow}, \&
  {York}}]{Welty20}
{Welty}, D.~E., {Sonnentrucker}, P., {Snow}, T.~P., \& {York}, D.~G. 2020,
  arXiv e-prints, arXiv:2005.10846

\bibitem[{{Weselak} {et~al.}(2008){Weselak}, {Galazutdinov}, {Musaev}, \&
  {Kre{\l}owski}}]{Weselak}
{Weselak}, T., {Galazutdinov}, G.~A., {Musaev}, F.~A., \& {Kre{\l}owski}, J.
  2008, \aap, 484, 381

\bibitem[{{Wright} {et~al.}(2010){Wright}, {Eisenhardt}, {Mainzer}, {Ressler},
  {Cutri}, {Jarrett}, {Kirkpatrick}, {Padgett}, {McMillan}, {Skrutskie},
  {Stanford}, {Cohen}, {Walker}, {Mather}, {Leisawitz}, {Gautier}, {McLean},
  {Benford}, {Lonsdale}, {Blain}, {Mendez}, {Irace}, {Duval}, {Liu}, {Royer},
  {Heinrichsen}, {Howard}, {Shannon}, {Kendall}, {Walsh}, {Larsen}, {Cardon},
  {Schick}, {Schwalm}, {Abid}, {Fabinsky}, {Naes}, \& {Tsai}}]{Wright10}
{Wright}, E.~L., {Eisenhardt}, P. R.~M., {Mainzer}, A.~K., {et~al.} 2010, \aj,
  140, 1868

\end{thebibliography}


\begin{appendix}

  \section{Tables and figures}
\begin{figure*} [!htbp] 
  \includegraphics[width=18.5cm,clip=true,trim=0cm 3cm 0cm 0.cm]{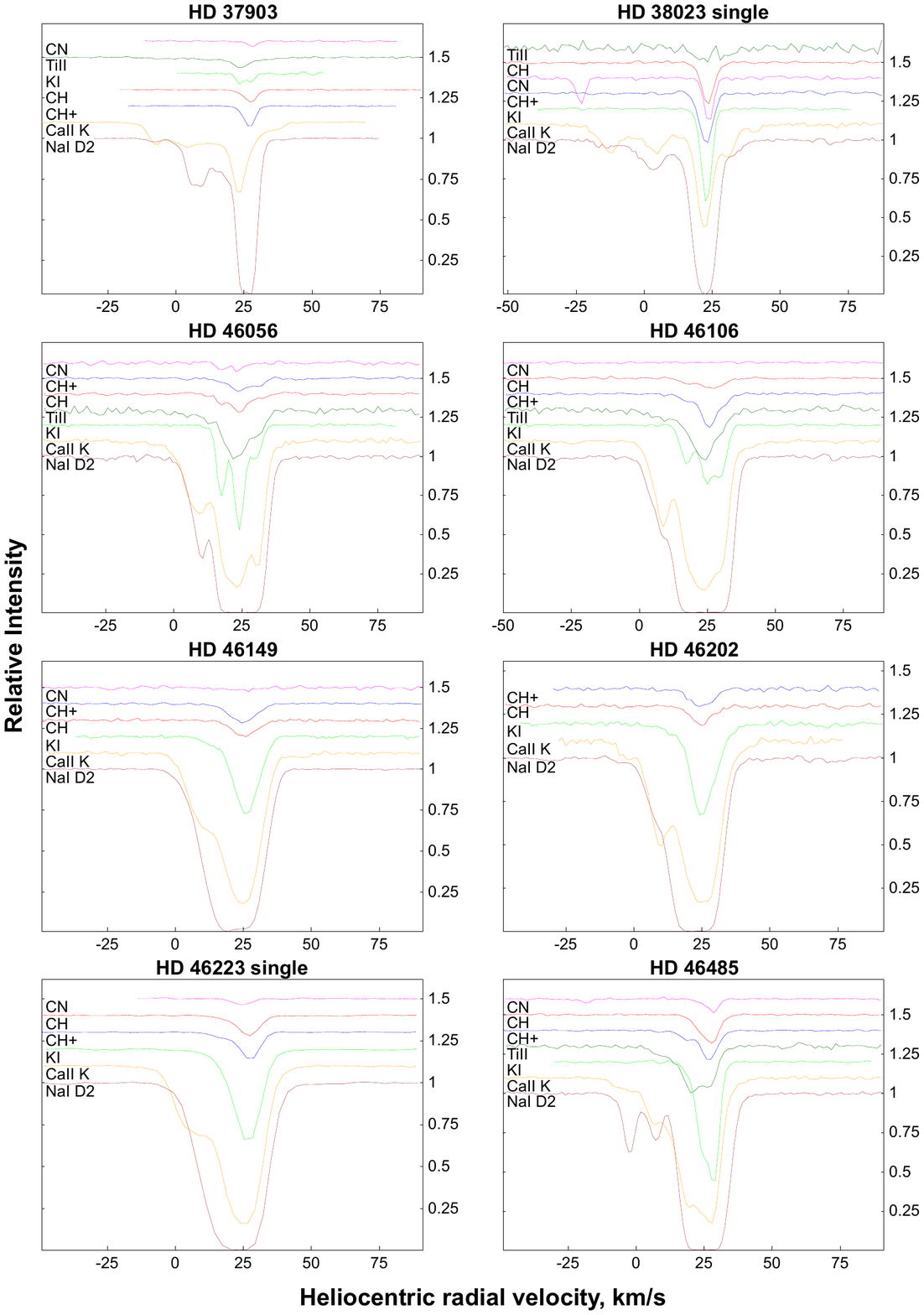}
  \caption{Heliocentric velocity profiles of
    Ti\,{\sc ii} 3383.8\,\AA\,(dark green), CN 3874.6\, \AA\,
    (magenta), CH 4300.3\,\AA\, (red), CH$^+$ 4232.5\,\AA\, (blue),
    K\,{\sc i} 7699\,\AA\ (green), Ca\,{\sc ii} K 3933.7\,\AA\,
    (orange), and Na\,{\sc i} D2 5889\,\AA\,
    (brown). {Single-cloud sightlines are marked ``single''.}  \label{appstart.fig}}
\end{figure*}

\begin{figure*} [!htbp] 
  \includegraphics[width=18.5cm,clip=true,trim=0cm 14cm 0cm 0.cm]{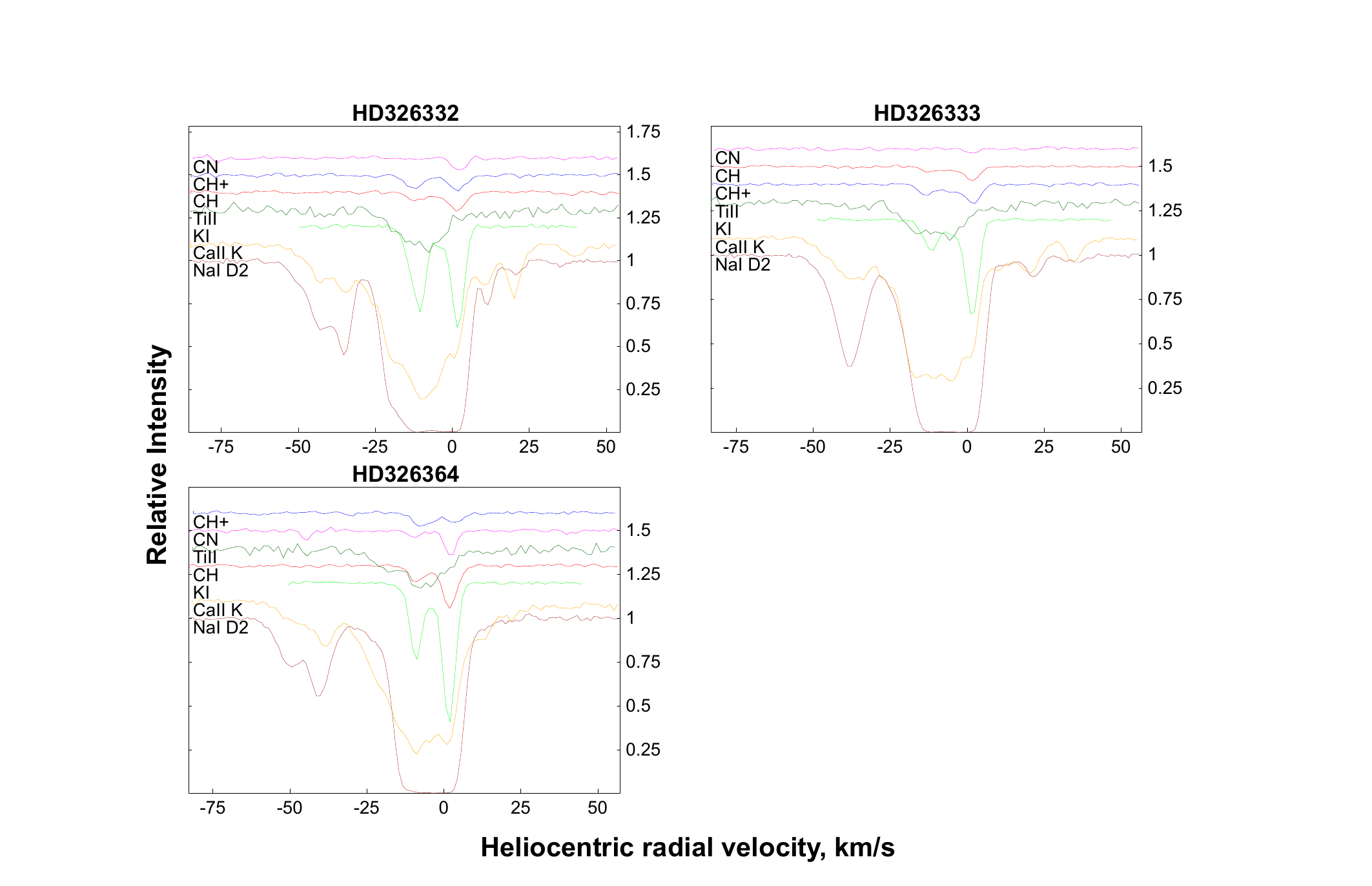}
  \caption{Velocity profiles of
    Ti\,{\sc ii} 3383.8\,\AA\,(dark green), CN 3874.6\, \AA\,
    (magenta), CH 4300.3\,\AA\, (red), CH$^+$ 4232.5\,\AA\, (blue),
    K\,{\sc i} 7699\,\AA\ (green), Ca\,{\sc ii} K 3933.7\,\AA\,
    (orange), and Na\,{\sc i} D2 5889\,\AA\,
    (brown).\label{append.fig} }
\end{figure*}

Complete entries of Table~\ref{sample.tab}, Table~\ref{KI.tab},
  and Table~\ref{res.tab} are provided in Table~\ref{appsample.tab},
  Table~\ref{appKI.tab}, and Table~\ref{appres.tab}, respectively. The
  remaining figures of the UVES-derived velocity profiles are
  displayed in Fig.~\ref{appstart.fig} to Fig.~\ref{append.fig}.

  \clearpage  

\begin{table*}[!htb]
\scriptsize
\begin{center}
  \caption {Study sample sorted by HD number\label{appsample.tab}.
    Column~(1) gives the identification number, Col.~(2) the HD
    identifier, Col.~(3) if different, the main name of the star as
    used in {\sc Simbad}, Col.~(4) the V magnitude, Col.~(5) the
    reddening E(B-V), Col.~(6) the visual extinction $A_{\rm V}$,
    Col.~(7) the total-to-selective extinction $R_{\rm V}$, Col.~(8)
    the spectral type from previous literature, Col.~(9) the spectral
    type as derived in this work, and Col.~(10) the instrument used to
    obtain high-resolution spectroscopy.}
\begin{tabular}{r l l c c c c l l l}
\hline\hline
1  & 2    & 3            & 4 &     5                   &     6                &            7           & 8     & 9           & 10 \\
\hline
ID & Name & {\sc Simbad} & V & $E_{\rm {(B-V)}}^{\imath}$ & $A_{\rm V}$$^{\imath}$ &  $R_{\rm V}$$^{\imath}$  &  SpTy  & SpTy$^{a}$  & Instrument \\
&      &              & mag  &                          &   mag                &                  &  literature & this work  &                 \\
\hline
  1 & HD~023180  & * omi Per  & 3.83 &  	0.29$\pm$  0.03 &   0.91$\pm$  0.08 &   3.14$\pm$  0.31	&  B1IV$^{\rm{F}}$       &  B1III              & UVES$^{b}$     \\
  2 & HD~024263  & * 31 $\tau$& 5.69 &  	0.21$\pm$  0.06 &   0.72$\pm$  0.22 &   3.44$\pm$  0.65	&  B5V$^{\rm{V}}$	&  B3.5V + binary    & UVES$^{a}$     \\
  3 & HD~024912  & ksi Per    & 4.02 &  	0.35$\pm$  0.06 &   1.00$\pm$  0.26 &   2.86$\pm$  0.71	&  O7V$^{\rm{V}}$	&  O8III            & ELODIE$^{l}$   \\
  4 & HD~027778  & 62 $\tau$  & 6.36 &  	0.39$\pm$  0.10 &   1.09$\pm$  0.09 &   2.79$\pm$  0.59	&  B3V$^{\rm{F}}$	&  B3V             & UVES$^{b}$     \\
  5 & HD~030123  &            & 8.61 &  	0.48$\pm$  0.03 &   1.58$\pm$  0.08 &   3.30$\pm$  0.27	&  B8III$^{\rm{F}}$   & B5IV + cloud      & UVES$^{a}$     \\
  6 & HD~030470  &            & 9.47 &  	0.35$\pm$  0.03 &   1.08$\pm$  0.08 &   3.09$\pm$  0.29	&  A0V$^{\rm{F}}$	&  B9.5V + cloud    & UVES$^{a}$     \\
  7 & HD~030492  &            & 8.96 &  	0.40$\pm$  0.04 &   1.17$\pm$  0.11 &   2.93$\pm$  0.35	&  A0IV$^{\rm{F}}$	&  B9.5V + cloud     & UVES$^{a}$     \\
  8 & HD~037130  &            & 9.97 &  	0.23$\pm$  0.03 &   1.26$\pm$  0.11 &   5.50$\pm$  0.44	&  B8-9IV$^{\rm{HS}}$   &  B9V + cloud      & UVES$^{a}$     \\
  9 & HD~037903  &            & 7.83 &  	0.36$\pm$  0.11 &   1.49$\pm$  0.11 &   4.11$\pm$  0.64	&  B1.5V$^{\rm{F}}$	&  B0V + cloud      & UVES$^{b}$     \\
 10 & HD~038023  &            & 8.86 &  	0.52$\pm$  0.03 &   1.64$\pm$  0.08 &   3.16$\pm$  0.26	&  B4V$^{\rm{F}}$	& {B2V}            & UVES$^{a}$     \\ 
 11 & HD~046056  &            & 8.15 &  	0.50$\pm$  0.10 &   1.41$\pm$  0.13 &   2.83$\pm$  0.40	&  O8V(n)$^{\rm{F}}$	&  O8V              & UVES$^{a}$     \\
 12 & HD~046106  &            & 7.92 &  	0.43$\pm$  0.03 &   1.26$\pm$  0.08 &   2.92$\pm$  0.27	&  B1V$^{\rm{F}}$	&  O9.7V          & UVES$^{a}$     \\
 13 & HD~046149  &            & 7.60 &  	0.46$\pm$  0.03 &   1.29$\pm$  0.08 &   2.81$\pm$  0.26	&  O8.5V$^{\rm{F}}$	&  O8.5V         & BOES$^{m}$    \\
 14 & HD~046202  &            & 8.18 &  	0.49$\pm$  0.11 &   1.53$\pm$  0.13 &   3.13$\pm$  0.43	&  O9V$^{\rm{F}}$	&  O9V            & UVES$^{a}$     \\
 15 & HD~046223  &            & 7.28 &  	0.54$\pm$  0.06 &   1.48$\pm$  0.19 &   2.73$\pm$  0.35 &  O5V$^{\rm{V}}$	&  O4V            & UVES$^{c}$     \\
 16 & HD~046485  &            & 8.26 &  	0.61$\pm$  0.03 &   1.83$\pm$  0.08 &   3.00$\pm$  0.25	&  O7Vn(e)$^{\rm{F}}$   &  O7V              & UVES$^{a}$     \\
 17 & HD~046660  &            & 8.04 &  	0.56$\pm$  0.03 &   1.65$\pm$  0.08 &   2.95$\pm$  0.25	&  B1V$^{\rm{F}}$	&  O9V              & UVES$^{a}$     \\
 18 & HD~046711  &            & 9.09 &  	1.05$\pm$  0.03 &   3.36$\pm$  0.08 &   3.20$\pm$  0.23	&  B3II$^{\rm{F}}$	&  B3I + binary      & UVES$^{a}$     \\
 19 & HD~046883  &            & 7.78 &  	0.63$\pm$  0.03 &   1.93$\pm$  0.08 &   3.06$\pm$  0.25	&  B0.5V$^{\rm{F}}$	&  {B3II}      & UVES$^{a}$     \\
 20 & HD~047107  &            & 8.01 &  	0.23$\pm$  0.03 &   0.69$\pm$  0.08 &   3.00$\pm$  0.33	&  B1.5Ia$^{\rm{F}}$	&  B1.5V              & UVES$^{a}$     \\
 21 & HD~047382  &            & 7.14 &  	0.44$\pm$  0.03 &   1.37$\pm$  0.08 &   3.12$\pm$  0.27	&  B0III$^{\rm{F}}$	&  B3II             & UVES$^{a}$     \\
 22 & HD~050562  &            & 8.60 &  	0.26$\pm$  0.08 &   0.64$\pm$  0.24 &   2.45$\pm$  0.65	&  B2III$^{\rm{V}}$	&  B3V              & UVES$^{a}$     \\
 23 & HD~054306  &            & 8.78 &  	0.23$\pm$  0.03 &   0.63$\pm$  0.08 &   2.73$\pm$  0.32	&  B2V$^{\rm{F}}$	&  B1.5V            & UVES$^{a}$     \\
 24 & HD~060479  &            & 8.41 &  	0.57$\pm$  0.06 &   1.70$\pm$  0.20 &   2.98$\pm$  0.37	&  B0II$^{\rm{V}}$	&  O9I + binary     & UVES$^{a}$     \\
 25 & HD~062542  &            & 8.04 &  	0.36$\pm$  0.03 &   0.99$\pm$  0.08 &   2.74$\pm$  0.28	&  B5V$^{\rm{F}}$	&  B3V              & UVES$^{d}$     \\
 26 & HD~068633  &            & 8.00 &  	0.49$\pm$  0.03 &   1.78$\pm$  0.09 &   3.64$\pm$  0.27	&  B5V$^{\rm{F}}$	&  B2V + bin & UVES$^{a}$     \\
 27 & HD~070614  &            & 9.27 &  	0.65$\pm$  0.06 &   1.91$\pm$  0.20 &   2.94$\pm$  0.34	&  B6$^{\rm{F}}$	&  B3III            & UVES$^{a}$     \\
 28 & HD~072648  &            & 7.61 &  	0.34$\pm$  0.03 &   1.21$\pm$  0.09 &   3.56$\pm$  0.30	&  B1-2Ib$^{\rm{F}}$   &  B3III (II)            & UVES$^{a}$     \\
 29 & HD~079186  &            & 5.04 &  	0.40$\pm$  0.06 &   1.28$\pm$  0.31 &   3.21$\pm$  0.76 &  B5Ia$^{\rm{V}}$	&  B5Ia             & UVES$^{b}$     \\
 30 & HD~083597  &            & 9.20 &  	0.34$\pm$  0.06 &   1.31$\pm$  0.24 &   3.84$\pm$  0.51	&  B2V$^{\rm{V}}$	&  B0Ve             & UVES$^{a}$     \\
 31 & HD~089137  &            & 7.98 &  	0.27$\pm$  0.06 &   0.72$\pm$  0.18 &   2.68$\pm$  0.48	&  B2II$^{\rm{V}}$	&  B2II             & UVES$^{a}$     \\
 32 & HD~091943  &            & 6.73 &  	0.25$\pm$  0.06 &   0.93$\pm$  0.23 &   3.73$\pm$  0.62	&  B0.5II$^{\rm{V}}$	&  B1Ia             & UVES$^{a}$     \\
 33 & HD~091969  &            & 6.53 &  	0.23$\pm$  0.06 &   0.85$\pm$  0.22 &   3.69$\pm$  0.61	&  B0Ib$^{\rm{V}}$	&  B0I + binary     & UVES$^{a}$     \\
 34 & HD~091983  &            & 8.58 &  	0.29$\pm$  0.03 &   0.93$\pm$  0.08 &   3.20$\pm$  0.31	&  B1III$^{\rm{F}}$	&  B2III            & UVES$^{a}$     \\
 35 & HD~092007  &            & 8.94 &  	0.32$\pm$  0.03 &   0.97$\pm$  0.08 &   3.03$\pm$  0.29	&  B0II$^{\rm{F}}$	&  B1III            & UVES$^{a}$     \\
 36 & HD~092044  &            & 8.25 &  	0.43$\pm$  0.03 &   1.42$\pm$  0.08 &   3.30$\pm$  0.28	&  B0.5II$^{\rm{F}}$	&  B1.5II            & UVES$^{a}$     \\
 37 & HD~093632  &            & 8.31 &  	0.56$\pm$  0.07 &   2.33$\pm$  0.30 &   4.17$\pm$  0.44	&  B0V$^{\rm{V}}$	&  O6I              & UVES$^{a}$     \\
 38 & HD~094663  &            & 9.35 &  	0.38$\pm$  0.08 &   1.29$\pm$  0.30 &   3.38$\pm$  0.58	&  O9.5III$^{\rm{V}}$   &  O9.5IV + bin          & UVES$^{a}$     \\
 39 & HD~096042  &            & 8.23 &  	0.43$\pm$  0.09 &   0.87$\pm$  0.30 &   2.01$\pm$  0.53	&  B1V$^{\rm{V}}$	&  B1V + bin              & UVES$^{a}$     \\
 40 & HD~096675  &            & 7.69 &  	0.30$\pm$  0.06 &   1.15$\pm$  0.24 &   3.83$\pm$  0.55	&  B6IV$^{\rm{V}}$    & {B5V} & UVES$^{a}$   \\ 
 41 & HD~097484  &V* EM Car   & 8.36 &  	0.60$\pm$  0.07 &   1.54$\pm$  0.21 &   2.57$\pm$  0.35	&  O8V$^{\rm{V}}$	&  O8V             & UVES$^{a}$     \\
 42 & HD~099264  &            & 5.58 &  	0.27$\pm$  0.03 &   0.85$\pm$  0.08 &   3.15$\pm$  0.32	&  B3III$^{\rm{F}}$	&  B2V            & UVES$^{a}$     \\
 43 & HD~099872  &            & 6.09 &  	0.36$\pm$  0.03 &   1.06$\pm$  0.10 &   2.95$\pm$  0.46	&  B3V$^{\rm{F}}$	&  B3V              & UVES$^{a}$     \\
 44 & HD~099890  &            & 8.28 &  	0.24$\pm$  0.16 &   0.75$\pm$  0.11 &   3.11$\pm$  0.71	&  B0.5V$^{\rm{G}}$	&  B1IV            & UVES$^{a}$     \\
 45 & HD~100213  &V* TU Mus   & 8.38 &  	0.37$\pm$  0.13 &   1.28$\pm$  0.14 &   3.47$\pm$  0.52	&  O8V$^{\rm{G}}$	&  O8V(n)z + B0V(n) & UVES$^{a}$     \\
 46 & HD~101008  &            & 9.15 &  	0.27$\pm$  0.03 &   0.93$\pm$  0.08 &   3.43$\pm$  0.33	&  O9V$^{\rm{F}}$	&  O9V              & UVES$^{a}$     \\
 47 & HD~104565  &            & 9.25 &  	0.60$\pm$  0.06 &   2.06$\pm$  0.22 &   3.43$\pm$  0.37	&  B0Ia$^{\rm{V}}$	&  O9.7II           & UVES$^{a}$     \\
 48 & HD~108927  &            & 7.78 &  	0.24$\pm$  0.03 &   0.74$\pm$  0.08 &   3.10$\pm$  0.33	&  B5V$^{\rm{F}}$	&  B5V              & UVES$^{a}$     \\
 49 & HD~110336  &            & 8.64 &  	0.45$\pm$  0.03 &   1.21$\pm$  0.08 &   2.69$\pm$  0.26	&  B9IV$^{\rm{F}}$	&  B9IV             & UVES$^{a}$     \\
 50 & HD~110715  &            & 8.65 &  	0.45$\pm$  0.03 &   1.31$\pm$  0.08 &   2.92$\pm$  0.26	&  B9V$^{\rm{F}}$	&  B9V   & UVES$^{a}$     \\
 51 & HD~110863  &            & 9.06 &  	0.54$\pm$  0.03 &   1.60$\pm$  0.08 &   2.97$\pm$  0.25	&  B1Vp$^{\rm{F}}$	&  B2II-III         & UVES$^{a}$     \\
 52 & HD~110946  &V* LN Mus   & 9.14 &  	0.50$\pm$  0.03 &   1.60$\pm$  0.08 &   3.20$\pm$  0.26	&  B1V$^{\rm{F}}$	&  B3III             & UVES$^{a}$     \\
 53 & HD~111973  &*$\kappa$Cru& 5.98 &  	0.39$\pm$  0.07 &   1.39$\pm$  0.26 &   3.55$\pm$  0.51	&  B5Ia$^{\rm{V}}$	&  B5I              & UVES$^{a}$     \\
 54 & HD~111990  &            & 6.80 &  	0.41$\pm$  0.06 &   1.45$\pm$  0.22 &   3.54$\pm$  0.45	&  B1I$^{\rm{V}}$	&  B1.5I + binary   & UVES$^{a}$     \\
 55 & HD~112607  &            & 8.06 &  	0.31$\pm$  0.04 &   0.85$\pm$  0.10 &   2.73$\pm$  0.38	&  B7-8III$^{\rm{F}}$  &  B5{IV}           & UVES$^{a}$     \\
 56 & HD~112954  &            & 8.37 &  	0.57$\pm$  0.03 &   1.75$\pm$  0.08 &   3.07$\pm$  0.25	&  B9IV$^{\rm{F}}$	&  A1.5V              & UVES$^{a}$     \\
 57 & HD~122669  &            & 8.96 &  	0.60$\pm$  0.08 &   2.46$\pm$  0.33 &   4.09$\pm$  0.45	&  B0V$^{\rm{V}}$	&  B0Ve             & UVES$^{a}$     \\
 58 & HD~123008  &            & 8.79 &  	0.62$\pm$  0.08 &   2.06$\pm$  0.29 &   3.32$\pm$  0.41	&  B0V$^{\rm{V}}$	&  O9.2Iab          & UVES$^{a}$     \\
 59 & HD~123335  &V* V883 Cen & 6.34 &  	0.23$\pm$  0.03 &   0.76$\pm$  0.08 &   3.30$\pm$  0.34	&  B5IV$^{\rm{F}}$	&  B3V+B8V          & UVES$^{a}$     \\
 60 & HD~125288  &* v Cen     & 4.36 &  	0.28$\pm$  0.06 &   0.61$\pm$  0.24 &   2.19$\pm$  0.55	&  B5II$^{\rm{V}}$	&  B5{II-III}             & FEROS$^{i}$   \\
 61 & HD~129557  &V* BU Cir   & 5.23 &  	0.23$\pm$  0.09 &   0.53$\pm$  0.28 &   2.29$\pm$  0.76	&  B2III$^{\rm{V}}$	&  B2III            & UVES$^{e}$     \\
 62 & HD~134591  &            & 8.37 &  	0.24$\pm$  0.07 &   0.60$\pm$  0.21 &   2.51$\pm$  0.60	&  B5III$^{\rm{V}}$	&  B8IV             & UVES$^{a}$     \\
 63 & HD~141318  &            & 5.77 &  	0.24$\pm$  0.05 &   0.83$\pm$  0.18 &   3.44$\pm$  0.54	&  B2II$^{\rm{V}}$	&  B2III            & UVES$^{a}$     \\
 64 & HD~141926  &            & 9.13 &  	0.80$\pm$  0.06 &   2.96$\pm$  0.22 &   3.69$\pm$  0.32	&  B2V$^{\rm{V}}$	&  {B1V}    & UVES$^{a}$     \\
 65 & HD~143054  &            & 9.39 &  	0.55$\pm$  0.03 &   1.54$\pm$  0.08 &   2.80$\pm$  0.25	&  B3II-III$^{\rm{F}}$  &  B2.5V             & UVES$^{a}$     \\
 66 & HD~146284  &            & 6.70 &  	0.25$\pm$  0.03 &   0.77$\pm$  0.08 &   3.10$\pm$  0.32	&  B8V$^{\rm{F}}$	&  B7IV             & UVES$^{a}$     \\
 67 & HD~146285  &            & 7.93 &  	0.32$\pm$  0.03 &   1.23$\pm$  0.09 &   3.83$\pm$  0.32	&  B8V$^{\rm{F}}$	&   B7V     & UVES$^{a}$     \\
 68 & HD~147165  &*$\sigma$Sco& 2.88 &  	0.41$\pm$  0.03 &   1.48$\pm$  0.09 &   3.60$\pm$  0.29	&  B1III$^{\rm{F}}$	&  B1III + binary   & UVES$^{b}$     \\
 69 & HD~147196  &            & 7.04 &  	0.27$\pm$  0.03 &   0.84$\pm$  0.08 &   3.10$\pm$  0.31	&  B5V$^{\rm{F}}$	&  B6.5V              & UVES$^{a}$     \\
 70 & HD~147331  &            & 8.75 &  	0.59$\pm$  0.06 &   1.75$\pm$  0.20 &   2.97$\pm$  0.36	&  B0Ia$^{\rm{V}}$      & O9I + binary      & UVES$^{a}$     \\
 71 & HD~147701  &            & 8.36 &  	0.69$\pm$  0.03 &   2.78$\pm$  0.09 &   4.03$\pm$  0.26	&  B5V$^{\rm{F}}$	&  B5III            & UVES$^{a}$     \\
 72 & HD~147888  &*$\rho$Oph D& 6.74 &  	0.48$\pm$  0.05 &   1.97$\pm$  0.09 &   4.08$\pm$  0.39	&  B3V$^{\rm{F}}$	&  B3V              & UVES$^{b}$     \\
 73 & HD~147889  &            & 7.90 &  	1.06$\pm$  0.03 &   4.44$\pm$  0.09 &   4.19$\pm$  0.24	&  B2V$^{\rm{F}}$	&  {B2V}    & UVES$^{b}$     \\
 74 & HD~147933  &*$\rho$Oph A& 4.59 &  	0.47$\pm$  0.03 &   2.07$\pm$  0.09 &   4.41$\pm$  0.29	&  B1.5V$^{\rm{F}}$	&  B1V              & UVES$^{b}$     \\
 75 & HD~148579  &            & 7.34 &  	0.35$\pm$  0.03 &   1.40$\pm$  0.09 &   4.01$\pm$  0.31	&  B9V$^{\rm{F}}$	&  {B8V}     & UVES$^{a}$     \\
 76 & HD~148594  &            & 6.89 &  	0.21$\pm$  0.03 &   0.65$\pm$  0.08 &   3.10$\pm$  0.35	&  B9V$^{\rm{F}}$	&  {B5.5V}  & UVES$^{a}$     \\
 77 & HD~149038  &* $\mu$ Nor & 4.92 &  	0.22$\pm$  0.07 &   1.08$\pm$  0.37 &   4.92$\pm$  1.11	&  B0Ia$^{\rm{V}}$	&  O9.5Ia           & UVES$^{b}$     \\
\hline
\end{tabular}
\end{center}
\end{table*}

\setcounter{table}{0}
\begin{table*}[!htb]
\scriptsize
\begin{center}
  \caption { -- continued --  }
  \begin{tabular}{r l l c c c c l l l}
\hline\hline
1  & 2    & 3            & 4 &     5                   &     6                &            7           & 8     & 9           & 10 \\
\hline
ID & Name & {\sc Simbad} & V & $E_{\rm {(B-V)}}^{\imath}$ & $A_{\rm V}$$^{\imath}$ &  $R_{\rm V}$$^{\imath}$  &  SpTy  & SpTy$^{a}$  & Instrument \\
   &      &              & mag  &                          &   mag                &                  &  literature & this work  &                 \\
\hline
78 & HD~149757  &* $\zeta$Oph& 2.57 &  	0.31$\pm$  0.03 &   0.95$\pm$  0.08 &   3.08$\pm$  0.30	&  O9.5Vnn$^{\rm{F}}$   &  O9.2IV           & UVES$^{b}$     \\
79 & HD~151346  &            & 7.91 &  	0.59$\pm$  0.03 &   2.25$\pm$  0.09 &   3.81$\pm$  0.26	&  B7p$^{\rm{F}}$	&  B8IV             & UVES$^{a}$     \\
80 & HD~152096  &            & 9.41 &  	0.30$\pm$  0.09 &   0.94$\pm$  0.31 &   3.15$\pm$  0.66	&  B2Ib$^{\rm{V}}$	&  B2I-II + binary  & UVES$^{a}$     \\
 81 & HD~152245  &            & 8.41 &  	0.36$\pm$  0.06 &   1.08$\pm$  0.22 &   2.99$\pm$  0.54	&  B0Ib$^{\rm{V}}$	&  B0V              & UVES$^{a}$     \\
 82 & HD~152247  &            & 7.17 &  	0.41$\pm$  0.06 &   1.43$\pm$  0.22 &   3.49$\pm$  0.44	&  O9.5Iab$^{\rm{V}}$   &  O9.2II           & UVES$^{e}$     \\
 83 & HD~152560  &            & 8.28 &  	0.39$\pm$  0.03 &   1.27$\pm$  0.08 &   3.25$\pm$  0.28	&  B0.5IV$^{\rm{F}}$	&  O9.5IV           & UVES$^{b}$     \\
 84 & HD~152667  & V* V861 Sco& 6.18 &  	0.49$\pm$  0.06 &   1.63$\pm$  0.21 &   3.33$\pm$  0.39	&  B0.5Ia$^{\rm{V}}$	&  O9I + binary     & UVES$^{a}$     \\
 85 & HD~155756  &            & 9.27 &  	0.78$\pm$  0.09 &   2.39$\pm$  0.31 &   3.07$\pm$  0.38	&  B0Ia$^{\rm{V}}$	&  O9I + binary     & UVES$^{a}$     \\
 86 & HD~156247  &V* U Oph    & 5.91 &  	0.25$\pm$  0.04 &   0.73$\pm$  0.11 &   2.92$\pm$  0.43	&  B5V$^{\rm{F}}$	&  B4V + binary     & UVES$^{a}$     \\
 87 & HD~161653  &            & 7.20 &  	0.28$\pm$  0.06 &   0.94$\pm$  0.22 &   3.37$\pm$  0.61	&  B0.5Iab$^{\rm{V}}$   &  B3II {+ binary}             & UVES$^{a}$     \\
 88 & HD~162978  & * 63 Oph   & 6.20 &  	0.34$\pm$  0.06 &   1.20$\pm$  0.23 &   3.54$\pm$  0.54	&  O8III$^{\rm{V}}$	&  O8II             & UVES$^{a}$     \\
 89 & HD~163181  &            & 6.49 &  	0.74$\pm$  0.08 &   2.40$\pm$  0.37 &   3.24$\pm$  0.52	&  B0Ia$^{\rm{V}}$	&  O9.5Iab          & UVES$^{a}$     \\
 90 & HD~164536  &            & 7.12 &  	0.25$\pm$  0.03 &   0.93$\pm$  0.09 &   3.73$\pm$  0.35	&  O7-8$^{\rm{F}}$	&  O7.5V            & UVES$^{a}$     \\
 91 & Herschel~36&            &10.30 &  	0.86$\pm$  0.13 &   4.48$\pm$  0.36 &   5.21$\pm$  0.81	&  O7.5V(n)$^{\rm{F}}$  &  O8V + cloud      & UVES$^{b}$     \\
 92 & HD~164816  &            & 7.11 &  	0.31$\pm$  0.12 &   0.99$\pm$  0.12 &   3.21$\pm$  0.68	&  B0V$^{\rm{F}}$	&  O9.5V            & UVES$^{a}$     \\
 93 & HD~164863  &            & 7.27 &  	0.26$\pm$  0.03 &   0.87$\pm$  0.08 &   3.34$\pm$  0.33	&  B0V$^{\rm{F}}$	&  B0V              & UVES$^{a}$     \\
 94 & HD~164906  &            & 7.47 &  	0.43$\pm$  0.11 &   2.17$\pm$  0.11 &   5.01$\pm$  0.65	&  B1IV$^{\rm G}$       &  B1IV + cloud     & UVES$^{a}$     \\
 95 & HD~164947A &            & 9.41 &  	0.29$\pm$  0.03 &   1.07$\pm$  0.09 &   3.70$\pm$  0.33	&  B2V$^{\rm{F}}$	&  B2V + binary     & UVES$^{a}$     \\
 95 & HD~164947B &            & 9.80 &  	0.29$\pm$  0.03 &   1.07$\pm$  0.09 &   3.70$\pm$  0.33	&  B2V$^{\rm{F}}$	&  B2V + binary     & UVES$^{a}$     \\
 96 & HD~167771  &            & 6.54 &  	0.42$\pm$  0.13 &   1.48$\pm$  0.14 &   3.53$\pm$  0.52	&  O7III$^{\rm{F}}$	&  O7III + O8III      & UVES$^{a}$     \\
 97 & HD~168750  &            & 8.27 &  	0.36$\pm$  0.07 &   1.27$\pm$  0.27 &   3.52$\pm$  0.58	&  B1Ib$^{\rm{V}}$	&  B2III            & UVES$^{a}$     \\
 98 & HD~168941  &            & 9.38 &  	0.34$\pm$  0.09 &   1.23$\pm$  0.12 &   3.56$\pm$  0.50	&  O9.5II$^{\rm{G}}$    &  O9III            & CES$^{f}$     \\
 99 & HD~169582  &            & 8.70 &  	0.87$\pm$  0.06 &   2.60$\pm$  0.21 &   2.99$\pm$  0.31	&  O5V$^{\rm{V}}$	& O6V (Ia,O3-4?)+ bin  & UVES$^{a}$     \\
100 & HD~170634  &            & 9.85 &  	0.69$\pm$  0.03 &   2.05$\pm$  0.08 &   2.97$\pm$  0.24	&  B7V$^{\rm{F}}$	&  B8-9IV         & UVES$^{a}$     \\
101 & HD~170740  &            & 5.72 &  	0.47$\pm$  0.03 &   1.37$\pm$  0.08 &   2.91$\pm$  0.26	&  B2V$^{\rm{F}}$	&  B2V + binary     & UVES$^{b}$     \\
102 & HD~172028  &            & 7.83 &  	0.78$\pm$  0.03 &   2.27$\pm$  0.08 &   2.91$\pm$  0.24	&  B3II$^{\rm{F}}$	&  B3II             & UVES$^{g}$     \\
103 & HD~172140  &            & 9.95 &  	0.23$\pm$  0.03 &   0.62$\pm$  0.08 &   2.71$\pm$  0.32	&  B0.5III$^{\rm{F}}$   &  B0.5III          & CES$^{f}$      \\
104 & HD~172175  &            & 9.40 &  	0.95$\pm$  0.06 &   2.63$\pm$  0.19 &   2.77$\pm$  0.28	&  O6V$^{\rm{V}}$	&  O6.5IV           & UVES$^{a}$     \\
105 & HD~177989  &            & 9.33 &  	0.22$\pm$  0.03 &   0.63$\pm$  0.08 &   2.85$\pm$  0.33	&  B2II$^{\rm{F}}$	&  B2II             & CES$^{f}$      \\
106 & HD~180968  &* 2 Vul     & 5.43 &  	0.21$\pm$  0.09 &   0.99$\pm$  0.36 &   3.31$\pm$  0.86	&  B0nnvar$^{\rm{F}}$   &  B3IV + binary    & UVES$^{h}$     \\
107 & HD~185418  &            & 7.45 &  	0.52$\pm$  0.09 &   1.39$\pm$  0.12 &   2.67$\pm$  0.41	&  B0.5V$^{\rm{F}}$	&  B0.5V            & UVES$^{b}$     \\
108 & HD~185859  &            & 6.49 &  	0.60$\pm$  0.06 &   1.65$\pm$  0.19 &   2.74$\pm$  0.33	&  B0.5Ia$^{\rm{V}}$    &  B3I + binary     & UVES$^{b}$     \\
109 & HD~198478  & * 55 Cyg   & 4.83 &  	0.57$\pm$  0.06 &   1.48$\pm$  0.22 &   2.60$\pm$  0.39	&  B3Ia$^{\rm{V}}$	&  B4Ia             & SAO1m$^{j}$    \\
110 & HD~200775  &            & 7.39 &  	0.57$\pm$  0.12 &   3.21$\pm$  0.10 &   5.61$\pm$  0.88	&  B2Ve$^{\rm{G}}$	&  B2Ve             & SOPHIE$^{k}$  \\
111 & HD~203532  &            & 6.37 &  	0.31$\pm$  0.03 &   0.92$\pm$  0.08 &   2.96$\pm$  0.30	&  B3IV$^{\rm{F}}$	&  B3.5V            & UVES$^{b}$     \\
112 & HD~204827  &            & 7.94 &  	1.08$\pm$  0.03 &   2.64$\pm$  0.07 &   2.44$\pm$  0.22	&  B0V$^{\rm{F}}$	&  O8IV             & BOES$^{m}$     \\
113 & HD~206267  &            & 5.62 &  	0.47$\pm$  0.08 &   1.26$\pm$  0.12 &   2.66$\pm$  0.45	&  O6V$^{\rm{G}}$	&  O5V              & SOPHIE$^{k}$  \\
114 & HD~207198  &            & 5.94 &  	0.58$\pm$  0.08 &   1.54$\pm$  0.09 &   2.68$\pm$  0.31	&  O9II$^{\rm{G}}$	&  O8.5II           & ELODIE$^{l}$  \\
115 & HD~209975  & * 19 Cep   & 5.11 &  	0.34$\pm$  0.06 &   0.96$\pm$  0.23 &   2.82$\pm$  0.58	&  O9Ib$^{\rm{V}}$	&  O9I + binary     & SOPHIE$^{k}$  \\
116 & HD~210121  &            & 7.67 &  	0.31$\pm$  0.07 &   0.75$\pm$  0.20 &   2.42$\pm$  0.49	&  B9V$^{\rm{V}}$	&  B3.5V              & UVES$^{b}$     \\
117 & HD~218376  & * 1 Cas    & 4.85 &  	0.23$\pm$  0.03 &   0.71$\pm$  0.08 &   3.10$\pm$  0.33	&  B1III$^{\rm{F}}$	&  B1III          & SOPHIE$^{k}$  \\
118 & HD~259105  &            & 9.37 &  	0.47$\pm$  0.03 &   1.42$\pm$  0.08 &   3.02$\pm$  0.26	&  B2V$^{\rm{F}}$	&  B1V              & UVES$^{a}$     \\
119 & HD~281159  &            & 8.52 &  	0.89$\pm$  0.03 &   2.79$\pm$  0.08 &   3.14$\pm$  0.24	&  B5V$^{\rm{F}}$	&  B5V              & SOPHIE$^{k}$  \\
120 & HD~284839  &            & 9.69 &  	0.33$\pm$  0.03 &   0.96$\pm$  0.08 &   2.92$\pm$  0.29	&  B9III$^{\rm{F}}$     &  B8-7IV        & ELODIE$^{l}$  \\
121 & HD~284841  &            & 9.34 &  	0.43$\pm$  0.03 &   1.28$\pm$  0.08 &   2.98$\pm$  0.27	&  B9II$^{\rm{F}}$	&  B6.5V + cloud    & UVES$^{a}$     \\
122 & HD~287150  &            & 9.28 &  	0.37$\pm$  0.03 &   1.22$\pm$  0.08 &   3.29$\pm$  0.29	&  A2V$^{\rm{F}}$	&  {B8V}    & UVES$^{a}$     \\
123 & HD~292167  &            & 9.25 &  	0.72$\pm$  0.03 &   2.29$\pm$  0.08 &   3.18$\pm$  0.24	&  O9III$^{\rm{F}}$	&  O9II             & UVES$^{a}$     \\
124 & HD~294264  &            & 9.53 &  	0.52$\pm$  0.03 &   2.85$\pm$  0.10 &   5.48$\pm$  0.31	&  B3Vn$^{\rm{F}}$	&  B2.5V              & UVES$^{a}$     \\
125 & HD~294304  &            &10.03 &  	0.42$\pm$  0.03 &   1.23$\pm$  0.08 &   2.92$\pm$  0.27	&  B8Ve$^{\rm{F}}$	&  B8?              & UVES$^{a}$     \\
126 & HD~315021  &            & 8.59 &  	0.34$\pm$  0.03 &   1.24$\pm$  0.09 &   3.65$\pm$  0.31	&  B2IVn$^{\rm{F}}$	&  B2.5V             & UVES$^{a}$     \\
127 & HD~315023  &            &10.09 &  	0.36$\pm$  0.03 &   1.48$\pm$  0.09 &   4.10$\pm$  0.31	&  B2.5Ve$^{\rm{F}}$    &  B3V + cloud      & UVES$^{a}$     \\
128 & HD~315024  &            & 9.55 &  	0.30$\pm$  0.04 &   1.20$\pm$  0.13 &   3.99$\pm$  0.47	&  B2.5Ve$^{\rm{F}}$    &  B3V + cloud      & UVES$^{a}$     \\
129 & HD~315031  &            & 8.29 &  	0.33$\pm$  0.03 &   1.20$\pm$  0.09 &   3.64$\pm$  0.31	&  B2IVn$^{\rm{F}}$	&  B0.5IV-V + B1V   & UVES$^{a}$     \\
130 & HD~315032  &            & 9.18 &  	0.28$\pm$  0.03 &   1.01$\pm$  0.09 &   3.61$\pm$  0.33	&  B2Vne$^{\rm{F}}$	&  B2IVe + cloud    & UVES$^{a}$     \\
131 & HD~315033  &            & 8.93 &  	0.35$\pm$  0.03 &   1.56$\pm$  0.09 &   4.45$\pm$  0.33	&  B2Vp$^{\rm{F}}$	&  B1V + cloud      & UVES$^{a}$     \\
132 & HD~326309  &            &10.00 &  	0.51$\pm$  0.03 &   1.57$\pm$  0.08 &   3.07$\pm$  0.26	&  B0V$^{\rm{F}}$	&  B3IV             & UVES$^{a}$     \\
133 & HD~326330  & V* V964 Sco& 9.61 &  	0.44$\pm$  0.03 &   1.42$\pm$  0.08 &   3.23$\pm$  0.27	&  B0.5V$^{\rm{F}}$	&  B1V              & UVES$^{a}$     \\
134 & HD~326332  &            & 9.66 &  	0.51$\pm$  0.03 &   1.62$\pm$  0.08 &   3.17$\pm$  0.26	&  B0.5V$^{\rm{F}}$	&  B2IVe +cloud     & UVES$^{a}$     \\
135 & HD~326333  & V* V920 Sco& 9.62 &  	0.45$\pm$  0.03 &   1.52$\pm$  0.08 &   3.38$\pm$  0.28	&  B1V$^{\rm{F}}$	&  B1V  + cloud     & UVES$^{a}$   \\
136 & HD~326364  &            & 9.60 &  	0.62$\pm$  0.03 &   1.80$\pm$  0.08 &   2.91$\pm$  0.25	&  B0IV$^{\rm{F}}$	&  {B1IV}    + cloud  & UVES$^{a}$     \\
\hline
  \end{tabular}
\end{center}
    {\bf Notes:} Cols.~5--7 extracted from \cite{Gordon09},
    \cite{FM07}, with uncertainties revised as explained in the text.
    Spectral types as derived by $^{\rm {F}}$ \cite{FM07}, $^{\rm
      {G}}$ \cite{Gordon09}, $^{\rm {H}}$ \cite{Houk75}, $^{\rm {HS}}$
    \cite{Houk99}, and $^{\rm {V}}$ \cite{Valencic}.  Spectroscopic
    data from: ${a}$: this work; ${b}$: \cite{Cox17}; ${c}$: UVES
    programme ID 096.D-0008(A) (unpublished ESO archive); ${d}$: UVES
    programme ID 099.C-0637(A) (unpublished ESO archive); ${e}$:
    Bagnulo et al. (2013); ${f}$: \citet{Sembach}; ${g}$:
    \cite{Welty10}; ${h}$: \cite{Tolstoy03}; ${i}$: \citet{K10};
    ${j}$: SAO 1.0\,m; ${k}$: SOPHIE archive; ${l}$: ELODIE archive,
    ${m}$: BOES \citep{Kim07}.
\end{table*}

\clearpage
\begin{table*}[!htb]
  \scriptsize
  \begin{center}
    \caption{UVES and FEROS archive spectra covering the K\,{\sc i}
      line. \label{appKI.tab} Column~(1) gives the identification
      number, Col.~(2) the HD identifier, Col.~(3) the ESO programme
      ID, Cols.(4-6) the fit parameters of the K\,{\sc i} line
      profile, and Col.~(7) the instrumental FWHM. Single-cloud
      sight-lines are highlighted in boldface. }
\begin{tabular}{clcrrrc}
\hline  \hline
1    & 2    & 3         & 4             &  5   & 6 & 7 \\
ID   & Star & Prog. ID  & v$_{\bigodot}$  & $b$ &  log($N$) & FWHM$_{\rm{ins}}$ \\
            &                         &                       &      km/s        &   km/s         &   cm$^{-2}$  & km/s         \\
\hline
       137  &{\bf {\bf {\bf CPD~592600}}}  &         071.C-0513(C)  &   -17.63$\pm$0.95  &   6.08$\pm$1.39  &  10.62$\pm$0.08  &  2.7 \\
            &                        &                        &    -4.73$\pm$0.32  &   1.70$\pm$0.31  &  10.99$\pm$0.02  &  2.7 \\
            &                        &                        &     1.81$\pm$0.30  &   1.50$\pm$0.20  &  11.74$\pm$0.03  &  2.7 \\
       138  &            CPD~573509  &         094.D-0355(A)  &    -3.25$\pm$0.30  &   1.34$\pm$0.24  &  12.05$\pm$0.10  &  2.7 \\
            &                        &                        &     9.22$\pm$0.39  &   3.24$\pm$0.40  &  11.14$\pm$0.04  &  2.7 \\
       139  & {\bf {\bf {\bf HD~037022}}}  &         194.C-0833(E)  &    18.85$\pm$0.31  &   1.50$\pm$0.30  &  10.44$\pm$0.02  &  2.7 \\
            &                        &                        &    22.97$\pm$0.35  &   0.12$\pm$0.12  &   9.86$\pm$1.05  &  2.7 \\
       140  & {\bf {\bf {\bf HD~037023}}}  &         067.C-0281(A)  &    20.36$\pm$0.35  &   1.31$\pm$0.46  &  10.29$\pm$0.05  &  2.7 \\
            &                        &                        &    24.15$\pm$0.40  &   0.45$\pm$0.45  &   9.99$\pm$0.07  &  2.7 \\
       141  & {\bf {\bf {\bf HD~037041}}}  &         194.C-0833(A)  &    20.02$\pm$0.36  &   1.05$\pm$0.50  &  10.53$\pm$0.06  &  2.7 \\
            &                        &                        &    23.93$\pm$1.30  &   1.44$\pm$1.44  &   9.80$\pm$0.32  &  2.7 \\
       142  &             HD~037367  &         194.C-0833(A)  &    13.47$\pm$0.33  &   1.12$\pm$0.36  &  11.29$\pm$0.04  &  2.7 \\
            &                        &                        &    17.00$\pm$0.32  &   0.94$\pm$0.37  &  11.41$\pm$0.04  &  2.7 \\
       143  &             HD~041117  &         194.C-0833(C)  &    10.01$\pm$0.32  &   2.33$\pm$0.19  &  11.75$\pm$0.03  &  2.7 \\
            &                        &                        &    14.90$\pm$0.30  &   0.58$\pm$0.58  &  12.86$\pm$1.68  &  2.7 \\
            &                        &                        &    16.81$\pm$0.88  &   2.74$\pm$0.55  &  11.51$\pm$0.15  &  2.7 \\
       144  &             HD~045314  &         194.C-0833(F)  &     3.72$\pm$0.42  &   0.60$\pm$0.60  &  10.34$\pm$0.08  &  2.7 \\
            &                        &                        &    14.83$\pm$0.33  &   1.44$\pm$0.23  &  11.46$\pm$0.04  &  2.7 \\
            &                        &                        &    18.89$\pm$0.31  &   1.39$\pm$0.29  &  11.99$\pm$0.06  &  2.7 \\
            &                        &                        &    24.71$\pm$0.31  &   0.23$\pm$0.23  &  11.39$\pm$0.45  &  2.7 \\
       145  & {\bf {\bf {\bf HD~054439}}}  &         194.C-0833(H)  &    14.68$\pm$0.96  &   3.28$\pm$3.28  &  10.50$\pm$0.53  &  2.7 \\
            &                        &                        &    23.14$\pm$0.96  &   3.23$\pm$2.14  &  10.71$\pm$0.36  &  2.7 \\
            &                        &                        &    27.41$\pm$1.07  &   0.27$\pm$0.27  &   9.93$\pm$0.46  &  2.7 \\
            &                        &                        &    33.19$\pm$0.30  &   1.79$\pm$0.12  &  11.58$\pm$0.03  &  2.7 \\
       146  &             HD~064315  &         082.C-0831(A)  &    36.27$\pm$0.30  &   3.06$\pm$0.52  &  11.65$\pm$0.01  &  6.7 \\
            &                        &                        &    72.45$\pm$0.30  &   0.60$\pm$0.60  &  12.04$\pm$0.51  &  6.7 \\
       147  &             HD~073882  &         194.C-0833(H)  &    21.12$\pm$0.30  &   2.18$\pm$0.16  &  11.37$\pm$0.01  &  2.7 \\
            &                        &                        &    27.72$\pm$0.31  &   1.16$\pm$0.35  &  10.76$\pm$0.02  &  2.7 \\
            &                        &                        &    34.98$\pm$0.42  &   1.52$\pm$0.66  &  10.23$\pm$0.07  &  2.7 \\
       148  &             HD~075309  &         194.C-0833(B)  &    11.30$\pm$1.04  &   5.24$\pm$1.50  &  10.57$\pm$0.08  &  2.7 \\
            &                        &                        &    18.27$\pm$0.30  &   1.17$\pm$0.28  &  11.17$\pm$0.02  &  2.7 \\
            &                        &                        &    23.66$\pm$0.32  &   1.58$\pm$0.41  &  10.86$\pm$0.03  &  2.7 \\
            &                        &                        &    28.40$\pm$0.36  &   0.20$\pm$0.20  &  10.40$\pm$0.08  &  2.7 \\
       149  &             HD~091824  &         194.C-0833(B)  &    -0.79$\pm$0.30  &   1.15$\pm$0.27  &  11.66$\pm$0.05  &  2.7 \\
            &                        &                        &     8.58$\pm$0.37  &   2.20$\pm$0.42  &  10.88$\pm$0.04  &  2.7 \\
       150  &            HD~093129A  &         100.D-0767(A)  &   -41.75$\pm$0.35  &   0.14$\pm$0.14  &  10.83$\pm$0.20  &  4.3 \\
            &                        &                        &   -21.38$\pm$0.54  &   0.09$\pm$0.09  &  10.46$\pm$0.13  &  4.3 \\
            &                        &                        &   -15.35$\pm$0.30  &   0.60$\pm$0.60  &  12.16$\pm$0.52  &  4.3 \\
            &                        &                        &    -6.96$\pm$0.49  &   0.14$\pm$0.14  &  10.66$\pm$0.15  &  4.3 \\
            &                        &                        &    -1.39$\pm$0.42  &   0.29$\pm$0.17  &  11.13$\pm$0.29  &  4.3 \\
            &                        &                        &     3.27$\pm$0.44  &   3.19$\pm$0.38  &  11.52$\pm$0.04  &  4.3 \\
       151  &             HD~093205  &         194.C-0833(B)  &    -2.85$\pm$0.42  &   2.79$\pm$0.55  &  10.75$\pm$0.05  &  2.7 \\
            &                        &                        &     4.69$\pm$0.39  &   1.59$\pm$0.43  &  11.16$\pm$0.06  &  2.7 \\
            &                        &                        &     8.42$\pm$0.38  &   1.12$\pm$0.48  &  11.08$\pm$0.06  &  2.7 \\
       152  &             HD~093222  &         194.C-0833(H)  &   -23.02$\pm$0.37  &   5.42$\pm$0.50  &  10.67$\pm$0.05  &  2.7 \\
            &                        &                        &    -7.07$\pm$0.30  &   1.89$\pm$0.09  &  10.89$\pm$0.05  &  2.7 \\
            &                        &                        &     5.40$\pm$0.31  &   1.51$\pm$1.51  &  10.79$\pm$0.18  &  2.7 \\
            &                        &                        &     6.75$\pm$0.49  &   3.92$\pm$1.64  &  10.75$\pm$0.26  &  2.7 \\
       153  &             HD~093250  &         088.D-0424(D)  &   -28.01$\pm$0.75  &   0.05$\pm$0.05  &  10.56$\pm$0.22  &  6.7 \\
            &                        &                        &   -15.93$\pm$0.31  &   3.32$\pm$0.54  &  11.56$\pm$0.01  &  6.7 \\
            &                        &                        &    -3.12$\pm$0.42  &   2.78$\pm$0.97  &  11.11$\pm$0.04  &  6.7 \\
            &                        &                        &     6.23$\pm$0.36  &   2.71$\pm$0.68  &  11.34$\pm$0.02  &  6.7 \\
       154  &             HD~093843  &         194.C-0833(F)  &   -42.50$\pm$0.37  &   0.89$\pm$0.71  &  10.15$\pm$0.06  &  2.7 \\
            &                        &                        &   -30.53$\pm$0.30  &   4.94$\pm$0.11  &  11.54$\pm$0.01  &  2.7 \\
            &                        &                        &   -17.72$\pm$0.38  &   4.46$\pm$0.37  &  11.09$\pm$0.03  &  2.7 \\
            &                        &                        &    -8.42$\pm$0.33  &   4.54$\pm$0.20  &  11.33$\pm$0.02  &  2.7 \\
       155  &             HD~094493  &         194.C-0833(B)  &   -10.20$\pm$0.38  &   5.52$\pm$0.35  &  10.88$\pm$0.02  &  2.7 \\
            &                        &                        &     5.33$\pm$0.46  &   1.34$\pm$1.34  &  10.19$\pm$0.43  &  2.7 \\
            &                        &                        &     9.98$\pm$0.31  &   0.94$\pm$0.17  &  10.97$\pm$0.12  &  2.7 \\
       156  &             HD~101190  &         095.D-0234(A)  &   -10.65$\pm$0.96  &   6.11$\pm$1.40  &  10.63$\pm$0.08  &  2.7 \\
            &                        &                        &     2.27$\pm$0.32  &   1.70$\pm$0.31  &  10.99$\pm$0.02  &  2.7 \\
            &                        &                        &     8.81$\pm$0.30  &   1.50$\pm$0.20  &  11.74$\pm$0.03  &  2.7 \\
       157  &             HD~101205  &         080.D-0855(A)  &   -16.56$\pm$0.35  &   0.27$\pm$0.27  &  10.97$\pm$0.31  &  5.5 \\
            &                        &                        &     7.88$\pm$0.30  &   1.01$\pm$0.87  &  11.67$\pm$0.14  &  5.5 \\
       158  &             HD~103779  &         194.C-0833(F)  &   -19.14$\pm$0.31  &   0.15$\pm$0.15  &  10.69$\pm$0.34  &  2.7 \\
            &                        &                        &   -16.12$\pm$0.53  &   7.53$\pm$0.56  &  10.74$\pm$0.03  &  2.7 \\
            &                        &                        &    -1.04$\pm$0.35  &   2.00$\pm$0.51  &  10.38$\pm$0.09  &  2.7 \\
            &                        &                        &     3.48$\pm$0.35  &   0.05$\pm$0.05  &  10.39$\pm$0.13  &  2.7 \\
            &                        &                        &     8.76$\pm$0.33  &   2.67$\pm$0.26  &  10.66$\pm$0.04  &  2.7 \\
       159  & {\bf {\bf {\bf HD~104705}}}  &         071.C-0513(C)  &   -24.66$\pm$0.36  &   2.37$\pm$0.37  &  10.55$\pm$0.04  &  2.7 \\
            &                        &                        &   -16.40$\pm$0.48  &   1.77$\pm$0.76  &  10.15$\pm$0.08  &  2.7 \\
            &                        &                        &     2.74$\pm$0.35  &   0.19$\pm$0.19  &  10.36$\pm$0.75  &  2.7 \\
            &                        &                        &     7.65$\pm$0.31  &   1.77$\pm$0.23  &  11.08$\pm$0.01  &  2.7 \\
       160  &             HD~105056  &         266.D-5655(A)  &     2.06$\pm$0.37  &   0.35$\pm$0.35  &  10.51$\pm$0.27  &  3.6  \\
            &                        &                        &     6.98$\pm$0.38  &   0.11$\pm$0.11  &  10.72$\pm$0.47  &  3.6  \\
            &                        &                        &    11.60$\pm$0.39  &   0.11$\pm$0.11  &  10.70$\pm$0.38  &  3.6  \\
            &                        &                        &    17.42$\pm$0.31  &   0.81$\pm$0.63  &  10.88$\pm$0.12  &  3.6  \\
       161  &             HD~112244  &         089.D-0975(A)  &   -17.09$\pm$0.61  &   1.44$\pm$1.39  &  11.07$\pm$0.07  &  6.3  \\
            &                        &                        &   -11.28$\pm$1.24  &   1.62$\pm$1.62  &  10.70$\pm$0.18  &  6.3  \\
            &                        &                        &     8.20$\pm$0.34  &   4.28$\pm$0.62  &  11.10$\pm$0.02  &  6.3  \\
            &                        &                        &    18.30$\pm$0.71  &   0.17$\pm$0.17  &  10.25$\pm$0.23  &  6.3  \\
\hline
\end{tabular}
\end{center}
\end{table*}

\setcounter{table}{1}
\begin{table*}[!htb]
  \scriptsize
  \begin{center}
  \caption { -- continued --  }
  \begin{tabular}{clcrrrc}
\hline  \hline
1  & 2    & 3            & 4 &     5   \\
ID   &        Star             &    Prog. ID           & v$_{\bigodot}$     &      $b$         &     log($N$)    & FWHM$_{\rm{ins}}$ \\
            &                         &                       &      km/s        &   km/s         &   cm$^{-2}$  & km/s         \\
\hline
       162  &             HD~122879  &         194.C-0833(B)  &   -21.38$\pm$0.55  &   4.95$\pm$0.75  &  10.55$\pm$0.05  &  2.7 \\
            &                        &                        &    -9.36$\pm$0.37  &   3.20$\pm$0.39  &  10.64$\pm$0.03  &  2.7 \\
            &                        &                        &     1.75$\pm$0.30  &   1.58$\pm$0.24  &  11.02$\pm$0.01  &  2.7 \\
            &                        &                        &     6.99$\pm$0.32  &   0.48$\pm$0.48  &  10.55$\pm$0.04  &  2.7 \\
            &                        &                        &    11.71$\pm$0.31  &   0.69$\pm$0.49  &  10.70$\pm$0.02  &  2.7 \\
       163  &             HD~144470  &         194.C-0833(C)  &   -11.22$\pm$0.45  &   0.32$\pm$0.32  &  10.79$\pm$0.26  &  2.7  \\
            &                        &                        &    -8.99$\pm$0.33  &   2.84$\pm$0.53  &  11.02$\pm$0.14  &  2.7  \\
       164  &             HD~148379  &         082.C-0566(A)  &   -26.71$\pm$0.52  &   2.46$\pm$0.54  &  11.10$\pm$0.08  &  2.7  \\
            &                        &                        &   -22.20$\pm$0.31  &   1.57$\pm$0.21  &  11.80$\pm$0.04  &  2.7  \\
            &                        &                        &   -16.07$\pm$0.34  &   2.93$\pm$0.30  &  11.27$\pm$0.02  &  2.7  \\
            &                        &                        &    -6.91$\pm$0.32  &   0.91$\pm$0.43  &  10.85$\pm$0.03  &  2.7  \\
            &                        &                        &     0.40$\pm$0.30  &   1.80$\pm$0.20  &  11.35$\pm$0.01  &  2.7  \\
       165  &             HD~149408  &         082.C-0566(A)  &   -22.34$\pm$0.30  &   1.28$\pm$0.24  &  11.75$\pm$0.05  &  2.7  \\
            &                        &                        &   -11.97$\pm$0.31  &   0.55$\pm$0.55  &  11.33$\pm$0.19  &  2.7  \\
            &                        &                        &    -9.40$\pm$0.40  &   7.63$\pm$0.31  &  11.74$\pm$0.02  &  2.7  \\
            &                        &                        &    -3.76$\pm$0.30  &   0.63$\pm$0.47  &  11.80$\pm$0.36  &  2.7  \\
       166  &             HD~151804  &         089.D-0975(A)  &   -15.65$\pm$0.52  &   0.28$\pm$0.28  &  10.45$\pm$0.07  &  6.3  \\
            &                        &                        &    -6.51$\pm$0.38  &   1.28$\pm$1.28  &  10.96$\pm$0.13  &  6.3  \\
            &                        &                        &     1.46$\pm$0.49  &   2.70$\pm$0.98  &  10.89$\pm$0.07  &  6.3  \\
       167  &             HD~152233  &         079.D-0718(B)  &    -5.97$\pm$0.58  &   3.65$\pm$2.95  &  11.08$\pm$0.28  &  5.5  \\
            &                        &                        &     2.28$\pm$0.31  &   0.99$\pm$0.92  &  13.09$\pm$0.72  &  5.5  \\
       168  &             HD~152234  &         083.D-0066(C)  &   -10.12$\pm$0.32  &   1.81$\pm$1.32  &  11.58$\pm$0.03  &  6.7 \\
            &                        &                        &     1.28$\pm$0.30  &   1.06$\pm$0.87  &  12.59$\pm$1.62  &  6.7 \\
       169  &             HD~152235  &         266.D-5655(A)  &   -48.50$\pm$0.44  &   2.54$\pm$0.64  &  10.88$\pm$0.05  &  3.7  \\
            &                        &                        &   -23.34$\pm$0.58  &   2.37$\pm$1.00  &  10.57$\pm$0.08  &  3.7  \\
            &                        &                        &    -5.09$\pm$0.33  &   1.75$\pm$0.38  &  11.86$\pm$0.04  &  3.7  \\
            &                        &                        &     0.31$\pm$0.33  &   2.23$\pm$0.32  &  11.94$\pm$0.03  &  3.7  \\
       170  &             HD~152236  &         082.C-0566(A)  &   -16.51$\pm$0.31  &   1.19$\pm$0.32  &  11.36$\pm$0.03  &  2.7 \\
            &                        &                        &    -8.92$\pm$0.31  &   2.34$\pm$0.11  &  11.69$\pm$0.02  &  2.7 \\
            &                        &                        &    -2.70$\pm$0.43  &   0.78$\pm$0.78  &  11.17$\pm$0.46  &  2.7 \\
            &                        &                        &     0.65$\pm$0.32  &   1.10$\pm$0.41  &  11.91$\pm$0.25  &  2.7 \\
       171  & {\bf {\bf {\bf HD~152249}}}  &         082.C-0566(A)  &   -39.54$\pm$0.32  &   2.76$\pm$0.21  &  10.96$\pm$0.02  &  2.7 \\
            &                        &                        &   -28.00$\pm$0.44  &   2.00$\pm$2.00  &  10.31$\pm$0.05  &  2.7 \\
            &                        &                        &   -21.47$\pm$0.46  &   2.00$\pm$2.00  &  10.29$\pm$0.06  &  2.7 \\
            &                        &                        &   -16.02$\pm$0.33  &   0.88$\pm$0.88  &  10.56$\pm$0.06  &  2.7 \\
            &                        &                        &    -9.77$\pm$0.33  &   0.71$\pm$0.71  &  10.49$\pm$0.13  &  2.7 \\
            &                        &                        &    -4.51$\pm$0.57  &   6.34$\pm$0.81  &  11.23$\pm$0.05  &  2.7 \\
            &                        &                        &     2.10$\pm$0.30  &   1.20$\pm$0.24  &  11.75$\pm$0.06  &  2.7 \\
       172  &             HD~153919  &         194.C-0833(F)  &   -33.39$\pm$0.34  &   2.16$\pm$0.28  &  11.07$\pm$0.03  &  2.7 \\
            &                        &                        &   -29.35$\pm$0.33  &   0.63$\pm$0.63  &  10.75$\pm$0.05  &  2.7 \\
            &                        &                        &   -24.42$\pm$0.33  &   2.02$\pm$0.32  &  10.84$\pm$0.03  &  2.7 \\
            &                        &                        &    -7.33$\pm$1.23  &   7.20$\pm$0.70  &  11.39$\pm$0.10  &  2.7 \\
            &                        &                        &    -4.60$\pm$3.25  &   1.78$\pm$1.63  &  11.33$\pm$1.10  &  2.7 \\
            &                        &                        &    -3.14$\pm$0.63  &   0.74$\pm$0.74  &  11.19$\pm$1.31  &  2.7 \\
            &                        &                        &     3.41$\pm$0.31  &   0.62$\pm$0.54  &  11.07$\pm$0.05  &  2.7 \\
       173  & {\bf {\bf {\bf HD~154445}}}  &         082.C-0566(A)  &   -15.87$\pm$0.30  &   0.61$\pm$0.51  &  12.08$\pm$0.84  &  2.7 \\
       174  & {\bf {\bf {\bf HD~164073}}}  &         194.C-0833(D)  &    -6.70$\pm$1.33  &  3.7$\pm$2.21  &   9.92$\pm$0.18  &  2.7 \\
            &                        &                        &    -0.22$\pm$0.30  &   1.02$\pm$0.32  &  11.00$\pm$0.02  &  2.7 \\
       175  &      {\bf {HD~164402}}  &         079.D-0567(A)  &   -25.97$\pm$0.42  &   0.13$\pm$0.13  &  10.44$\pm$0.19  &  6.1  \\
            &                        &                        &    -5.55$\pm$1.37  &   0.91$\pm$0.88  &  11.50$\pm$0.48  &  6.1  \\
            &                        &                        &    -9.10$\pm$1.60  &   0.61$\pm$0.61  &  10.42$\pm$0.97  &  6.1  \\
       176  &             HD~166734  &         073.D-0609(A)  &   -11.12$\pm$0.33  &   4.55$\pm$0.37  &  12.15$\pm$0.01  &  6.3  \\
            &                        &                        &    -2.53$\pm$0.33  &   1.84$\pm$0.79  &  11.75$\pm$0.05  &  6.3  \\
            &                        &                        &     7.78$\pm$0.36  &   4.52$\pm$0.50  &  11.50$\pm$0.02  &  6.3  \\
       177  & {\bf {\bf {\bf HD~167264}}}  &         194.C-0833(A)  &   -30.22$\pm$0.43  &   0.02$\pm$0.02  &  11.11$\pm$0.23  &  2.7  \\
            &                        &                        &   -11.54$\pm$0.41  &   0.16$\pm$0.16  &  10.53$\pm$0.40  &  2.7  \\
            &                        &                        &    -6.71$\pm$0.30  &   0.56$\pm$0.56  &  12.45$\pm$0.81  &  2.7  \\
            &                        &                        &     6.23$\pm$0.40  &   0.07$\pm$0.07  &  10.53$\pm$0.16  &  2.7  \\
       178  &             HD~167838  &         194.C-0833(F)  &   -24.93$\pm$0.38  &   1.50$\pm$0.55  &  10.43$\pm$0.06  &  2.7 \\
            &                        &                        &   -16.55$\pm$0.30  &   1.40$\pm$0.24  &  11.45$\pm$0.02  &  2.7 \\
            &                        &                        &   -11.98$\pm$0.30  &   0.74$\pm$0.38  &  11.65$\pm$0.14  &  2.7 \\
            &                        &                        &    -6.71$\pm$0.58  &   4.00$\pm$0.85  &  10.90$\pm$0.07  &  2.7 \\
            &                        &                        &     1.95$\pm$0.33  &   1.07$\pm$0.49  &  10.74$\pm$0.06  &  2.7 \\
            &                        &                        &     8.77$\pm$0.87  &   4.25$\pm$0.78  &  10.92$\pm$0.07  &  2.7 \\
            &                        &                        &    11.07$\pm$0.31  &   0.57$\pm$0.57  &  11.08$\pm$0.07  &  2.7 \\
       179  &             HD~167971  &         087.D-0264(F)  &   -12.15$\pm$0.31  &   2.04$\pm$0.71  &  12.06$\pm$0.10  &  6.7 \\
            &                        &                        &    -0.47$\pm$0.33  &   3.78$\pm$0.60  &  11.64$\pm$0.02  &  6.7 \\
            &                        &                        &    11.01$\pm$0.31  &   1.75$\pm$0.80  &  12.03$\pm$0.13  &  6.7 \\
       180  &             HD~168075  &         083.D-0066(C)  &   -16.34$\pm$2.86  &   2.46$\pm$1.97  &  11.66$\pm$0.58  &  6.7 \\
            &                        &                        &   -11.35$\pm$1.01  &   1.20$\pm$0.55  &  12.84$\pm$0.92  &  6.7 \\
            &                        &                        &     5.20$\pm$0.31  &   3.58$\pm$0.56  &  11.75$\pm$0.02  &  6.7 \\
       181  &             HD~168076  &         071.C-0513(C)  &   -17.20$\pm$0.33  &   1.34$\pm$0.35  &  11.23$\pm$0.03  &  2.7 \\
            &                        &                        &   -12.99$\pm$0.31  &   0.77$\pm$0.49  &  12.00$\pm$0.39  &  2.7 \\
            &                        &                        &    -8.67$\pm$0.31  &   1.84$\pm$0.22  &  11.95$\pm$0.05  &  2.7 \\
            &                        &                        &    -3.10$\pm$1.41  &   3.25$\pm$2.15  &  10.78$\pm$0.40  &  2.7 \\
            &                        &                        &     7.23$\pm$0.52  &   2.51$\pm$1.30  &  11.10$\pm$0.51  &  2.7 \\
            &                        &                        &     7.54$\pm$0.51  &   5.07$\pm$0.69  &  11.63$\pm$0.16  &  2.7 \\
       182  &             HD~168137  &         090.D-0600(A)  &   -13.12$\pm$0.96  &   3.14$\pm$1.11  &  11.91$\pm$0.13  &  6.7 \\
            &                        &                        &    -9.97$\pm$1.85  &   1.22$\pm$1.11  &  12.30$\pm$0.50  &  6.7 \\
            &                        &                        &     2.09$\pm$0.33  &   0.78$\pm$0.78  &  11.79$\pm$0.33  &  6.7 \\
            &                        &                        &     6.47$\pm$1.47  &   9.04$\pm$2.23  &  11.25$\pm$0.15  &  6.7 \\
\hline
\end{tabular}
\end{center}
\end{table*}

\setcounter{table}{1}
\begin{table*}[!htb]
  \scriptsize
  \begin{center}
  \caption { -- continued --  }
\begin{tabular}{clcrrrc}
\hline  \hline
1  & 2    & 3            & 4 &     5   \\
ID   &        Star             &    Prog. ID           & v$_{\bigodot}$     &      $b$         &     log($N$)    & FWHM$_{\rm{ins}}$ \\
            &                         &                       &      km/s        &   km/s         &   cm$^{-2}$  & km/s         \\
\hline
       183  & {\bf {\bf {\bf HD~169454}}}  &         194.C-0833(F)  &   -25.96$\pm$0.41  &   3.03$\pm$0.55  &  10.70$\pm$0.05  &  2.7 \\
            &                        &                        &   -13.41$\pm$0.56  &   3.55$\pm$0.96  &  11.70$\pm$0.17  &  2.7 \\
            &                        &                        &    -9.40$\pm$0.31  &   1.78$\pm$0.09  &  12.20$\pm$0.12  &  2.7 \\
            &                        &                        &     0.28$\pm$0.52  &   4.23$\pm$1.26  &  10.96$\pm$0.11  &  2.7 \\
            &                        &                        &     7.66$\pm$0.42  &   2.14$\pm$0.91  &  10.79$\pm$0.12  &  2.7 \\
            &                        &                        &    12.35$\pm$0.42  &   0.96$\pm$0.73  &  10.58$\pm$0.09  &  2.7 \\
       184  &    {\bf {HD~175156}$^a$}  &         079.D-0567(A)  &   -11.74$\pm$0.41  &   1.54$\pm$0.70  &  11.01$\pm$0.04  &    5.0 \\
            &                        &                        &    -5.60$\pm$0.30  &   1.52$\pm$0.50  &  12.27$\pm$0.20  &    5.0 \\
       185  &             HD~303308  &         194.C-0833(D)  &   -31.74$\pm$0.88  &  13.07$\pm$1.24  &  10.87$\pm$0.07  &  2.7 \\
            &                        &                        &   -10.51$\pm$0.36  &   0.13$\pm$0.13  &  10.20$\pm$0.21  &  2.7 \\
            &                        &                        &     6.04$\pm$0.36  &   3.98$\pm$0.21  &  11.49$\pm$0.06  &  2.7 \\
            &                        &                        &    10.14$\pm$0.31  &   1.19$\pm$0.32  &  11.12$\pm$0.16  &  2.7 \\
       186  &   {\bf {Walker~67}}  &         092.C-0019(A)  &    26.82$\pm$0.31  &   0.67$\pm$0.67  &  12.30$\pm$0.49  &  2.7 \\
\hline
\end{tabular}
\end{center}
      {\bf Notes: } $^{a}${HD 175156 is classified as single although
        the spectrum has resolution of 5\,km/s (FWHM)}.
        
\end{table*}

\clearpage

\begin{table*}[!htb]
\scriptsize
\begin{center}
 \caption {Result summary. \label{appres.tab} Column~(1) lists the
   identification number as of Table~\ref{sample.tab}, Col.~(2) the HD
   identifier, the equivalent width in Col.~(3) of the Ca\,{\sc ii}~K
   and in Col.~(4) of the Ca\,{\sc ii}~H doublet used in Col.~(5) for
   the distance estimate \citep{Megier09}. Distances obtained from the
   GAIA parallax are listed in Col.~(6) and from the spectral
   type-luminosity ratio in Col.~(7). Appearance of single (S) or
   multiple (M) velocity components of profiles for the Ca\,{\sc ii}~H
   and K, the Na\,{\sc i}, and the K\,{\sc i} lines in Cols.~(8-10).
   The fit parameters of the K\,{\sc i} line profile are listed for
   the position, the column density, and the derived equivalent width
   in Cols.~(11-13). Column~(14) lists the classification of cloud
   environment as cold (c) or warm (w) with or without a CN\,{\sc i}
   detection. Column~(15) lists the infrared morphology of the WISE
   (W4) image. Single-cloud sightlines are highlighted in boldface. }
  \begin{tabular}{r | l |c  c | r r r | c c c | r r r | c c }
    \hline
    \hline
1  &  2     & 3       & 4         & 5           & 6                   & 7                               & 8      & 9    &10   & 11                   & 12         &  13      & 14 & 15 \\
\hline
ID & Target & \multicolumn{2}{c|}{Ca\,{\sc ii}} & \multicolumn{3}{c|}{Distance} & \multicolumn{3}{c|}{Components}  & \multicolumn{3}{c|}{K\,{\sc i}} & Environ. & WISE \\
   &        &   EW(K) &  EW(H)    & Ca~{\sc ii} & GAIA                & Sp/L        & Ca~{\sc ii}~K & Na~{\sc i} & K~{\sc i} & v$_{\bigodot}$  & log N  & EW  &class  & class \\
   &        &  (m\AA)  &  (m\AA)  & \multicolumn{3}{c|}{(pc)}         &  &  &   & (km/s) & ($10^9$\,/cm$^2$) & (m\AA) &  &  \\
\hline
  1 & {HD~023180} &  86$\pm$ 2&  47$\pm$ 1  & $340^{+20}_{-15}$     & $  256^{+72}_{- 47}$ &  320  &  S & S & {M}  &  13.08   &  464 $\pm$ 11   &   69.1 $\pm$  1.6  & c+CN & a \\                           
  2 & {\bf HD~024263} & 129$\pm$ 3&  84$\pm$ 2  & $670^{+65}_{-50}$     & $  222^{+ 8}_{-  8}$ &  210  &  M & S & S  &  21.93   &  706 $\pm$ 10   &   90.6 $\pm$  1.2  & - & a \\                              
  3 &   HD~024912  & 121$\pm$ 6&  73$\pm$ 5  & $540^{+130}_{-85}$    & $ 1134^{+1142}_{-605}$ &  650  &  M & M & M  &   7.27   &    239 $\pm$  17   &  36.8 $\pm$  2.6  & w-CN & a \\
   &  ''             &           &             &              &                      &     &   &   &    &  10.61   &  103 $\pm$  12  &   16.7 $\pm$  1.9  &    &   \\
   &  ''             &           &             &              &                      &     &   &   &    &  15.00   &   39 $\pm$  13   &   6.7 $\pm$  2.2  &    &   \\
  4 & {HD~027778} &  62$\pm$ 1&  36$\pm$ 1  & $300^{+15}_{-20}$       & $  223^{+ 3}_{-  3}$ &  250  &  S & - &  {M}  &  11.87   &  596 $\pm$  4   &   86.8 $\pm$  0.6  & c+CN & a \\
  5 & {HD~030123} & 177$\pm$ 4& 127$\pm$ 3  &$1145^{+140}_{-115}$     & $  594^{+20}_{- 19}$ &  500 &  S & S &  {M} &  15.86   &  316 $\pm$  3   &   45.1 $\pm$  0.4  & c+CN & a \\
  6 & {\bf HD~030470} & 149$\pm$ 6& 112$\pm$ 4  &$1120^{\imath}_{-185}$   & $  778^{+60}_{- 52}$ &  440  &  S & S & S  &  20.98   & 1002 $\pm$  8   &  112.8 $\pm$  1.0  & - & a \\
  7 & {\bf HD~030492} & 156$\pm$ 7& 117$\pm$ 4  &$1160^{\imath}_{-190}$   & $  597^{+33}_{- 30}$ &  390  &  S & S & S  &  21.24   &  893 $\pm$  6   &  100.8 $\pm$  0.7  & c+CN & a \\
  8 & {\bf HD~037130} & 100$\pm$ 7&  57$\pm$ 5  & $410^{+110}_{-75}$      & $  400^{+ 9}_{-  8}$ &  705  &  S & - & S  &  28.57   &   67 $\pm$  7   &   11.3 $\pm$  1.1  & no CH & a \\
  9 & HD~037903  & 101$\pm$ 4&  55$\pm$ 3  & $390^{+50}_{-45}$       & $  397^{+ 9}_{-  9}$ & 1225  &  M & M & M  &  23.47   &   29 $\pm$  3   &    4.9 $\pm$  0.6  & w+CN & c \\
    &  ''             &           &             &              &                      &      &   &   &    &  27.42   &   33 $\pm$  4   &    5.6 $\pm$  0.7  &    &   \\
 10 & {\bf HD~038023} & 141$\pm$ 9&  78$\pm$ 7  & $525^{+160}_{-90}$      & $  404^{+ 7}_{-  7}$ & {860}  &  M & S & S  &  23.03   &  562 $\pm$  6   &   73.2 $\pm$  0.8  & c+CN & a \\
 11 &      HD~046056  & 279$\pm$14& 193$\pm$ 9  &$1590^{+430}_{-280}$     & $ 1469^{+148}_{-124}$ & 2030  &  M & M & M  &  17.64   &  371 $\pm$  4   &   52.7 $\pm$  0.6  & - & c \\
    &   ''            &           &             &              &                      &      &   &   &    &  24.07   &  636 $\pm$  5   &   79.2 $\pm$  0.6  &    &   \\
    &   ''            &           &             &              &                      &      &   &   &    &  29.95   &  198 $\pm$  5   &   31.7 $\pm$  0.8  &    &   \\
 12 & HD~046106  & 279$\pm$13& 202$\pm$ 7  &$1810^{\imath}_{-290}$   & $ 1623^{+287}_{-214}$ &  1540  &  M & M & M  &   5.20   &   24 $\pm$  8   &    4.1 $\pm$  1.3  & w-CN & b \\
    &   ''            &           &             &              &                      &      &   &   &    &  17.55   &  233 $\pm$  5   &   36.8 $\pm$  0.8  &    &   \\
    &   ''            &           &             &              &                      &      &   &   &    &  24.76   &  346 $\pm$  5   &   52.0 $\pm$  0.7  &    &   \\
    &   ''            &           &             &              &                      &      &   &   &    &  29.90   &  291 $\pm$  5   &   44.6 $\pm$  0.7  &    &   \\
 13 & HD~046149$^{{LR}}$ & 264$\pm$13& 182$\pm$ 7  &$1500^{+340}_{-240}$     & $ 1473^{+172}_{-140}$ & 1815 &  M & S & S  &  {26.12}   & {975 $\pm$ 25}   &  {136.1 $\pm$  3.4}  & w-CN & a \\
    &   ''            &           &             &              &      &      &   &   &    &  {13.88}   &  {74} $\pm$  19   &   {12.7 $\pm$  3.1}  &    &   \\ 
 14 & {HD~046202} & 290$\pm$23& 202$\pm$21  &$1680^{\imath}_{-480}$   & $ 1347^{+383}_{-247}$ & 1805  &  M & S & S  &  {23.84}   &{717 $\pm$ 29}   &  {99.7 $\pm$  4.0}  & - & d \\
    &   ''            &           &             &              &       &      &   &   &    &  {29.60}   &  {359 $\pm$  26}   &   {55.8 $\pm$  4.0}  &    &   \\ 
 15 & {\bf HD~046223} & 297$\pm$ 5& 204$\pm$ 3  &$1660^{+120}_{-110}$     & $ 1663^{+169}_{-141}$ & 1895  &  M & S & S  &  26.36   & 1314 $\pm$ 18   &  178.0 $\pm$  2.4  & c+CN & b \\
 16 &      HD~046485  & 251$\pm$12& 175$\pm$ 8  &$1470^{+390}_{-250}$     & $ 1735^{+255}_{-199}$ & 2420  &  M & S & M  &  21.42   &  147 $\pm$  6   &   24.2 $\pm$  1.0  & c+CN & a \\
    &   ''            &           &             &              &                      &      &   &   &    &  24.20   &  332 $\pm$  4   &   46.6 $\pm$  0.6  &    &   \\
    &   ''            &           &             &              &                      &      &   &   &    &  28.32   &  995 $\pm$  7   &  111.1 $\pm$  0.8  &    &   \\
 17 & {HD~046660} & 286$\pm$ 7& 184$\pm$ 5  &$1360^{+145}_{-125}$     & $ 1153^{+73}_{- 65}$ & 1510  &  M & S & M  &  19.66   & 1788 $\pm$  9   &  175.6 $\pm$  0.9  & c+CN & a \\
 18 &      HD~046711  & 387$\pm$32& 289$\pm$22  &$2730^{\imath}_{-820}$   & $ 1740^{+157}_{-133}$ & 2130  &  M & M & M  &  22.20   & 1462 $\pm$ 21   &  163.2 $\pm$  2.3  & c+CN & a \\
    &   ''            &           &             &              &                      &      &   &   &    &  28.52   & 1015 $\pm$ 13   &  107.9 $\pm$  1.4  & w-CN &   \\
    &   ''            &           &             &              &                      &      &   &   &    &  33.87   &  202 $\pm$ 10   &   31.8 $\pm$  1.5  &    &   \\
 19 &      HD~046883  & 198$\pm$ 8& 138$\pm$ 5  &$1170^{+192}_{-135}$     & $ 1381^{+129}_{-109}$ & 1420  &  M & M & M  &  17.37   &  516 $\pm$  3   &   67.4 $\pm$  0.4  & c+CN & a \\
    &   ''            &           &             &              &                      &      &   &   &    &  22.48   &  875 $\pm$  4   &   97.8 $\pm$  0.4  &    &   \\
 20 &      HD~047107  & 260$\pm$12& 167$\pm$ 9  &$1235^{+220}_{-150}$     & $  803^{+180}_{-125}$ & 1160  &  M & - & M  &   4.49   &   18 $\pm$  4   &    3.2 $\pm$  0.8  & no CH & b \\
    &   ''            &           &             &              &                      &      &   &   &    &   9.39   &   53 $\pm$  6   &    9.0 $\pm$  0.9  &    &   \\
    &   ''            &           &             &              &                      &      &   &   &    &  19.98   &   78 $\pm$  6   &   13.2 $\pm$  1.1  &    &   \\
    &   ''            &           &             &              &                      &      &   &   &    &  26.62   &   80 $\pm$  7   &   13.5 $\pm$  1.1  &    &   \\
 21 &      HD~047382  & 290$\pm$ 9& 197$\pm$ 5  &$1570^{+210}_{-160}$     & $ 4063^{+1616}_{-1000}$ & 1410  &  M & M & M  &  19.63   &  343 $\pm$ 32   &   55.3 $\pm$  5.2  & - & a \\
    &   ''            &           &             &              &                      &      &   &   &    &  27.70   &  703 $\pm$ 34   &  104.5 $\pm$  5.1  &    &   \\
 22 &      HD~050562  & 305$\pm$15& 178$\pm$14  &$1160^{+310}_{-210}$     & $ 3395^{+634}_{-470}$ & 1100  &  M & M & M  &  23.97   &   14 $\pm$  6   &    2.5 $\pm$  1.0  & no CH & a \\
    &   ''            &           &             &              &                      &      &   &   &    &  31.43   &   31 $\pm$  5   &    5.3 $\pm$  0.9  &    &   \\
    &   ''            &           &             &              &                      &      &   &   &    &  39.43   &   66 $\pm$  8   &   11.3 $\pm$  1.3  &    &   \\
    &   ''            &           &             &              &                      &      &   &   &    &  47.59   &   49 $\pm$  4   &    8.3 $\pm$  0.6  &    &   \\
    &   ''            &           &             &              &                      &      &   &   &    &  52.01   &   42 $\pm$  4   &    7.1 $\pm$  0.7  &    &   \\
    &   ''            &           &             &              &                      &      &   &   &    &  72.21   &   24 $\pm$  7   &    4.1 $\pm$  1.2  &    &   \\
 23 &      HD~054306  & 198$\pm$ 5& 114$\pm$ 4  & $759^{+68}_{-59}$      & $ 1930^{+408}_{-290}$ &  1540  &  M & M & M  &  21.21   &   25 $\pm$  7   &    4.3 $\pm$  1.3  & w-CN & a \\
    &   ''            &           &             &              &                      &      &   &   &    &  33.64   &  141 $\pm$  9   &   23.3 $\pm$  1.5  &    &   \\
 24 &      HD~060479  & 353$\pm$43& 253$\pm$23  &$2200^{\imath}_{-720}$   & $ 3419^{+602}_{-450}$ & 4000  &  M & M & M  &  35.42   & 2150 $\pm$ 15   &  219.6 $\pm$  1.5  & -$^{v}$ & a \\
    &   ''            &           &             &               &                      &      &   &   &    &  45.49   &  690 $\pm$  9   &   96.8 $\pm$  1.3  &    &   \\
 25 & {\bf HD~062542} & 53$\pm$2.3 & 36.7$\pm$2.3 & $365^{+95}_{-59}$     & $  386^{+ 5}_{-  5}$ &  555  &  M & S & S  &  13.65   &  497 $\pm$  5   &   70.6 $\pm$  0.7  & c+CN & a \\
 26 &      HD~068633  &  96$\pm$ 3&  68$\pm$ 3  & $630^{+130}_{-115}$     & $  677^{+41}_{- 37}$ &  795  &  M & M & M  &   9.29   &  594 $\pm$  5   &   74.7 $\pm$  0.6  & c+CN & a \\
    &   ''            &           &             &              &                      &      &   &   &    &  19.09   &  333 $\pm$  5   &   48.2 $\pm$  0.7  & c+CN &   \\
 27 &      HD~070614  & 272$\pm$18& 187$\pm$12  & 1530$^{\imath}_{-340}$  & $ 1868^{+137}_{-120}$ & {1400}  &  S & S & M  &  22.52   &  527 $\pm$ 10   &   71.3 $\pm$  1.4  & - & c \\
 &   ''            &           &             &              &                      &      &   &   &    &  30.64   &  259 $\pm$ 16   &   42.2 $\pm$  2.6  &    &   \\
 28 &      HD~072648  & 476$\pm$17& 347$\pm$ 10  &$3360^{\imath}_{-460}$  & $ 1233^{+66}_{- 60}$ &  {825}   &  M & S & M  &  21.61   &  528 $\pm$  8   &   76.6 $\pm$  1.1  & w-CN & b \\
    &   ''            &           &             &              &                      &      &   &   &    &  44.40   &   54 $\pm$  6   &    9.3 $\pm$  1.1  &    &   \\
 29 & {\bf HD~079186} & 223$\pm$3 & 145$\pm$ 2  &$1100^{+60}_{-50}$       & $ 1279^{+369}_{-239}$ & 1650  &  M & M & S  &  11.75   &  439 $\pm$  3   &   57.3 $\pm$  0.4  & - & a \\
 30 &      HD~083597  & 247$\pm$11& 160$\pm$ 9  &$1200^{+290}_{-200}$     & $ 2631^{+283}_{-234}$ & 2120  &  M & M & M  &   7.86   &   91 $\pm$ 14   &   15.6 $\pm$  2.4  & - & a \\
    &   ''            &           &             &              &                      &      &   &   &    &  16.64   &  331 $\pm$  7   &   48.8 $\pm$  1.1  &    &   \\
 31 & {\bf HD~089137} & 253$\pm$8& 154$\pm$ 6  &1070$^{+140}_{-105}$   & $ 3417^{+534}_{-412}$ & 3090  &  M & M & S  &   8.70   &  106 $\pm$  5   &   17.5 $\pm$  0.9  & w-CN & c \\
 32 &      HD~091943  & 296$\pm$ 6& 182$\pm$ 6  &$1265^{+135}_{-115}$  & $ 3422^{+455}_{-363}$ & 3585  &  M & S & M  & -10.91   &  477 $\pm$  5   &   64.0 $\pm$  0.7  & - & a \\
    &   ''            &           &             &              &                      &      &   &   &    &  -3.26   &   81 $\pm$  5   &   13.6 $\pm$  0.9  &    &   \\
    &   ''            &           &             &              &                      &      &   &   &    &   8.43   &   68 $\pm$  9   &   11.6 $\pm$  1.5  &    &   \\
 33 &      HD~091969  & 307$\pm$16& 188$\pm$12  &$1300^{+330}_{-230}$     & $ 2567^{+374}_{-292}$& 2290  &  M & - & M  & -12.30   &   29 $\pm$ 11   &    5.0 $\pm$  1.9  & no CH & a \\
    &   ''            &           &             &              &                      &      &   &   &    &  -3.83   &   97 $\pm$ 10   &   16.2 $\pm$  1.6  &    &   \\
    &   ''            &           &             &              &                      &      &   &   &    &   2.25   &  110 $\pm$ 10   &   18.2 $\pm$  1.6  &    &   \\
    &   ''            &           &             &              &                      &      &   &   &    &   9.04   &  102 $\pm$ 10   &   17.0 $\pm$  1.6  &    &   \\
 34 & {\bf HD~091983} & 308$\pm$11& 202$\pm$ 8  &$1525^{+280}_{-210}$    & $ 3698^{+853}_{-599}$ & 2110  &  M & S & S  &  -3.01   &  437 $\pm$  6   &   62.0 $\pm$  0.8  & - & a \\
 35 &      HD~092007  & 283$\pm$12& 195$\pm$10  &$1600^{+410}_{-280}$    & $ 2841^{+417}_{-325}$ & 3080  &  M & M & M  &  -3.51   &  445 $\pm$  6   &   61.2 $\pm$  0.8  & c+CN & b \\
    &   ''            &           &             &              &                       &      &   &   &    &   6.34   &  130 $\pm$  8   &   21.8 $\pm$  1.4  &    &   \\
 36 & {\bf HD~092044} & 238$\pm$ 7& 173$\pm$ 6  &$1570^{\imath}_{-220}$  & $ 2852^{+539}_{-397}$ & 2880  &  M & M & S  &  -4.28   &  531 $\pm$ 10   &   67.7 $\pm$  1.3  & c+CN & b \\
 \hline
\end{tabular}
\end{center}
\end{table*}

\setcounter{table}{2}
\begin{table*}[!htb]
\scriptsize
\begin{center}
  \caption { -- continued -- }
 \begin{tabular}{r | l |c  c | r r r | c c c | r r r | c c }
   \hline
   \hline
   1  &  2     & 3     & 4     & 5      & 6     & 7    & 8      & 9    &10   & 11                   & 12         &  13      & 14 & 15 \\
\hline
ID & Target & \multicolumn{2}{c|}{Ca\,{\sc ii}} & \multicolumn{3}{c|}{Distance} & \multicolumn{3}{c|}{Components}  & \multicolumn{3}{c|}{K\,{\sc i}} & Environ. & WISE \\
&        &   EW(K)    &  EW(H) & Ca~{\sc ii}  & GAIA  & Sp/L & Ca~{\sc ii}~K & Na~{\sc i} & K~{\sc i} & v$_{\bigodot}$  & log N  & EW  &class  & class \\
   &        &  (m\AA)  &  (m\AA) & \multicolumn{3}{c|}{(pc)}  &  &  &   & (km/s) & ($10^9$\,/cm$^2$) & (m\AA) &  &  \\
\hline
 37 &      HD~093632  &   795$\pm$31& 511$\pm$19 &3520$^{+630}_{-470}$ & $ 2486^{+230}_{-195}$ &  3465   &  M & M & M  & -17.88   &  234 $\pm$  7   &   35.6 $\pm$  1.1  & w-CN & d \\
    &   ''            &             &            &                   &                     &       &    &   &    & -12.18   &   90 $\pm$  7   &   14.9 $\pm$  1.1  &    &   \\
    &   ''            &             &            &                   &                     &       &    &   &    &  -1.81   &   66 $\pm$ 11   &   11.2 $\pm$  1.9  &    &   \\
    &   ''            &             &            &                   &                     &       &    &   &    &   7.29   &  119 $\pm$  7   &   19.5 $\pm$  1.2  &    &   \\
 38 &      HD~094663  &   415$\pm$36& 292$\pm$19 &2440$^{\imath}_{-610}$ & $ 3311^{+385}_{-314}$ & 4040  &  M & M & M  & -10.20   &  505 $\pm$  9   &   64.4 $\pm$  1.2  & - & d \\
    &   ''            &             &            &                    &                      &       &    &   &    &  -6.68   &  555 $\pm$ 24   &   86.0 $\pm$  3.8  &    &   \\
    &   ''            &             &            &                    &                      &       &    &   &    &   6.95   &  147 $\pm$ 13   &   24.2 $\pm$  2.2  &    &   \\
 39 &      HD~096042  &   459$\pm$32& 326$\pm$19 &2760$^{\imath}_{-650}$ & $ 4007^{+649}_{-497}$ & 1080  &  M & M & M  & -12.88   &  183 $\pm$  7   &   29.0 $\pm$  1.1  & - & c \\
    &   ''            &             &            &                     &                      &       &    &   &    &  -5.85   &  330 $\pm$  7   &   48.3 $\pm$  1.1  &    &   \\
    &   ''            &             &            &                     &                      &       &    &   &    &  -0.47   &  513 $\pm$  8   &   69.2 $\pm$  1.1  &    &   \\
    &   ''            &             &            &                     &                      &       &    &   &    &   9.68   &   97 $\pm$  9   &   16.4 $\pm$  1.6  &    &   \\
 40 & {\bf HD~096675} &    32$\pm$ 3&  19$\pm$ 5 & 193$^{+38}_{-25}$   & $  162^{+ 2}_{-  2}$ &   {360}  &  S & S & S  &  14.97   &  418 $\pm$  4   &   57.1 $\pm$  0.5  & c+CN & a \\
 41 &      HD~097484  &   396$\pm$24& 295$\pm$11 &2770$^{\imath}_{-550}$ & $ 2760^{+387}_{-304}$ & 2575  &  M & S & M  &  -8.38   &  332 $\pm$  9   &   51.4 $\pm$  1.5  & c+CN & b \\
    &   ''            &             &            &                     &                      &       &    &   &    &  -0.79   & 1823 $\pm$ 16   &  194.0 $\pm$  1.7  &    &   \\
 42 &      HD~099264  &    50$\pm$ 2&  28$\pm$ 2 & 240$^{+38}_{-27}$   & $  234^{+ 6}_{-  6}$ &  300  &  M & S & M  &   7.06   &   21 $\pm$  4   &    3.7 $\pm$  0.7  & w-CN & a \\
 &   ''            &             &            &                     &                      &       &    &   &    &  14.00   &  155 $\pm$  6   &   25.2 $\pm$  1.0  &    &   \\
 43 &      HD~099872  &    59$\pm$ 2&  37$\pm$ 2 & 320$^{+24}_{-20}$   & $  240^{+24}_{-20}$ & 240 &  S & S & M  &   8.81   &   72 $\pm$  5   &   12.3 $\pm$  0.8  & w+CN & a \\
    &   ''            &             &            &                     &                      &       &    &   &    &  13.43   &  292 $\pm$  4   &   42.9 $\pm$  0.5  &    &   \\
 44 &      HD~099890  &   458$\pm$13& 294$\pm$ 9 &2120$^{+270}_{-220}$ & $ 1861^{+192}_{-160}$ & 1760  &  M & M & M  & -18.53   &  153 $\pm$  5   &   24.9 $\pm$  0.9  & no CH & b \\
    &   ''            &             &            &                     &                      &       &    &   &    &  -9.03   &  125 $\pm$  5   &   20.5 $\pm$  0.9  &    &   \\
 45 &      HD~100213  &   314$\pm$ 5& 191$\pm$ 5 &1310$^{+100}_{-85}$ & $ 2230^{+222}_{-186}$ & 4590  &  M & M & M  & -17.38   &  179 $\pm$  4   &   28.1 $\pm$  0.6  & w-CN & a \\
    &   ''            &             &            &                     &                      &       &    &   &    &   6.12   &  305 $\pm$  4   &   44.1 $\pm$  0.6  &    &   \\
 46 & {\bf HD~101008} &   295$\pm$ 8& 177$\pm$ 7 &1195$^{+152}_{-123}$ & $ 3703^{+667}_{-497}$ & 3580  &  M & M & S  &   8.69   &  203 $\pm$  8   &   32.3 $\pm$  1.2  & w-CN & a \\
 47 &      HD~104565  &   491$\pm$15& 372$\pm$11 &3620$^{\imath}_{-520}$ & $ 5181^{+1154}_{-828}$ & 4615 &  M & S & M  & -17.06   &  375 $\pm$  4   &   51.9 $\pm$  0.5  & c+CN & a \\
    &   ''            &             &            &                     &                      &       &    &   &    &   9.49   &   75 $\pm$ 4 &   12.6 $\pm$  0.6  & w-CN &   \\
 48 & {\bf HD~108927} &    62$\pm$ 3&  41$\pm$ 3 & 375$^{+104}_{-65}$  & $  338^{+ 5}_{-  5}$ & 450   &  S & S & S  &  11.56   &  462 $\pm$  4   &   61.7 $\pm$  0.5  & c+CN & a \\
 49 & {\bf HD~110336} &    60$\pm$ 3&  37$\pm$ 3 & 320$^{+80}_{-52}$   & $  314^{+ 4}_{-  4}$ & 305   &  M & S & S  &  12.04   &  638 $\pm$  7   &   77.0 $\pm$  0.9  & c+CN & a \\
 50 & {\bf HD~110715} &    98$\pm$ 3&  63$\pm$ 3 & 504$^{+70}_{-54}$   & ${532^{+12}_{- 12}}$ & 267  &  S & S & S  &   4.78   &  960 $\pm$  7   &  116.6 $\pm$  0.8  & c+CN & a \\
 51 &      HD~110863  &   340$\pm$ 9& 208$\pm$ 9 &1430$^{+200}_{-170}$ & $ 1944^{+162}_{-140}$ & 1770  &  M & M & M  & -30.36   &  536 $\pm$  6   &   68.4 $\pm$  0.7  & c+CN & a \\
    &   ''            &             &            &                     &                      &       &    &   &    &  -9.84   &  142 $\pm$  8   &   23.6 $\pm$  1.3  & w-CN &   \\
    &   ''            &             &            &                     &                      &       &    &   &    &   7.52   &  297 $\pm$  6   &   45.2 $\pm$  1.0  &    &   \\
 52 & {\bf HD~110946} &   288$\pm$12& 180$\pm$ 8 &1280$^{+230}_{-150}$ & $ 2078^{+164}_{-142}$ & 1480  &  M & M & S  &   3.13   &  755 $\pm$ 16   &  103.1 $\pm$  2.1  & c+CN & a \\
 53 &      HD~111973  &   426$\pm$13& 242$\pm$ 8 &1510$^{+390}_{-190}$ & $ 2171^{+262}_{-212}$ & 2340  &  M & M & M  & -14.20   &   70 $\pm$  5   &   11.9 $\pm$  0.8  & w-CN & a \\
    &   ''            &             &            &                     &                      &       &    &   &    &   7.55   &  160 $\pm$  4   &   25.6 $\pm$  0.7  &    &   \\
 54 &      HD~111990  &   435$\pm$15& 256$\pm$ 8 &1660$^{+200}_{-160}$ & $ 3643^{+591}_{-452}$ & 1800  &  M & M & M  & -14.10   &   87 $\pm$  6   &   14.5 $\pm$  1.0  & c+CN & a \\
    &   ''            &             &            &                     &                      &       &    &   &    &   7.86   &  205 $\pm$  6   &   31.9 $\pm$  1.0  &    &   \\
 55 & {\bf HD~112607} &    86$\pm$ 4&  54$\pm$ 4 & 430$^{+120}_{- 64}$ &$  592^{+15}_{- 14}$ &   {633}  &  S & S & S  &   5.38   &  730 $\pm$  7   &   90.1 $\pm$  0.9  & c+CN & a \\
 56 & {\bf HD~112954} &    72$\pm$ 3&  53$\pm$ 2 & 556$^{\imath}_{-61}$   &$  387^{+ 6}_{-  6}$ &   {140}  &  S & S & S  &   4.18   & 1176 $\pm$  7   &  124.8 $\pm$  0.8  & c+CN & a \\
 57 &      HD~122669  &   343$\pm$15& 223$\pm$12 &1650$^{+400}_{-270}$ &$ 2258^{+172}_{-150}$ & 1610  &  M & M & M  &  18.14   &  163 $\pm$  4   &   26.1 $\pm$  0.7  & w-CN & a \\
    &   ''            &             &            &                     &                      &       &    &   &    &  -3.30   &   40 $\pm$  5   &    6.9 $\pm$  0.8  &    &   \\
    &   ''            &             &            &                     &                      &       &    &   &    &   2.18   &  146 $\pm$  6   &   24.1 $\pm$  1.0  &    &   \\
 58 &      HD~123008  &   594$\pm$19& 383$\pm$11 &2750$^{+365}_{-290}$  & $ 4115^{+780}_{-575}$ &  {5191}  &  M & M & M  & -15.74   &  177 $\pm$  8   &   28.3 $\pm$  1.3  & w-CN & a \\
    &   ''            &             &            &                     &                      &       &    &   &    &  -4.45   &   95 $\pm$  7   &   15.8 $\pm$  1.2  &    &   \\
    &   ''            &             &            &                     &                      &       &    &   &    &   0.16   &   93 $\pm$  7   &   15.4 $\pm$  1.2  &    &   \\
    &   ''            &             &            &                     &                      &       &    &   &    &   5.14   &   41 $\pm$  7   &    7.0 $\pm$  1.2  &    &   \\
 60 & HD~125288$^{{LR}}$ &    37$\pm$ 1&  16$\pm$ 1 & 150$^{+12}_{- 7}$   & $  239^{+57}_{- 39}$ &   {271}  &  S & S & S  &   2.00   &  262 $\pm$  7   &   42.2 $\pm$  1.1  & - & a \\
 61 & {\bf HD~129557} &    60$\pm$2 &  26$\pm$2  & 200$^{+12}_{-10}$   & $  430^{+11}_{- 11}$ &  295  &  S & S & S  &  -1.00   &    0 $\pm$  0   &   -1.0 $\pm$ -1.0  & - & a \\
 63 &      HD~141318  &   111$\pm$ 4&  62$\pm$ 4 & 440$^{+70}_{-60}$   & $  589^{+33}_{- 30}$ &  660  &  M & M & M  &  -1.82   &  132 $\pm$  4   &   21.3 $\pm$  0.7  & - & a \\
    &   ''            &             &            &                     &                      &       &    &   &    &   2.53   &   99 $\pm$  6   &   16.5 $\pm$  1.0  &    &   \\
 64 &      HD~141926  &   285$\pm$ 7& 172$\pm$ 5 &1191$^{+287}_{-174}$   & $ 1345^{+88}_{- 78}$ & 1110  &  M & M & M  & -15.82   &  165 $\pm$ 10   &   27.1 $\pm$  1.7  & - & c \\
    &  ''             &             &            &                     &                      &       &    &   &    &     -5.72   &  548 $\pm$ 20   &   74.8 $\pm$  1.3  &    &   \\
    &  ''             &             &            &                     &                      &       &    &   &    &      2.30   &  138 $\pm$  8   &   22.6 $\pm$  1.4  &    &   \\
 65 &      HD~143054  &   215$\pm$ 9& 147$\pm$ 9 &1210$^{+220}_{-160}$ & $  290^{+50}_{- 38}$ & 1070  &  M & M & M  & -23.97   &  241 $\pm$  7   &   36.5 $\pm$  1.1  & - & b \\
    &   ''            &             &            &                     &                      &       &    &   &    & -15.81   &  183 $\pm$  8   &   29.1 $\pm$  1.3  &    &   \\
    &   ''            &             &            &                     &                      &       &    &   &    &  -7.24   &  141 $\pm$  9   &   23.2 $\pm$  1.5  &    &   \\
    &   ''            &             &            &                     &                      &       &    &   &    &   0.10   &  232 $\pm$  7   &   35.5 $\pm$  1.1  &    &   \\
 66 & {\bf HD~146284} &    61$\pm$ 3&  33$\pm$ 2 & 260$^{+40}_{-30}$   & $  205^{+ 3}_{-  3}$ &  250  &  M & M & S  &  -7.48   &  372 $\pm$  4   &   51.0 $\pm$  0.5  & - & b \\
 67 & {\bf HD~146285} &    69$\pm$ 4&  41$\pm$ 4 & 326$^{+85}_{-54}$   & $  156^{+ 2}_{-  2}$ &  275  &  M & S & S  &  -5.39   &  137 $\pm$  5   &   22.2 $\pm$  0.8  & w-CN & b \\
 68 & {\bf HD~147165} &    55$\pm$ 1&  30$\pm$ 1 & 247$^{+16}_{-14}$   & $  103^{+14}_{- 11}$ &  130  &  S & S & S  &  -5.89   &  143 $\pm$  6   &   22.9 $\pm$  0.9  & - & c \\
 69 & {\bf HD~147196} &    32$\pm$ 4&  20$\pm$ 3 & 213$^{+62}_{-58}$   & $  139^{+ 2}_{-  2}$ &  175  &  M & S & S  &  -6.46   &  359 $\pm$  5   &   50.0 $\pm$  0.7  & c+CN & a \\
 71 & {\bf HD~147701} &    50$\pm$ 2&  40$\pm$ 1 & $^{\imath}-$        & $  139^{+ 1}_{-  1}$ &  140  &  S & S & S  &  -6.70   &  862 $\pm$  4   &   93.1 $\pm$  0.4  & c+CN & b \\
 72 & {\bf HD~147888} &    66$\pm$ 2&  53$\pm$ 2 & $^{\imath}-$        & $   92^{+ 4}_{-  3}$ &  185  &  S & S & S  &  -8.57   &  828 $\pm$  5   &   86.7 $\pm$  0.5  & c+CN & a \\
 73 & {\bf HD~147889} &   43$\pm$2  &  37$\pm$ 2 & $^{\imath}-$        & $  138^{+ 2}_{-  2}$ &   {140}  &  S & S & S  &  -7.73   &  675 $\pm$  4   &   72.6 $\pm$  0.5  & c+CN & a \\
 74 & {\bf HD~147933} &    65$\pm$ 4&  49$\pm$ 3 & $^{\imath}-$        & $  140^{+ 4}_{-  4}$ &  155  &  S & S & S  &  -8.17   &  685 $\pm$  3   &   88.0 $\pm$  0.4  & c+CN & a \\
 75 & {\bf HD~148579} &    55$\pm$ 5&  33$\pm$ 4 & 290$^{+120}_{-65}$  & $  139^{+ 3}_{-  3}$ &   {165}  &  S & S & S  &  -5.79   &  344 $\pm$  3   &   49.3 $\pm$  0.5  & w-CN & a \\
 76 & {\bf HD~148594} &    64$\pm$ 5&  38$\pm$ 4 & 310$^{+115}_{-60}$  & $  192^{+ 4}_{-  4}$ & 285  &  M & S & S  &  -4.62   &  255 $\pm$  4   &   38.5 $\pm$  0.6  & w-CN & a \\
 77 &      HD~149038  &   238$\pm$ 3& 147$\pm$ 2 &1040$^{+50}_{-45}$   & $ 1032^{+1746}_{-425}$ &  980  &  M & - & M  &  -8.97   &   42 $\pm$  6   &    7.1 $\pm$  1.1  & w-CN & a \\
    &  ''             &             &            &                     &                      &       &    &   &    &  -5.35   &   76 $\pm$  5   &   12.4 $\pm$  0.8  &    &   \\
    &  ''             &             &            &                     &                      &       &    &   &    &  -2.32   &  115 $\pm$  6   &   18.5 $\pm$  1.0  &    &   \\
    &  ''             &             &            &                     &                      &       &    &   &    &   1.46   &  140 $\pm$  5   &   21.6 $\pm$  0.8  &    &   \\
 78 & {\bf HD~149757} &    49$\pm$ 1&  27$\pm$ 1 & 230$^{+8}_{-12}$    & $  182^{+53}_{- 34}$ &  230  &  M & M & S  & -14.23   &  474 $\pm$  4   &   58.4 $\pm$  0.5  & c+CN & a \\
 80 &      HD~152096  &   439$\pm$25& 296$\pm$14 &2300$^{+640}_{-360}$ & $ 1434^{+100}_{- 88}$ & --    &  M & M & M  &  -4.58   &  142 $\pm$  6   &   22.8 $\pm$  1.0  & w-CN & a \\
    &  ''             &             &            &                     &                      &       &    &   &    &   1.27   &  102 $\pm$  6   &   16.9 $\pm$  1.1  &    &   \\
 81 &      HD~152245  &   406$\pm$18& 252$\pm$14 &1740$^{+390}_{-270}$ & $ 1776^{+282}_{-216}$ & 1765  &  M & M & M  & -12.95   &   81 $\pm$  4   &   13.4 $\pm$  0.6  & w-CN & a \\
    &  ''             &             &            &                     &                      &       &    &   &    &  -5.60   &  217 $\pm$  6   &   34.7 $\pm$  1.0  &    &   \\
    &  ''             &             &            &                     &                      &       &    &   &    &   0.65   &  107 $\pm$  4   &   17.5 $\pm$  0.6  &    &   \\
    &  ''             &             &            &                     &                      &       &    &   &    &   4.71   &   42 $\pm$  4   &    7.1 $\pm$  0.7  &    &   \\
 82 &      HD~152247  &   430$\pm$34& 291$\pm$15 &2270$^{+820}_{-320}$ & $ 1850^{+220}_{-179}$ & 1960  &  M & M & M  &  -8.17   &  174 $\pm$ 35   &   29.8 $\pm$  6.1  & - & a \\
    &  ''             &             &            &                     &                      &       &    &   &    &   2.14   &  532 $\pm$ 15   &   80.2 $\pm$  2.2  &    &   \\
 \hline
\end{tabular}
\end{center}
\end{table*}

\setcounter{table}{2}
\begin{table*}[!htb]
\scriptsize
\begin{center}
  \caption { -- continued -- }
 \begin{tabular}{r | l |c  c | r r r | c c c | r r r | c c }
   \hline
   \hline
1  &  2     & 3     & 4     & 5      & 6     & 7    & 8      & 9    &10   & 11                   & 12         &  13      & 14 & 15 \\
\hline
ID & Target & \multicolumn{2}{c|}{Ca\,{\sc ii}} & \multicolumn{3}{c|}{Distance} & \multicolumn{3}{c|}{Components}  & \multicolumn{3}{c|}{K\,{\sc i}} & Environ. & WISE \\
&        &   EW(K)    &  EW(H) & Ca~{\sc ii} & GAIA  & Sp/L & Ca~{\sc ii}~K & Na~{\sc i} & K~{\sc i} & v$_{\bigodot}$  & log N  & EW  &class  & class \\
   &        &  (m\AA)  &  (m\AA) & \multicolumn{3}{c|}{(pc)}  &  &  &   & (km/s) & ($10^9$\,/cm$^2$) & (m\AA) &  &  \\
\hline
83 &      HD~152560  &   411$\pm$14& 283$\pm$ 9 &2280$^{+380}_{-100}$ & $ 1754^{+249}_{-195}$ & 2340  &  M & - & M  &  -4.26   &   69 $\pm$  2   &   11.4 $\pm$  0.4  & - & a  \\
&  ''             &             &            &                     &                      &       &    &   &    &   1.09   &  456 $\pm$  4   &   63.2 $\pm$  0.5  &  \\
 84 &      HD~152667  &   399$\pm$13& 267$\pm$ 9 &2055$^{+330}_{-250}$ & $ 1584^{+177}_{-145}$ & 1760  &  M & M & M  &  -1.41   &  490 $\pm$  4   &   66.2 $\pm$  0.5  & c+CN &-$^e$ \\
    &  ''             &             &            &                     &                      &       &    &   &    &   3.29   &  304 $\pm$  4   &   45.0 $\pm$  0.5  &    &   \\
 85 &      HD~155756  &   718$\pm$22& 494$\pm$14 &3910$^{+595}_{-450}$ & $ 3307^{+823}_{-561}$ & 6870  &  M & M & M  & -14.88   &  358 $\pm$  5   &   49.7 $\pm$  0.7  & c+CN & b \\
    &  ''             &             &            &                     &                      &       &    &   &    &   4.20   &  165 $\pm$  6   &   26.4 $\pm$  0.9  & w-CN &   \\
 86 & {\bf HD~156247} &    73$\pm$ 3&  46$\pm$ 3 & 390$^{+80}_{-60}$   & $  266^{+10}_{-  9}$ & 425  &  M & S & S  & -15.73   &  206 $\pm$  6   &   31.8 $\pm$  0.9  & w+CN & b \\
 87 &      HD~161653  &   243$\pm$ 6& 147$\pm$ 5 &1020$^{+110}_{-100}$ & $ 1418^{+192}_{-152}$ & 1230  &  M & M & S  &  -4.28   &  254 $\pm$  4   &   38.5 $\pm$  0.6  & w-CN & a \\
 88 &      HD~162978  &   200$\pm$ 6& 122$\pm$ 5 & 870$^{+110}_{-100}$ & $ 1082^{+120}_{- 99}$ & 1220  &  M & M & S  &  -5.17   &  229 $\pm$  4   &   35.4 $\pm$  0.7  & w-CN & a \\
 89 &      HD~163181  &   262$\pm$10& 168$\pm$ 7 &1240$^{+220}_{-160}$ & $ 1879^{+240}_{-192}$ & 1710  &  M & M & M  &  -9.41   &  135 $\pm$  3   &   21.6 $\pm$  0.4  & w+CN & a \\
    &  ''             &             &            &                   &                    &        &   &   &     &  -3.52   &  453 $\pm$  3   &   58.7 $\pm$  0.4  &    &   \\
 90 & {\bf HD~164536} &   274$\pm$13& 167$\pm$ 9 &1155$^{+240}_{-170}$ & $ 1402^{+1452}_{-488}$ & 1980  &  M & M & S  &  -5.07   &  293 $\pm$  6   &   43.5 $\pm$  0.9  & w-CN & d \\
 91 & {\bf Hershel~36}&   211$\pm$ 1& 129$\pm$ 1 & 910$^{+200}_{-200}$   & $ 1110^{+360}_{-110}$ &  -    &  M & S & S  &  -5.00   &  616 $\pm$  6   &   82.9 $\pm$  0.8  & - & -$^e$ \\
 92 & {\bf HD~164816} &   182$\pm$10& 111$\pm$10 & 790$^{+260}_{-160}$ & $ 1155^{+106}_{- 90}$ & 1020  &  M & M & S  &  -4.73   &  311 $\pm$ 30   &   49.7 $\pm$  4.8  & - & a \\
 93 &      HD~164863  &   169$\pm$ 6& 102$\pm$ 4 & 730$^{+100}_{-80}$  & $ 2747^{+1214}_{-674}$ & 1270  &  M & M & M  &  -8.92   &   74 $\pm$  4   &   12.4 $\pm$  0.7  & w-CN & a \\
    &  ''             &             &            &                     &                      &       &    &   &    &  -5.59   &  108 $\pm$  4   &   17.5 $\pm$  0.6  &    &   \\
 94 & {\bf HD~164906} &   222$\pm$ 4& 133$\pm$ 4 & 900$^{+100}_{-55}$  & $ 1170^{+609}_{-301}$ &  940  &  M & M & S  &  -5.40   &  459 $\pm$  5   &   58.7 $\pm$  0.6  & - & d \\
 95 & {\bf HD~164947A}&   262$\pm$14& 138$\pm$ 9 & 830$^{+268}_{-150}$ & $ 1264^{+58}_{- 54}$ & 1490  &  M & M & S  &  -4.95   &  290 $\pm$ 13   &   43.3 $\pm$  1.9  & w-CN & a \\
 95 & {\bf HD~164947B}&   238$\pm$23& 149$\pm$15 &1075$^{+440}_{-240}$ & $ 1225^{+89}_{- 78}$ & 1810  &  M & M & S  &  -4.96   &  292 $\pm$ 11   &   43.5 $\pm$  1.6  & w-CN & a \\
 96 & {\bf HD~167771} &   299$\pm$13& 193$\pm$10 &1430$^{+320}_{-230}$ & $ 1833^{+192}_{-159}$ & 3030  &  M & S & S  &  -6.30   &  603 $\pm$  6   &   79.2 $\pm$  0.8  & c+CN & a \\
 97 &      HD~168750  &   317$\pm$24& 193$\pm$15 &1320$^{+470}_{-280}$ & $ 1670^{+171}_{-142}$ & 1750  &  M & M & M  &  -3.83   &  501 $\pm$ 10   &   72.5 $\pm$  1.5  & - &  -$^e$ \\
    &  ''             &             &            &                     &                      &       &    &   &    &   1.85   &  131 $\pm$  5   &   20.9 $\pm$  0.9  &    &   \\
 98 &      HD~168941  &   423$\pm$36& 255$\pm$28 &1700$^{+850}_{-450}$ & $ 2394^{+637}_{-421}$ & 6940  &  S & M & M  &  -2.21   &  387 $\pm$ 24   &   62.5 $\pm$  3.9  & - & c \\
    &  ''             &             &            &                     &                      &       &    &   &    &  19.41   &   27 $\pm$ 23   &    4.7 $\pm$  3.9  &    &   \\
 99 &      HD~169582  &   624$\pm$22& 484$\pm$13 & $^{\imath}-$        & $ 1681^{+197}_{-160}$ & 3500  &  M & M & S  & -10.17   & 1422 $\pm$ 12   &  166.3 $\pm$  1.4  & c+CN & a \\
100 & {HD~170634} &   153$\pm$ 7& 122$\pm$ 6 & $^{\imath}-$        & $  417^{+ 9}_{-  9}$ &  420  &  S & S &  {M}  & -11.89   & 2436 $\pm$ 14   &  220.8 $\pm$  1.3  & c+CN & c \\
101 & {HD~170740} &   69$\pm$  1&  42$\pm$1  & 350$^{+17}_{-13}$   & $  230^{+ 7}_{-  7}$ &  320  &  M & S &  {M}  & -10.83   &  567 $\pm$  4   &   77.2 $\pm$  0.6  & - & b \\
104 &      HD~172175  &   658$\pm$22& 488$\pm$13 & 4480$^{\imath}_{-990}$& $ 2147^{+300}_{-236}$ & 4200  &  M & M & M  & -16.00   &  273 $\pm$  6   &   40.7 $\pm$  0.9  & c+CN & a \\
    &  ''             &             &            &                    &                         &      &    &   &    &  -9.12   & 2051 $\pm$ 15   &  181.9 $\pm$  1.3  & c+CN   &   \\
    &  ''             &             &            &                    &                         &      &    &   &    &  -0.89   &  892 $\pm$ 11   &  114.5 $\pm$  1.4  &    &   \\
105 & HD~177989$^{{LR}}$ &   281$\pm$26& 170$\pm$13 &1100$^{+520}_{-190}$ & $ 2398^{+405}_{-305}$ & --    &    &   &    &  -1.86   &  640 $\pm$ 29   &   91.4 $\pm$  4.2  & - & -$^e$ \\
106 & {\bf HD~180968} &   102$\pm$ 6&  67$\pm$ 6 & 560$^{+215}_{-120}$ & $  561^{+38}_{- 34}$ &  400  &  S & S & S  & -11.64   &  569 $\pm$ 28   &   82.8 $\pm$  4.1  & - & a \\
107 &      HD~185418  &   154$\pm$ 4& 112$\pm$ 3 &1050$^{+150}_{-120}$ & $  740^{+21}_{- 20}$ & 1050  &  S & - & M  & -11.23   &  406 $\pm$  5   &   52.9 $\pm$  0.6  & -$^{v}$ & a \\
    &  ''             &             &            &                     &                       &      &    &   &    &  -8.04   &  343 $\pm$  6   &   51.0 $\pm$  0.9  &    &   \\
108 & {\bf HD~185859} &   233$\pm$ 4& 179$\pm$ 3 &1830$^{\imath}_{-160}$ & $ 1086^{+49}_{- 45}$ & 1350  &  M & S & S  &  -7.38   & 1090 $\pm$  5   &  114.8 $\pm$  0.6  & c+CN & b \\
109 & HD~198478$^{{LR}}$ &   203$\pm$10& 149$\pm$ 8 &1390$^{\imath}_{-290}$ & $ 1164^{+238}_{-171}$ & 1260  &  S & S & S  & -13.99   &  868 $\pm$ 17   &  129.5 $\pm$  2.6  & - & c \\
110 & HD~200775$^{{LR}}$ &    80$\pm$3 &  52$\pm$ 2 & 445$^{+65}_{-50}$   & $  357^{+ 6}_{-  6}$ &  400  &  S & S & S  & -13.30   &  112 $\pm$  3   &   18.8 $\pm$  0.6  & c+CN & c \\
111 & {\bf HD~203532} &   50.5$\pm$1& 30.5$\pm$1 & 270$^{+20}_{-17}$   & $  289^{+ 4}_{-  4}$ &  250  &  S & S & S  &  14.55   &  968 $\pm$  4   &  108.6 $\pm$  0.4  & c+CN & a \\
112 &      HD~204827$^{{LR}}$  &   280$\pm$12& 208$\pm$ 8 &1960$^{\imath}_{-340}$ & $ 1269^{+183}_{-143}$ & 1630  &  M & - & M  & -17.29   & 2672 $\pm$ 23   &  279.6 $\pm$  2.4  & c+CN & a \\
    &  ''             &             &            &                     &                       &      &    &   &    &  -7.74   &  222 $\pm$  9   &   35.7 $\pm$  1.4  &    &   \\
113 &      HD~206267  &   249$\pm$17& 159$\pm$ 8 &1170$^{+320}_{-200}$ & $ 1191^{+751}_{-337}$ & 1030  &  M & - & M  & -17.77   &  401 $\pm$  8   &   56.5 $\pm$  1.1  & -$^{v}$ & a \\
    &  ''             &             &            &                     &                       &      &    &   &    & -13.61   & 1293 $\pm$ 18   &  164.0 $\pm$  2.3  &    &   \\
114 & HD~207198$^{{LR}}$ &   316$\pm$18& 226$\pm$ 9 &1970$^{\imath}_{-100}$ & $  999^{+58}_{- 52}$ & 1185  &  S & - & S  & -13.32   & 1930 $\pm$ 21   &  241.3 $\pm$  2.6  & c+CN & b \\
115 & HD~209975$^{{LR}}$ &   241$\pm$13& 173$\pm$ 7 &1530$^{\imath}_{-270}$ & $  857^{+132}_{-101}$ & 1210  &  S & - & S  & -13.40   &  680 $\pm$ 14   &  105.2 $\pm$  2.2  & - & a \\
116 & {\bf HD~210121} &   93$\pm$2  &  56$\pm$2  & 430$^{+45}_{-30}$   & $  339^{+16}_{- 14}$ &  475  &  M & S & S  & -14.57   &  896 $\pm$  3   &   98.8 $\pm$  0.3  & c+CN & a \\
117 &      HD~218376  &   150$\pm$ 3&  97$\pm$ 3 & 760$^{+80}_{-70}$   & $  375^{+41}_{- 34}$ &  470  &  M & - & M  & -16.87   &  238 $\pm$ 14   &   38.5 $\pm$  2.3  & - & a \\
    &  ''             &             &            &                     &                       &      &    &   &    &  -9.38   &  265 $\pm$  9   &   40.3 $\pm$  1.4  &    &   \\
    &  ''             &             &            &                     &                       &      &    &   &    &  -2.88   &   16 $\pm$  6   &    2.8 $\pm$  1.1  &    &   \\
118 &      HD~259105  &   271$\pm$ 7& 189$\pm$ 5 &1580$^{+210}_{-155}$ & $ 1471^{+191}_{-153}$ & 1800  &  M & M & M  &  17.89   &  131 $\pm$  8   &   21.5 $\pm$  1.4  & w-CN & d \\
    &  ''             &             &            &                     &                       &      &    &   &    &  24.22   &  392 $\pm$  9   &   56.8 $\pm$  1.3  &    &   \\
    &  ''             &             &            &                     &                       &      &    &   &    &  31.01   &  331 $\pm$  9   &   49.6 $\pm$  1.4  &    &   \\
119 & HD~281159$^{{LR}}$ &    95$\pm$ 8&  67$\pm$ 6 & 620$^{\imath}_{-160}$ &$ 1062^{+826}_{-401}$ &  370  &  S & S & S  &  13.28   & 1044 $\pm$ 19   &  143.4 $\pm$  2.6  & c+CN & c \\
120 &      HD~284839  &   202$\pm$ 8& 145$\pm$ 5 &1300$^{\imath}_{-190}$ &$  648^{+17}_{- 16}$ &  680  &  S & S & M  &  14.35   &  250 $\pm$  6   &   37.4 $\pm$  1.0  & c+CN & a \\
    &  ''             &             &            &                     &                       &      &    &   &    &  20.85   & 1461 $\pm$ 16   &  165.2 $\pm$  1.8  &    &   \\
121 &      HD~284841  &   187$\pm$ 7& 131$\pm$ 6 &1130$^{+260}_{-180}$ &$  606^{+15}_{- 14}$ & 557  &  S & - & M  &  15.14   &  289 $\pm$  5   &   42.7 $\pm$  0.8  & c+CN & a \\
    &  ''             &             &            &                     &                       &      &    &   &    &  20.45   & 1364 $\pm$ 11   &  145.1 $\pm$  1.1  &    &   \\
122 & {\bf HD~287150} &   174$\pm$ 9& 127$\pm$ 6 &1116$^{\imath}_{-250}$&$  432^{+ 9}_{-  8}$ &   {377}  &  M & S & S  &  20.94   &  883 $\pm$  7   &  102.2 $\pm$  0.8  & w-CN & a \\
123 &      HD~292167  &   587$\pm$22 & 422$\pm$15 &3640$^{\imath}_{-560}$&$ 4782^{+1375}_{-927}$&5000  &  M & M & M  &  26.37   &  916 $\pm$ 10   &  115.9 $\pm$  1.2  & c+CN & b \\
    &  ''             &             &            &                     &                       &      &    &   &    &  45.22   &  420 $\pm$  9   &   63.6 $\pm$  1.4  &    &   \\
124 &      HD~294264  &   137$\pm$ 8&  89$\pm$ 6 & 705$^{+215}_{-130}$ & $  445^{+ 9}_{-  9}$   & 755  &  M & S & M  &  26.10   &   97 $\pm$  7   &   16.3 $\pm$  1.2  & w-CN & d \\
    &  ''             &             &            &                     &                       &      &    &   &    &  29.42   &  239 $\pm$  4   &   35.7 $\pm$  0.6  &    &   \\
125 & {\bf HD~294304} &   177$\pm$ 9& 122$\pm$ 9 &1030$^{\imath}_{-230}$ &$  802^{+31}_{- 29}$ &  940  &  M & - & S  &  22.03   &  557 $\pm$ 10   &   77.8 $\pm$  1.4  & c+CN & d \\
126 & {\bf HD~315021} &   209$\pm$15& 131$\pm$13 & 955$^{+425}_{-230}$ & $ 1259^{+134}_{-111}$ & 870  &  M & - & S  &  -5.35   &  196 $\pm$  9   &   30.9 $\pm$  1.4  & - & d \\
127 & {\bf HD~315023} &   229$\pm$17& 127$\pm$11 & 810$^{+240}_{-160}$ & $ 1095^{+75}_{- 66}$  & 1135  &  M & - & M  & -25.27   &   80 $\pm$  8   &   13.3 $\pm$  1.4  & - & d \\
    &  ''             &             &            &                   &                          &      &    &   &    &  -4.47   &  315 $\pm$ 10   &   45.5 $\pm$  1.4  &    &   \\
128 & {\bf HD~315024} &   212$\pm$13& 131$\pm$ 9 & 940$^{+270}_{-180}$ & $ 1326^{+107}_{- 92}$ & 1225  &  M & M & S  &  -4.80   &  382 $\pm$  5   &   42.7 $\pm$  0.9  & w-CN & d \\
129 & {\bf HD~315031} &   229$\pm$ 9& 138$\pm$ 6 & 975$^{+130}_{-125}$ & $ 1407^{+161}_{-132}$ & 1650  &  M & M & S  &  -4.51   &  382 $\pm$  5   &   54.1 $\pm$  0.7  & w-CN & d \\
130 & {\bf HD~315032} &   260$\pm$11& 170$\pm$ 9 &1290$^{+300}_{-210}$ & $  996^{+87}_{- 74}$  & 1835  &  M & M & S  &  -5.18   &  438 $\pm$ 11   &   60.4 $\pm$  1.5  & w-CN & d \\
131 & {\bf HD~315033} &   261$\pm$13& 161$\pm$ 8 &1133$^{+258}_{-160}$ & $ 1578^{+146}_{-124}$ &  {1354}  &  M & M & S  &  -4.57   &  402 $\pm$  6   &   55.9 $\pm$  0.9  & w-CN & d \\
132 &      HD~326309  &   518$\pm$54& 333$\pm$34 &2390$^{\imath}_{-690}$ &$ 1599^{+152}_{-128}$& 1600  &  M & M & M  &  -3.54   &  270 $\pm$ 16   &   41.4 $\pm$  2.4  & c+CN & a \\
    &  ''             &             &            &                     &                       &      &    &   &    &   3.29   &  512 $\pm$ 14   &   66.3 $\pm$  1.9  &    &   \\
133 &      HD~326330  &   417$\pm$15& 264$\pm$14 &1870$^{+390}_{-280}$ & $ 1774^{+217}_{-175}$ & 2050  &  M & M & M  & -12.55   &   19 $\pm$  8   &    3.3 $\pm$  1.4  & w-CN & a \\
    &  ''             &             &            &                     &                       &      &    &   &    &  -5.74   &  192 $\pm$  9   &   30.5 $\pm$  1.5  &    &   \\
    &  ''             &             &            &                     &                       &      &    &   &    &   2.09   &  706 $\pm$ 11   &   87.1 $\pm$  1.4  &    &   \\
134 &      HD~326332  &   403$\pm$35& 246$\pm$18 &1670$^{+630}_{-360}$ & $ 1575^{+143}_{-121}$ & 1380  &  M & M & M  & -10.85   &  587 $\pm$ 17   &   83.8 $\pm$  2.4  & c+CN & a \\
    &  ''             &             &            &                     &                       &      &    &   &    &   1.74   &  619 $\pm$ 14   &   81.9 $\pm$  1.8  & w-CN   &   \\
135 &      HD~326333  &   391$\pm$16& 260$\pm$13 &1980$^{+460}_{-220}$ & $ 1500^{+134}_{-114}$ & 1960  &  M & M & M  & -10.64   &  237 $\pm$ 12   &   39.1 $\pm$  1.9  & c+CN & a \\
    &  ''             &             &            &                     &                       &      &    &   &    &   1.64   &  570 $\pm$  7   &   76.9 $\pm$  0.9  & w-CN   &   \\
\hline
  \end{tabular}
\end{center}
\end{table*}

\setcounter{table}{2}
\begin{table*}[!htb]
\scriptsize
\begin{center}
  \caption { -- continued -- }
 \begin{tabular}{r | l |c  c | r r r | c c c | r r r | c c }
   \hline
   \hline
1  &  2     & 3       & 4      & 5                & 6     & 7    & 8             & 9          &10         & 11             & 12    &  13  & 14    & 15 \\
\hline
ID & Target & \multicolumn{2}{c|}{Ca\,{\sc ii}} & \multicolumn{3}{c|}{Distance} & \multicolumn{3}{c|}{Components}  & \multicolumn{3}{c|}{K\,{\sc i}} & Environ. & WISE \\
   &        &   EW(K) &  EW(H) & Ca~{\sc ii} & GAIA  & Sp/L & Ca~{\sc ii}~K & Na~{\sc i} & K~{\sc i} & v$_{\bigodot}$  & log N  & EW  &class  & class \\
   &        &  (m\AA)  &  (m\AA) & \multicolumn{3}{c|}{(pc)}  &  &  &   & (km/s) & ($10^9$\,/cm$^2$) & (m\AA) &  &  \\
\hline
136 &      HD~326364  &   378$\pm$23& 196$\pm$13 &{$1597^{+379}_{-233}$} & $ 1679^{+149}_{-127}$ & {1859} &  M & M & M  &  -8.67   &  437 $\pm$  7   &   64.3 $\pm$  1.2  & c+CN & a \\
    &  ''             &             &            &                     &                       &      &    &   &    &   1.69   &  945 $\pm$  8   &  106.9 $\pm$  0.9  & w-CN   &   \\
\hline
\hline
  \end{tabular}
\end{center}
    {\bf Notes: } $^{{LR}}${{low-resolution spectrum has insufficient resolving power for
        classification as single-cloud sightline,}} $^{\imath}$
    saturated line, $^{v}$: ignored because lines have different
    velocities, $^e$: WISE image shows artefact. 
\end{table*}

\clearpage

\begin{figure*} [!htbp] 
  \includegraphics[width=18.5cm,clip=true,trim=0cm 3cm 0cm 0.cm]{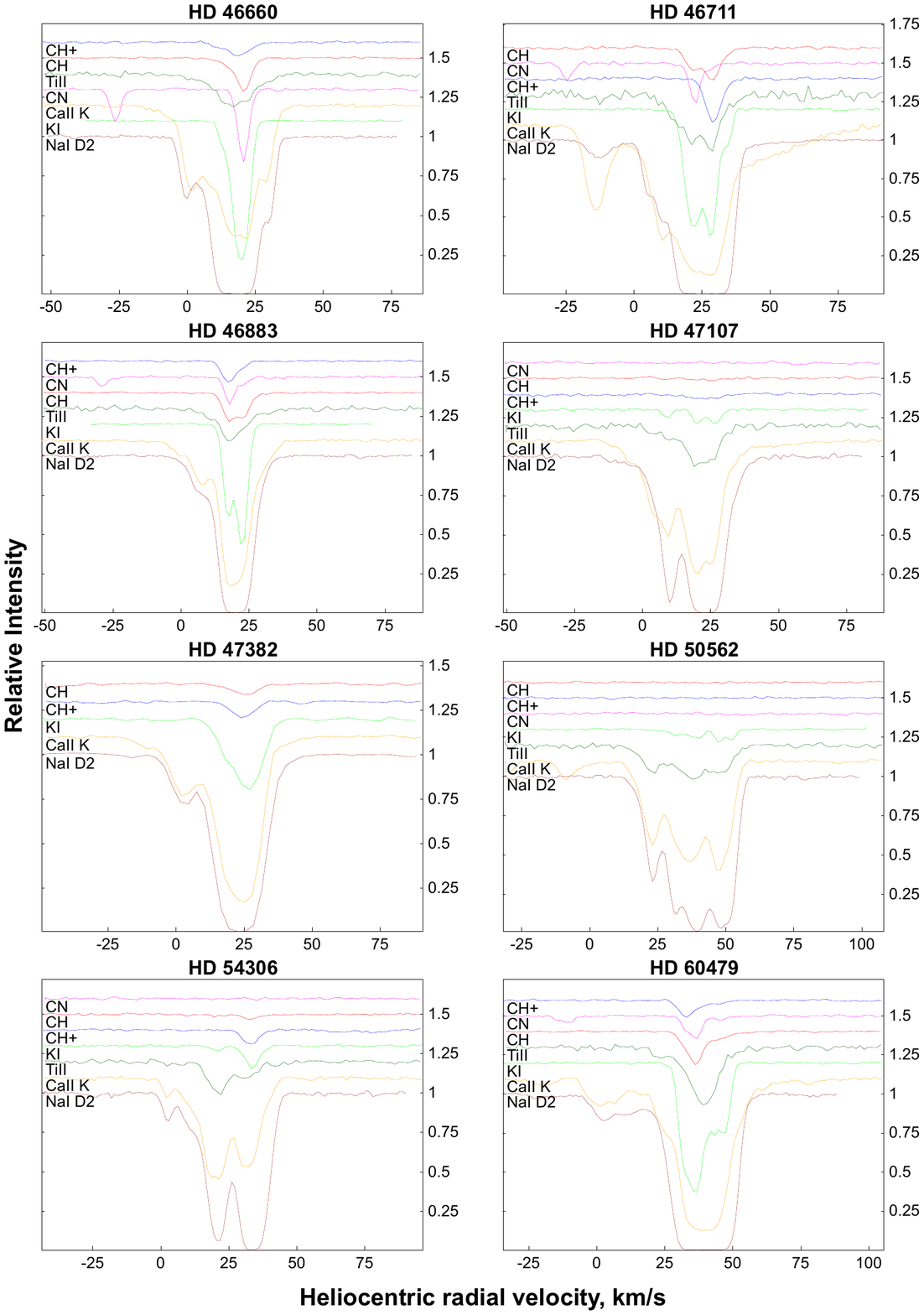}
  \caption{Velocity profiles of
    Ti\,{\sc ii} 3383.8\,\AA\,(dark green), CN 3874.6\, \AA\,
    (magenta), CH 4300.3\,\AA\, (red), CH$^+$ 4232.5\,\AA\, (blue),
    K\,{\sc i} 7699\,\AA\ (green), Ca\,{\sc ii} K 3933.7\,\AA\,
    (orange), and Na\,{\sc i} D2 5889\,\AA\,
    (brown). }
\end{figure*}

\begin{figure*} [!htbp] 
  \includegraphics[width=18.5cm,clip=true,trim=0cm 3cm 0cm 0.cm]{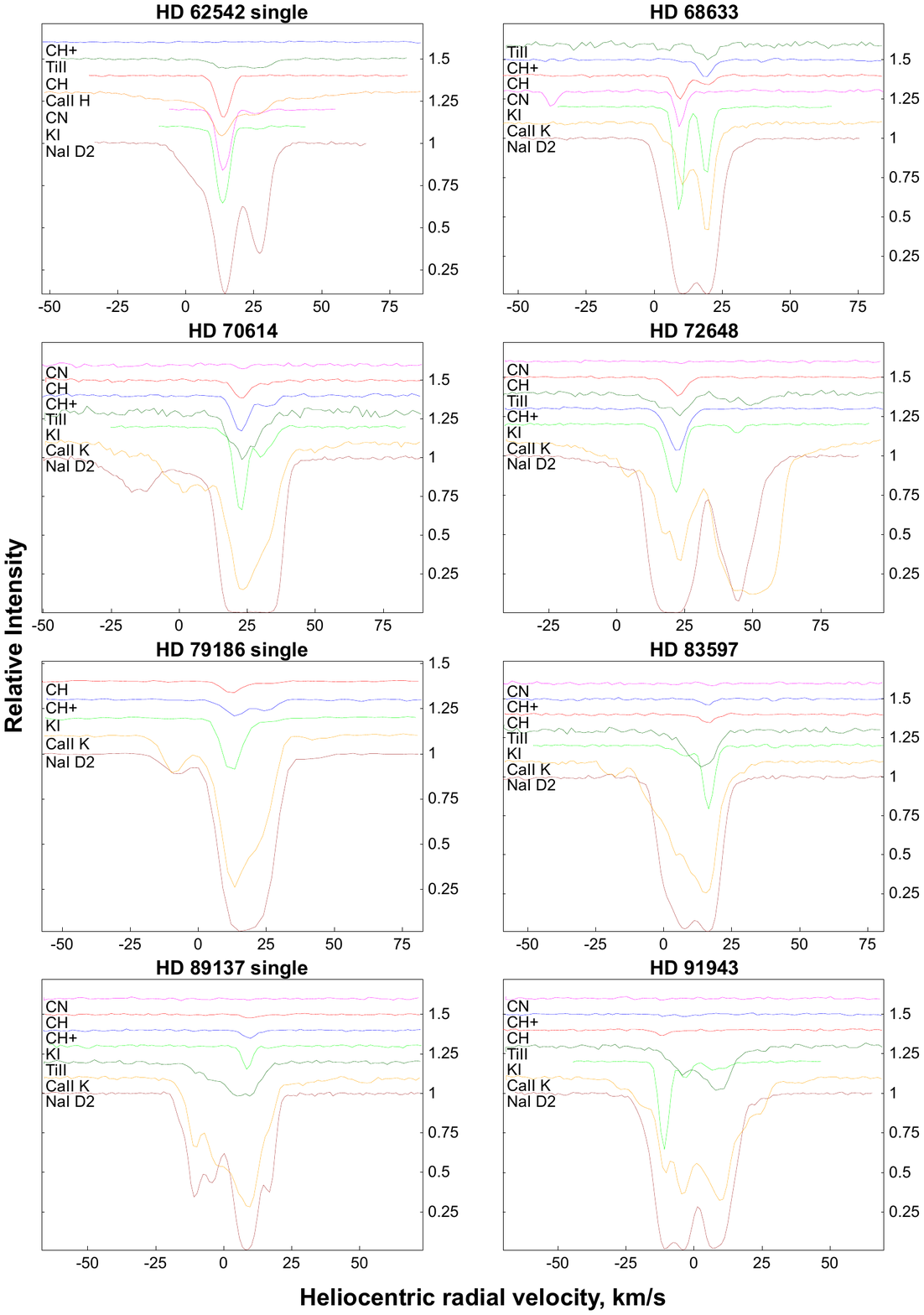}
  \caption{Velocity profiles of
    Ti\,{\sc ii} 3383.8\,\AA\,(dark green), CN 3874.6\, \AA\,
    (magenta), CH 4300.3\,\AA\, (red), CH$^+$ 4232.5\,\AA\, (blue),
    K\,{\sc i} 7699\,\AA\ (green), Ca\,{\sc ii} K 3933.7\,\AA\,
    (orange), and Na\,{\sc i} D2 5889\,\AA\,
    (brown). {Single-cloud sightlines are marked ``single''.} }
\end{figure*}

\begin{figure*} [!htbp] 
  \includegraphics[width=18.5cm,clip=true,trim=0cm 3cm 0cm 0.cm]{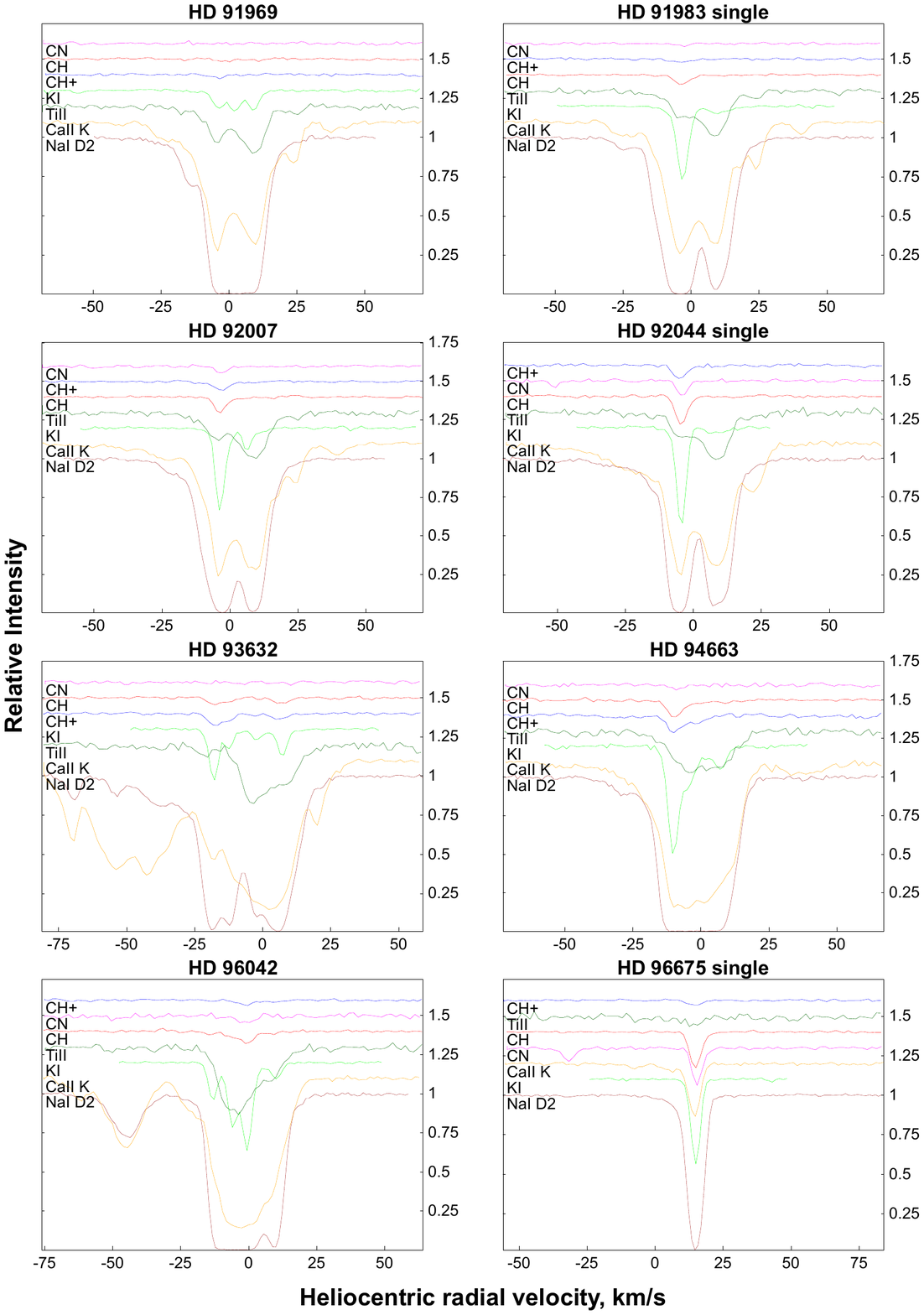}
  \caption{Velocity profiles of
    Ti\,{\sc ii} 3383.8\,\AA\,(dark green), CN 3874.6\, \AA\,
    (magenta), CH 4300.3\,\AA\, (red), CH$^+$ 4232.5\,\AA\, (blue),
    K\,{\sc i} 7699\,\AA\ (green), Ca\,{\sc ii} K 3933.7\,\AA\,
    (orange), and Na\,{\sc i} D2 5889\,\AA\,
    (brown). {Single-cloud sightlines are marked ``single''.} }
\end{figure*}

\begin{figure*} [!htbp] 
  \includegraphics[width=18.5cm,clip=true,trim=0cm 3cm 0cm 0.cm]{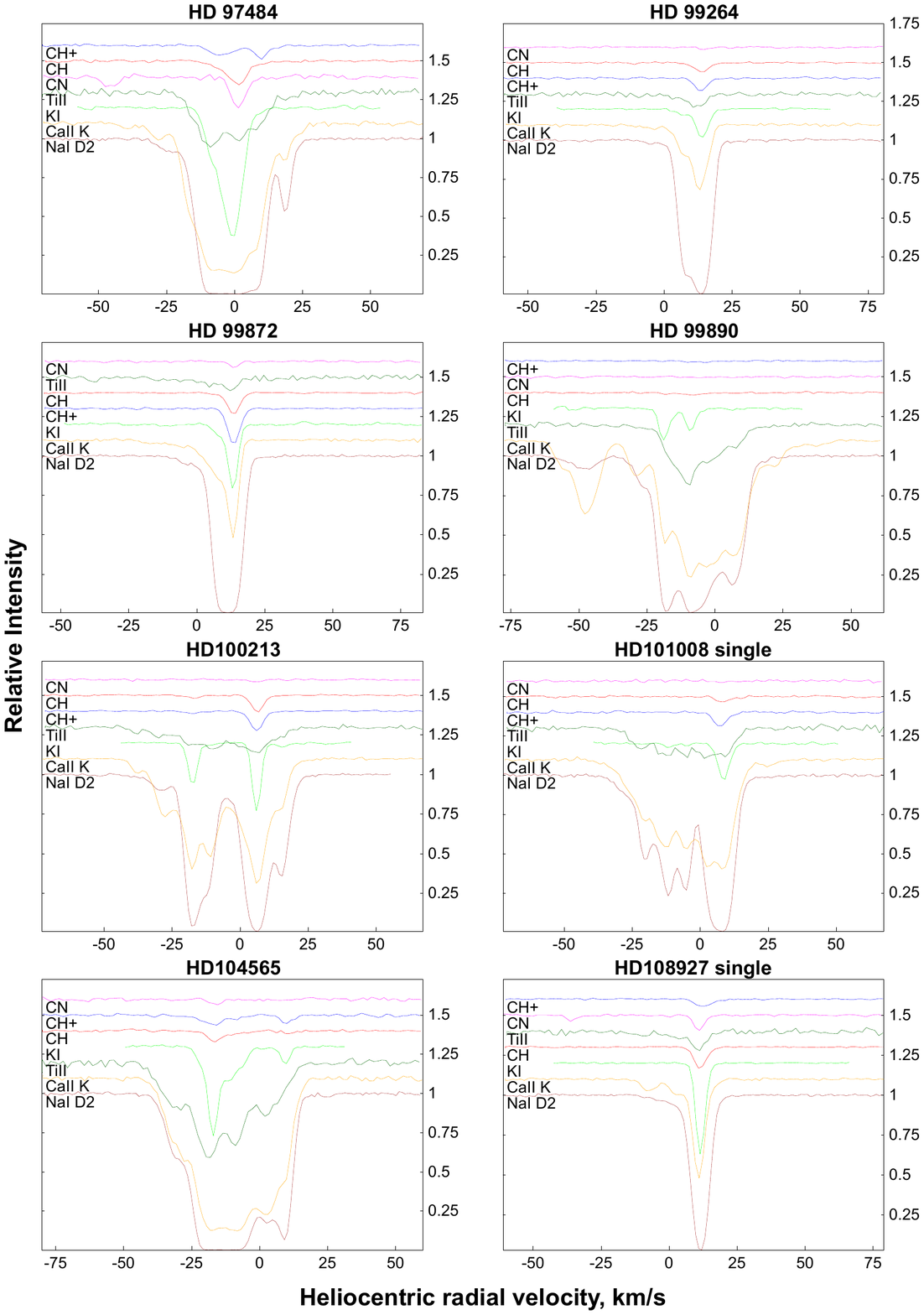}
  \caption{Velocity profiles of
    Ti\,{\sc ii} 3383.8\,\AA\,(dark green), CN 3874.6\, \AA\,
    (magenta), CH 4300.3\,\AA\, (red), CH$^+$ 4232.5\,\AA\, (blue),
    K\,{\sc i} 7699\,\AA\ (green), Ca\,{\sc ii} K 3933.7\,\AA\,
    (orange), and Na\,{\sc i} D2 5889\,\AA\,
    (brown). {Single-cloud sightlines are marked ``single''.} }
\end{figure*}

\begin{figure*} [!htbp] 
  \includegraphics[width=18.5cm,clip=true,trim=0cm 3cm 0cm 0.cm]{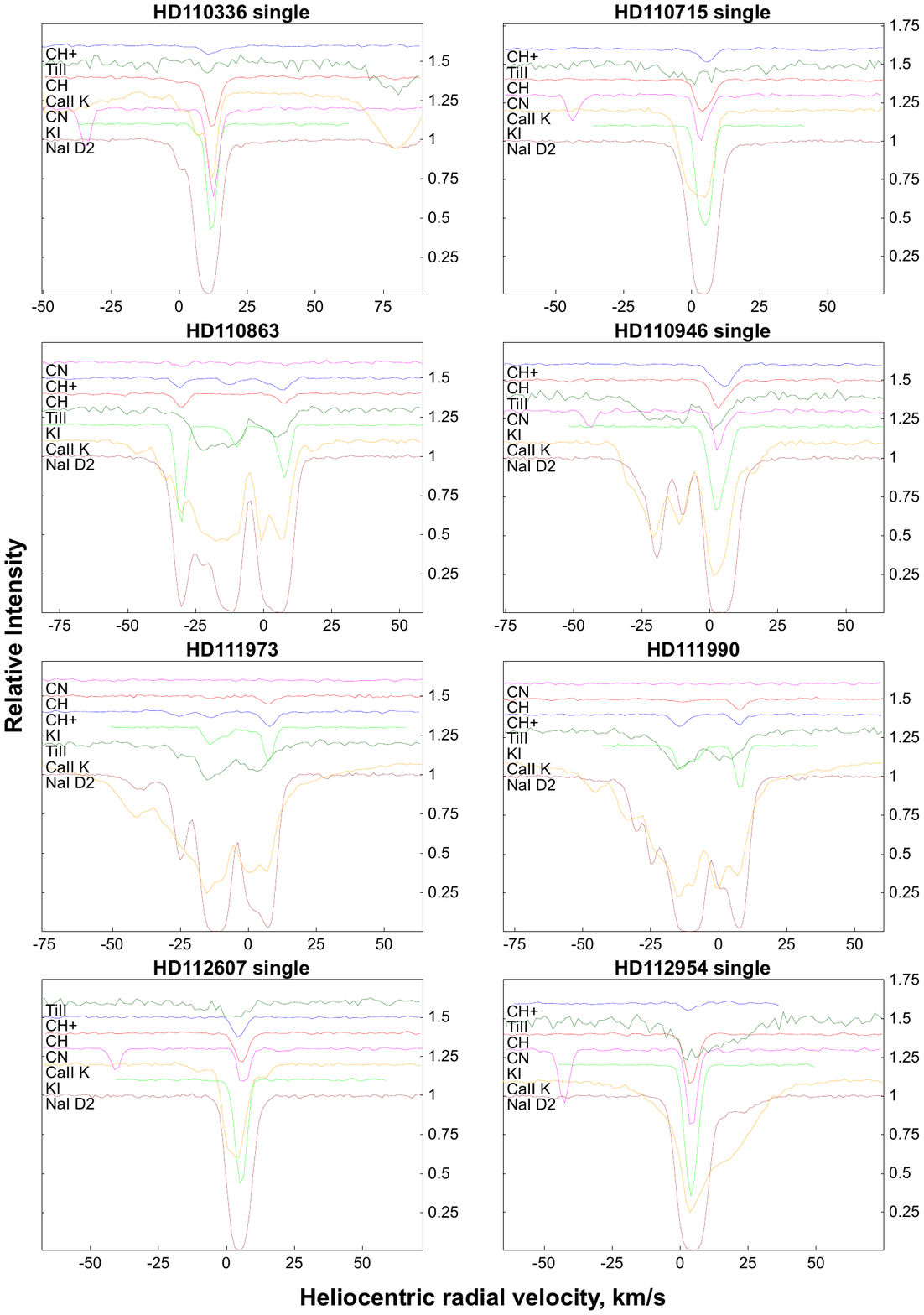}
  \caption{Velocity profiles of
    Ti\,{\sc ii} 3383.8\,\AA\,(dark green), CN 3874.6\, \AA\,
    (magenta), CH 4300.3\,\AA\, (red), CH$^+$ 4232.5\,\AA\, (blue),
    K\,{\sc i} 7699\,\AA\ (green), Ca\,{\sc ii} K 3933.7\,\AA\,
    (orange), and Na\,{\sc i} D2 5889\,\AA\,
    (brown). {Single-cloud sightlines are marked ``single''.} }
\end{figure*}

\begin{figure*} [!htbp] 
  \includegraphics[width=18.5cm,clip=true,trim=0cm 3cm 0cm 0.cm]{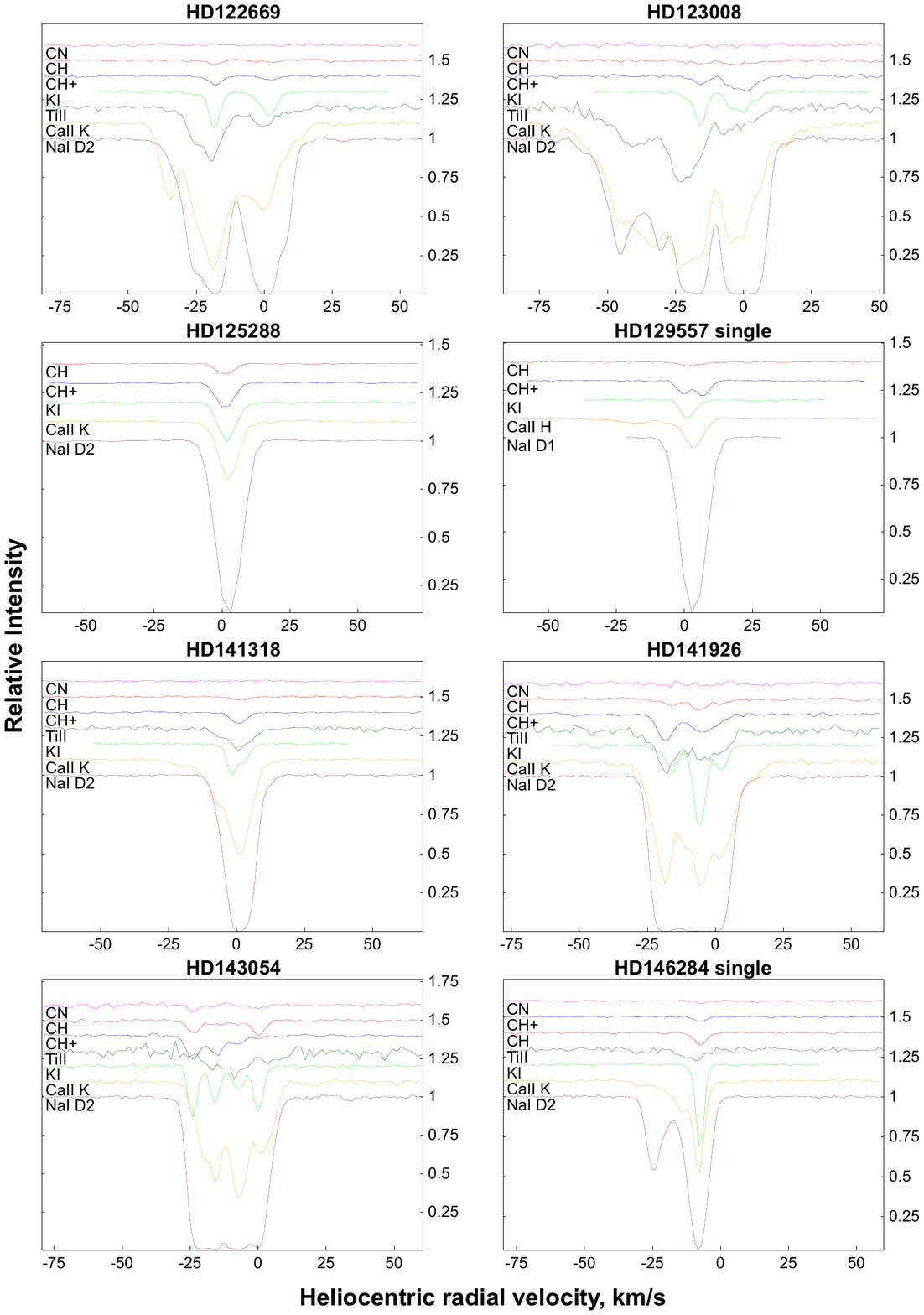}
  \caption{Velocity profiles of
    Ti\,{\sc ii} 3383.8\,\AA\,(dark green), CN 3874.6\, \AA\,
    (magenta), CH 4300.3\,\AA\, (red), CH$^+$ 4232.5\,\AA\, (blue),
    K\,{\sc i} 7699\,\AA\ (green), Ca\,{\sc ii} K 3933.7\,\AA\,
    (orange), and Na\,{\sc i} D2 5889\,\AA\,
    (brown). {Single-cloud sightlines are marked ``single''.} }
\end{figure*}

\begin{figure*} [!htbp] 
  \includegraphics[width=18.5cm,clip=true,trim=0cm 3cm 0cm 0.cm]{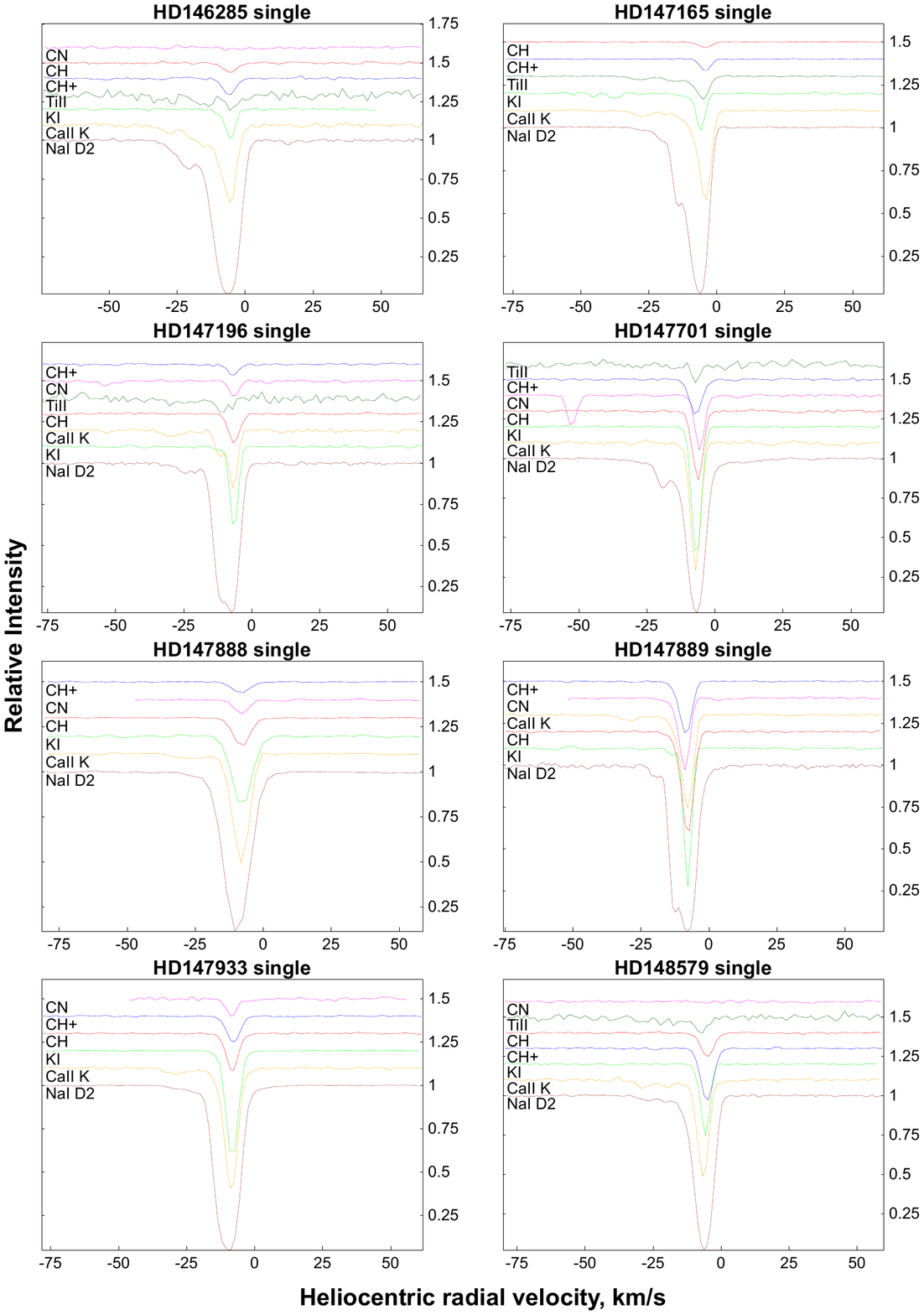}
  \caption{Velocity profiles of
    Ti\,{\sc ii} 3383.8\,\AA\,(dark green), CN 3874.6\, \AA\,
    (magenta), CH 4300.3\,\AA\, (red), CH$^+$ 4232.5\,\AA\, (blue),
    K\,{\sc i} 7699\,\AA\ (green), Ca\,{\sc ii} K 3933.7\,\AA\,
    (orange), and Na\,{\sc i} D2 5889\,\AA\,
    (brown). {Single-cloud sightlines are marked ``single''.} }
\end{figure*}

\begin{figure*} [!htbp] 
  \includegraphics[width=18.5cm,clip=true,trim=0cm 3cm 0cm 0.cm]{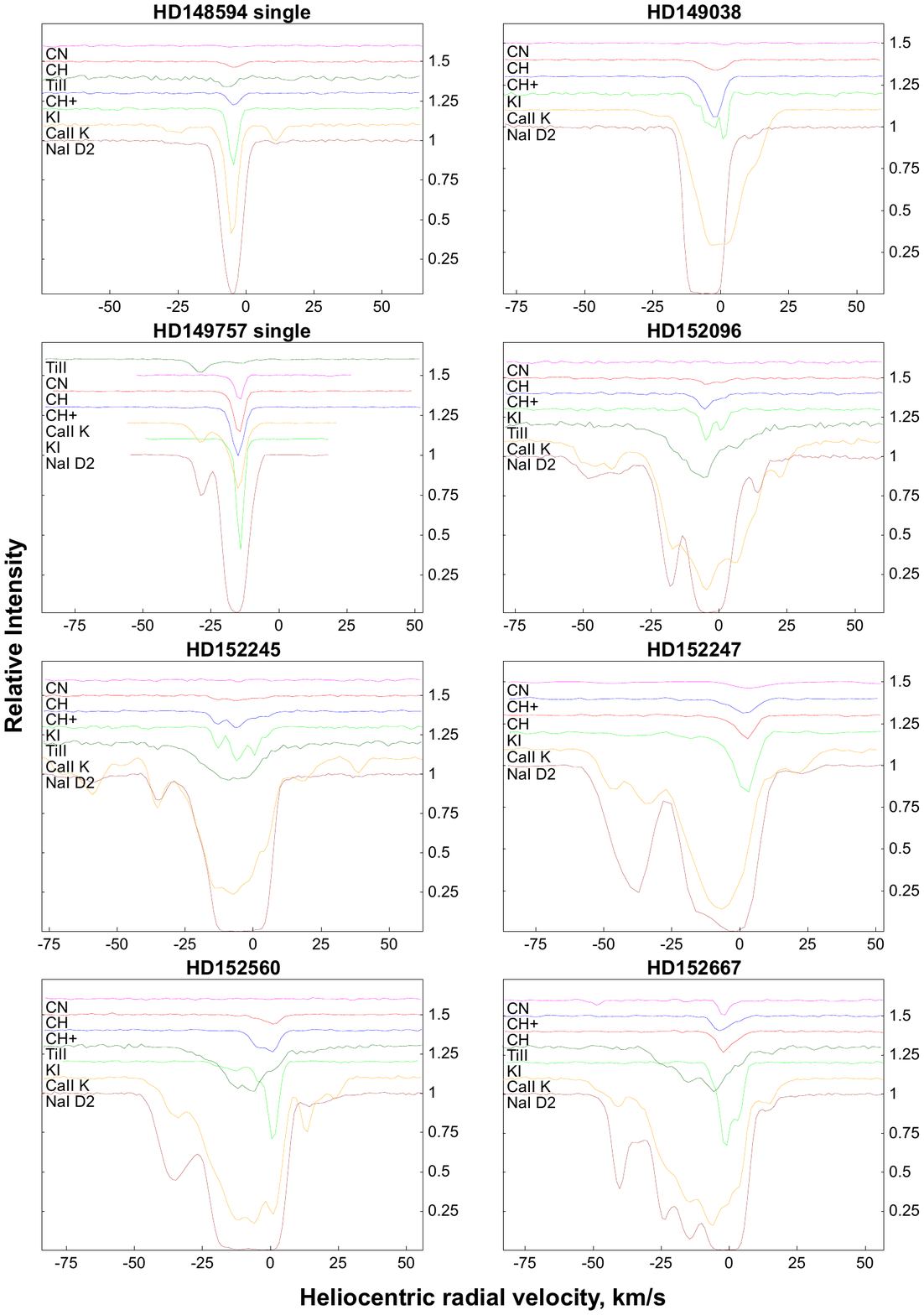}
  \caption{Velocity profiles of
    Ti\,{\sc ii} 3383.8\,\AA\,(dark green), CN 3874.6\, \AA\,
    (magenta), CH 4300.3\,\AA\, (red), CH$^+$ 4232.5\,\AA\, (blue),
    K\,{\sc i} 7699\,\AA\ (green), Ca\,{\sc ii} K 3933.7\,\AA\,
    (orange), and Na\,{\sc i} D2 5889\,\AA\,
    (brown). {Single-cloud sightlines are marked ``single''.} }
\end{figure*}

\clearpage

\begin{figure*} [!htbp] 
  \includegraphics[width=18.5cm,clip=true,trim=0cm 3cm 0cm 0.cm]{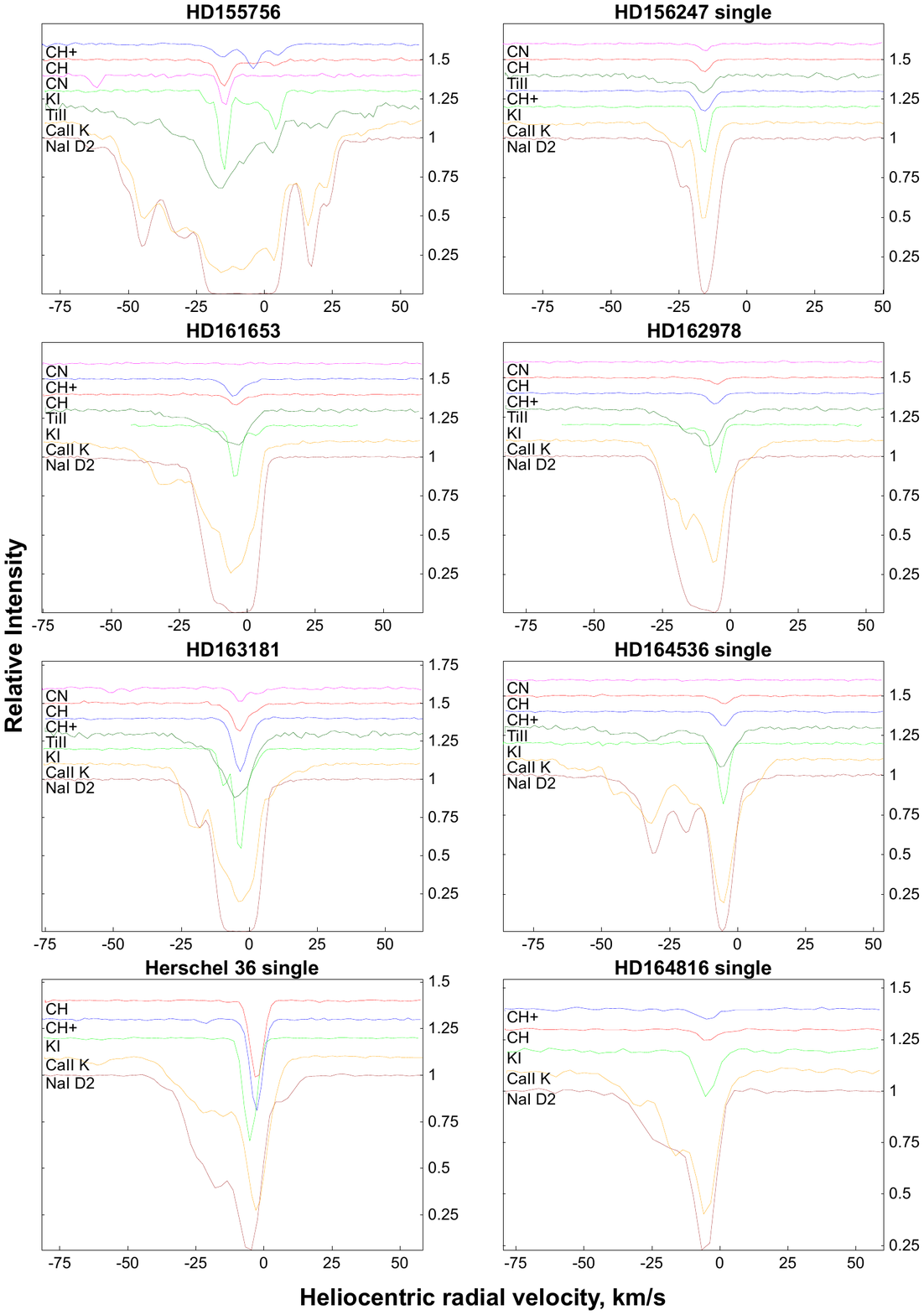}
  \caption{Velocity profiles of
    Ti\,{\sc ii} 3383.8\,\AA\,(dark green), CN 3874.6\, \AA\,
    (magenta), CH 4300.3\,\AA\, (red), CH$^+$ 4232.5\,\AA\, (blue),
    K\,{\sc i} 7699\,\AA\ (green), Ca\,{\sc ii} K 3933.7\,\AA\,
    (orange), and Na\,{\sc i} D2 5889\,\AA\,
    (brown). {Single-cloud sightlines are marked ``single''.} }
\end{figure*}

\begin{figure*} [!htbp] 
  \includegraphics[width=18.5cm,clip=true,trim=0cm 3cm 0cm 0.cm]{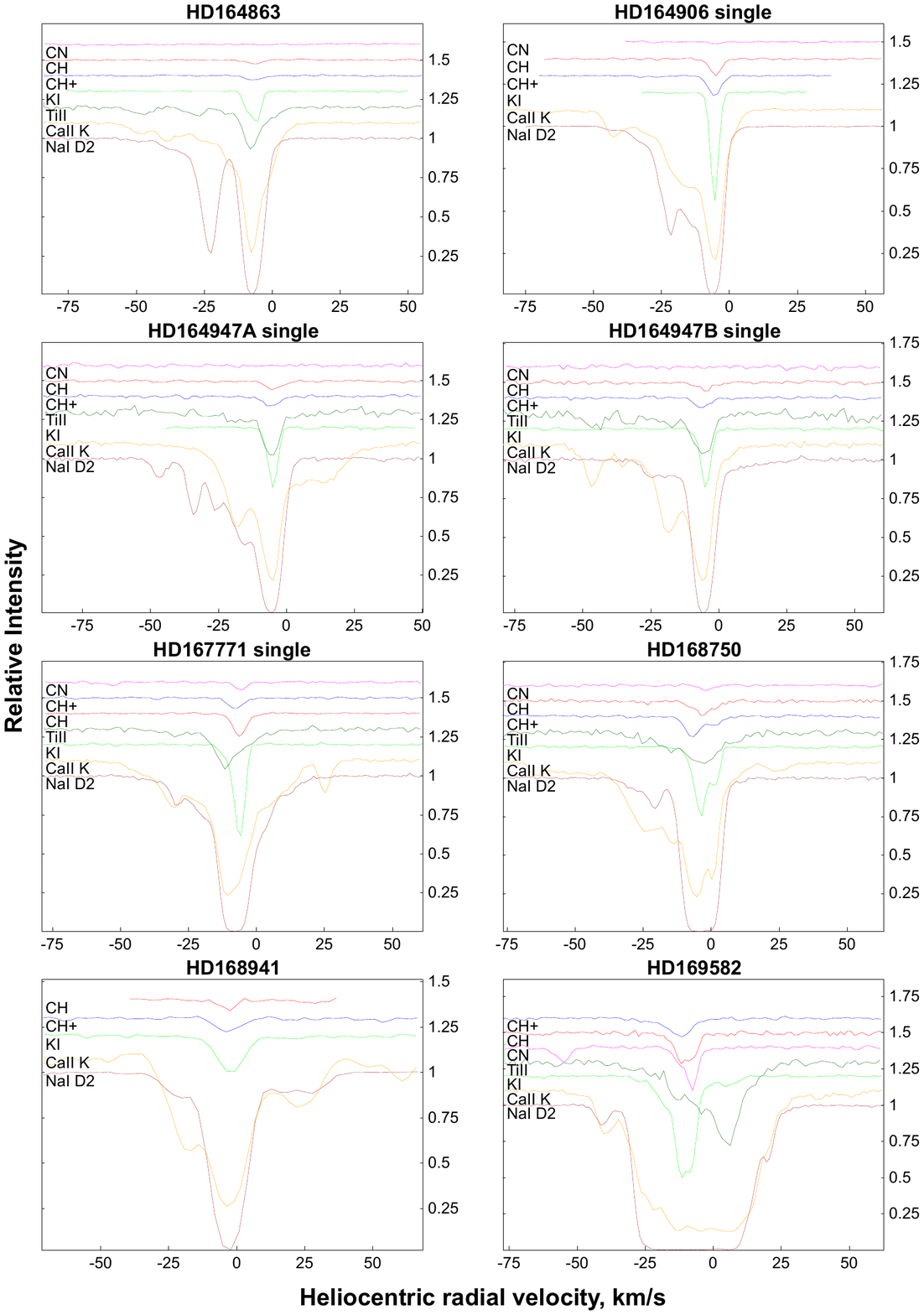}
  \caption{Velocity profiles of
    Ti\,{\sc ii} 3383.8\,\AA\,(dark green), CN 3874.6\, \AA\,
    (magenta), CH 4300.3\,\AA\, (red), CH$^+$ 4232.5\,\AA\, (blue),
    K\,{\sc i} 7699\,\AA\ (green), Ca\,{\sc ii} K 3933.7\,\AA\,
    (orange), and Na\,{\sc i} D2 5889\,\AA\,
    (brown). {Single-cloud sightlines are marked ``single''.} }
\end{figure*}

\begin{figure*} [!htbp] 
  \includegraphics[width=18.5cm,clip=true,trim=0cm 3cm 0cm 0.cm]{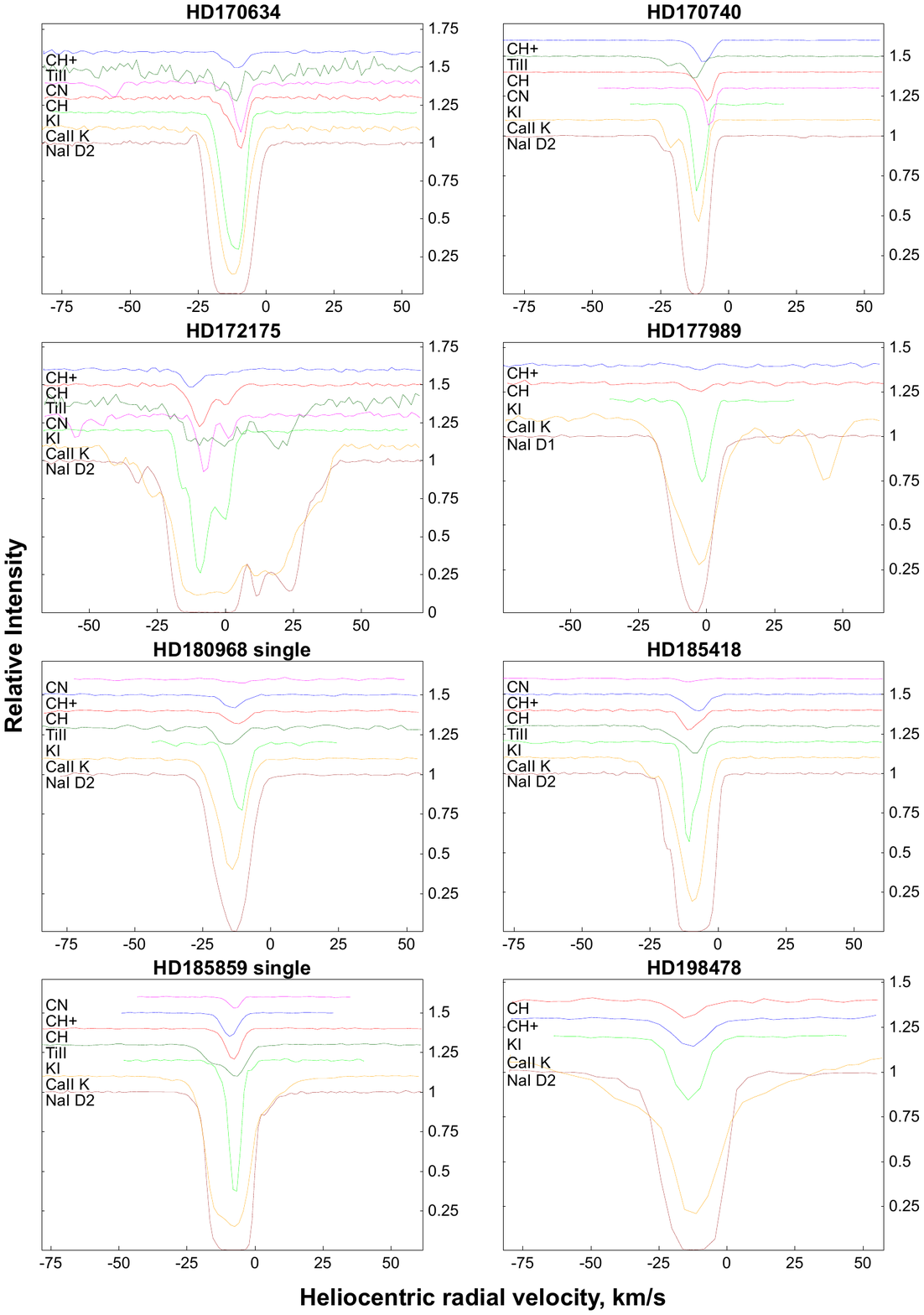}
  \caption{Velocity profiles of
    Ti\,{\sc ii} 3383.8\,\AA\,(dark green), CN 3874.6\, \AA\,
    (magenta), CH 4300.3\,\AA\, (red), CH$^+$ 4232.5\,\AA\, (blue),
    K\,{\sc i} 7699\,\AA\ (green), Ca\,{\sc ii} K 3933.7\,\AA\,
    (orange), and Na\,{\sc i} D2 5889\,\AA\,
    (brown). {Single-cloud sightlines are marked ``single''.} }
\end{figure*}

\begin{figure*} [!htbp] 
  \includegraphics[width=18.5cm,clip=true,trim=0cm 3cm 0cm 0.cm]{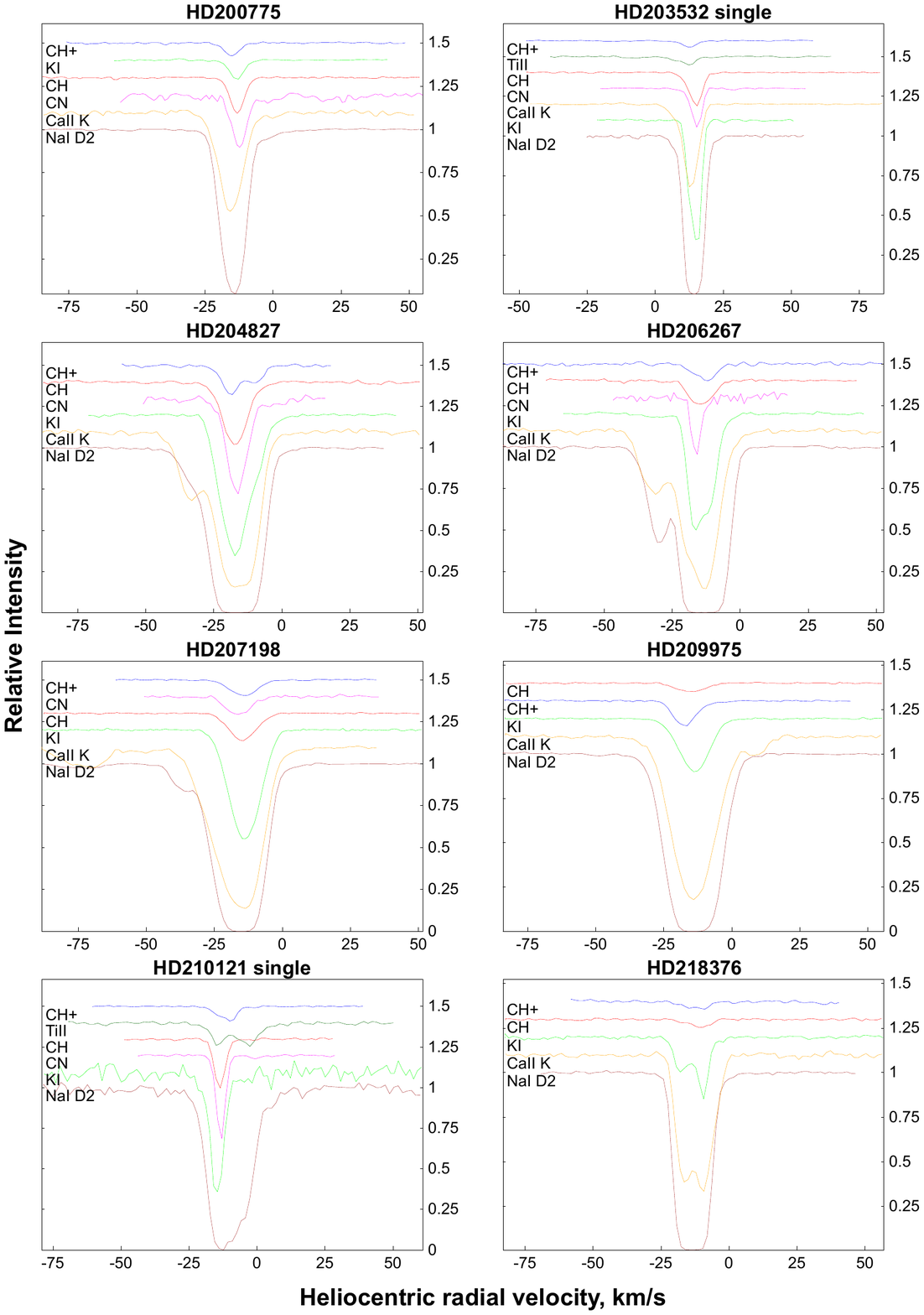}
  \caption{Velocity profiles of
    Ti\,{\sc ii} 3383.8\,\AA\,(dark green), CN 3874.6\, \AA\,
    (magenta), CH 4300.3\,\AA\, (red), CH$^+$ 4232.5\,\AA\, (blue),
    K\,{\sc i} 7699\,\AA\ (green), Ca\,{\sc ii} K 3933.7\,\AA\,
    (orange), and Na\,{\sc i} D2 5889\,\AA\,
    (brown). {Single-cloud sightlines are marked ``single''.} }
\end{figure*}

\begin{figure*} [!htbp] 
  \includegraphics[width=18.5cm,clip=true,trim=0cm 3cm 0cm 0.cm]{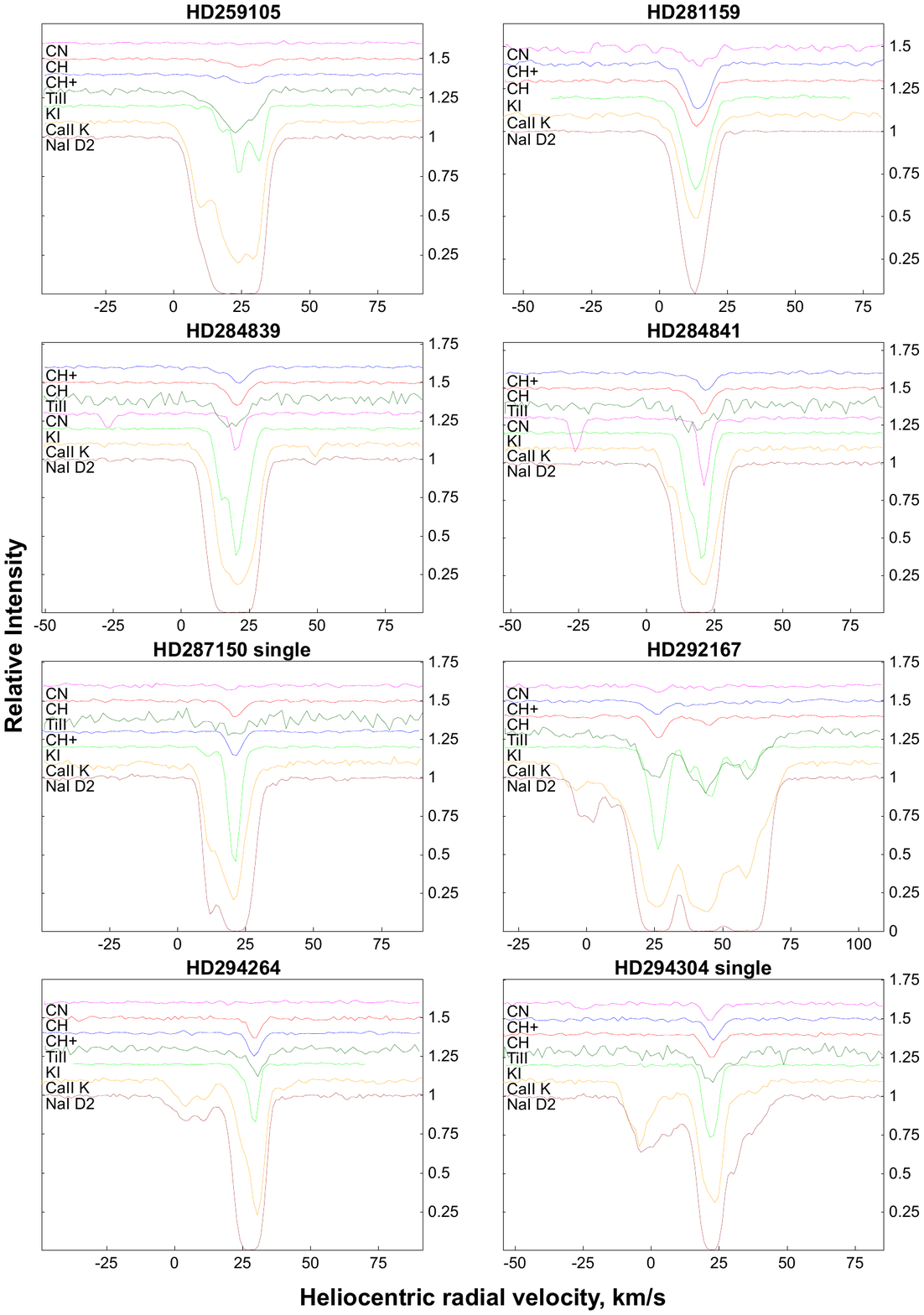}
  \caption{Velocity profiles of
    Ti\,{\sc ii} 3383.8\,\AA\,(dark green), CN 3874.6\, \AA\,
    (magenta), CH 4300.3\,\AA\, (red), CH$^+$ 4232.5\,\AA\, (blue),
    K\,{\sc i} 7699\,\AA\ (green), Ca\,{\sc ii} K 3933.7\,\AA\,
    (orange), and Na\,{\sc i} D2 5889\,\AA\,
    (brown). {Single-cloud sightlines are marked ``single''.} }
\end{figure*}

\begin{figure*} [!htbp] 
  \includegraphics[width=18.5cm,clip=true,trim=0cm 3cm 0cm 0.cm]{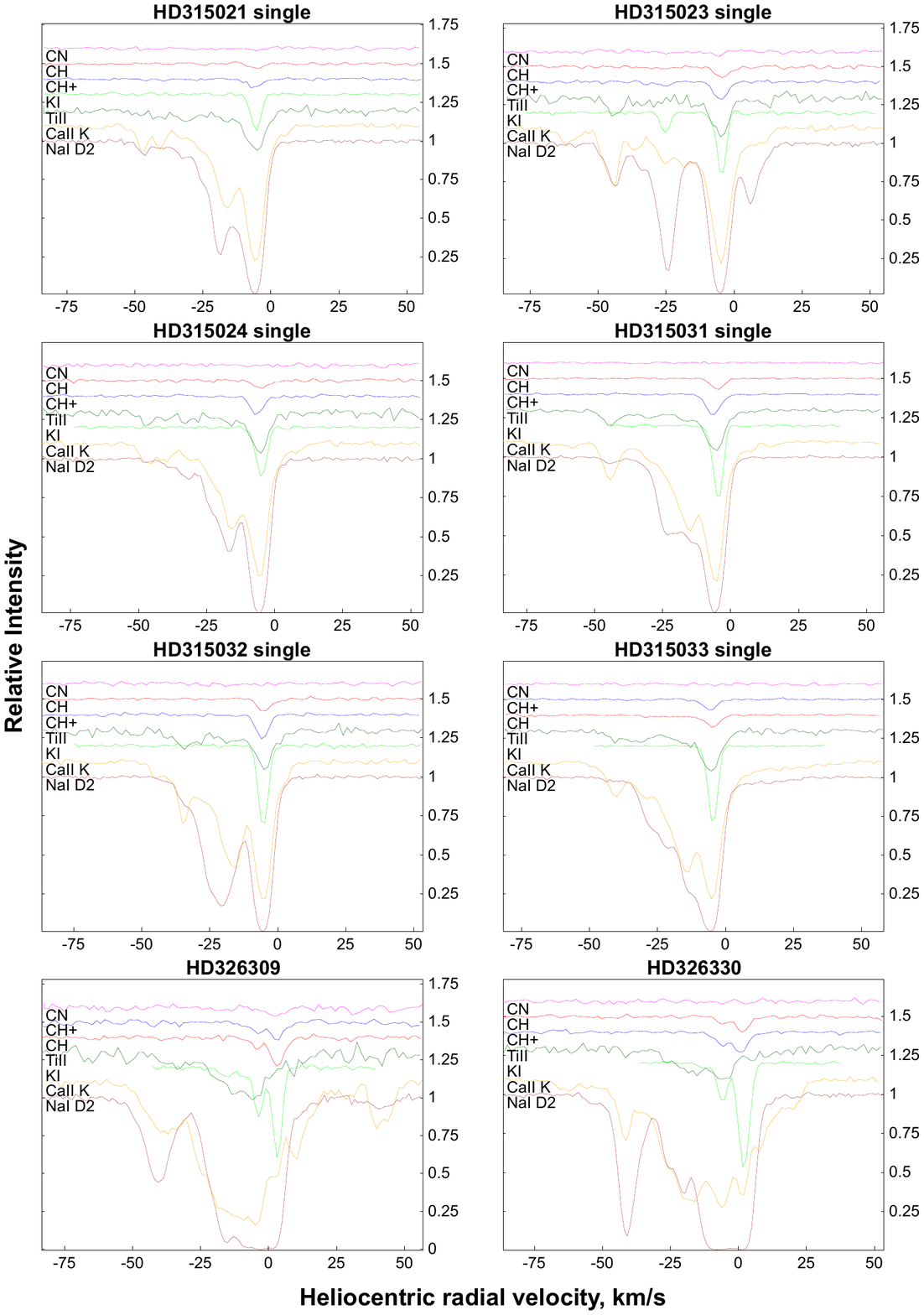}
  \caption{Velocity profiles of
    Ti\,{\sc ii} 3383.8\,\AA\,(dark green), CN 3874.6\, \AA\,
    (magenta), CH 4300.3\,\AA\, (red), CH$^+$ 4232.5\,\AA\, (blue),
    K\,{\sc i} 7699\,\AA\ (green), Ca\,{\sc ii} K 3933.7\,\AA\,
    (orange), and Na\,{\sc i} D2 5889\,\AA\,
    (brown). {Single-cloud sightlines are marked ``single''.} }
\end{figure*}

\end{appendix}

\end{document}